\pdfoutput=1
\documentclass[prd,aps,twocolumn,superscriptaddress,showpacs,showkeys,floatfix]{revtex4-2}

\usepackage{lineno, blindtext}
\usepackage{graphicx}
\usepackage{amsmath}
\usepackage{amsmath}
\usepackage{enumerate}
\usepackage{placeins}
\usepackage{appendix}
\usepackage[utf8]{inputenc}
\usepackage{chngcntr}
\usepackage{microtype}
\usepackage{color}
\usepackage{epstopdf}
\usepackage{colortbl}
\definecolor{darkred}{rgb}{0.5,0,0}
\definecolor{darkblue}{rgb}{0,0,0.5}
\definecolor{firebrick}{rgb}{0.75,0.125,0.125}
\definecolor{darkgreen}{rgb}{0,0.5,0}

\usepackage{booktabs}
\usepackage{tabularx}
\usepackage{dcolumn}
\usepackage{bm}
\usepackage{xspace} 
\usepackage{url}
\usepackage{lmodern} 
\usepackage[T1]{fontenc} 
\usepackage[utf8]{inputenc} 
\usepackage[caption=false]{subfig}

\usepackage{colortbl}
\definecolor{darkred}{rgb}{0.5,0,0}
\definecolor{darkblue}{rgb}{0,0,0.5}
\definecolor{firebrick}{rgb}{0.75,0.125,0.125}
\definecolor{darkgreen}{rgb}{0,0.5,0}

\usepackage[colorlinks=true,linkcolor=firebrick,citecolor=darkgreen,urlcolor=darkblue]{hyperref}

\usepackage{blindtext}

\usepackage{lineno}

\begin{document}


\newcommand{\eV}{\ensuremath{\mbox{e\kern-0.1em V}}\xspace}
\newcommand{\GeV}{\ensuremath{\mbox{Ge\kern-0.1em V}}\xspace}
\newcommand{\MeV}{\ensuremath{\mbox{Me\kern-0.1em V}}\xspace}
\newcommand{\GeVc}{\ensuremath{\mbox{Ge\kern-0.1em V}\!/\!c}\xspace}
\newcommand{\GeVcc}{\ensuremath{\mbox{Ge\kern-0.1em V}\!/\!c^2}\xspace}
\newcommand{\AGeV}{\ensuremath{A\,\mbox{Ge\kern-0.1em V}}\xspace}
\newcommand{\AGeVc}{\ensuremath{A\,\mbox{Ge\kern-0.1em V}\!/\!c}\xspace}
\newcommand{\MeVc}{\ensuremath{\mbox{Me\kern-0.1em V}/c}\xspace}
\newcommand{\T}{\ensuremath{\mbox{T}}\xspace}
\newcommand{\cmsq}{\ensuremath{\mbox{cm}^2}\xspace}
\newcommand{\msq}{\ensuremath{\mbox{m}^2}\xspace}
\newcommand{\cm}{\ensuremath{\mbox{cm}}\xspace}
\newcommand{\mm}{\ensuremath{\mbox{mm}}\xspace}
\newcommand{\micron}{\ensuremath{\mu\mbox{m}}\xspace}
\newcommand{\mrad}{\ensuremath{\mbox{mrad}}\xspace}
\newcommand{\ns}{\ensuremath{\mbox{ns}}\xspace}
\newcommand{\m}{\ensuremath{\mbox{m}}\xspace}
\newcommand{\s}{\ensuremath{\mbox{s}}\xspace}
\newcommand{\ms}{\ensuremath{\mbox{ms}}\xspace}
\newcommand{\ps}{\ensuremath{\mbox{ps}}\xspace}
\newcommand{\dd}{\ensuremath{{\mathrm d}}\xspace}
\newcommand{\dedx}{\ensuremath{\dd E\!/\!\dd x}\xspace}
\newcommand{\tof}{\ensuremath{\textup{\emph{tof}}}\xspace}
\newcommand{\pt}{\ensuremath{p_{\rm T}}\xspace}
\newcommand{\PT}{\ensuremath{P_\textup{T}}\xspace}
\newcommand{\mt}{\ensuremath{m_{\rm T}}\xspace}

\newcommand{\pbar}{\ensuremath{\overline{\textup{p}}}\xspace}
\newcommand{\pim}{\ensuremath{\pi^-}\xspace}
\newcommand{\pip}{\ensuremath{\pi^+}\xspace}
\newcommand{\km}{\ensuremath{\textup{K}^-}\xspace}
\newcommand{\kp}{\ensuremath{\textup{K}^+}\xspace}

\newcommand{\FlukaLong}{{\scshape Fluka2008}\xspace}
\newcommand{\FlukaEleven}{{\scshape Fluka2011}\xspace}
\newcommand{\GiBUU}{{\scshape GiBUU}\xspace}
\newcommand{\GiBUULong}{{\scshape GiBUU1.6}\xspace}
\newcommand{\FlukaNewLong}{{\scshape Fluka2011.2\_17}\xspace}
\newcommand{\Geant}{{\scshape Geant}\xspace}
\newcommand{\GeantThree}{{\scshape Geant3}\xspace}
\newcommand{\GeantFour}{{\scshape Geant4}\xspace}

\def\red#1{{\color{red}#1}}
\def\avg#1{\langle{#1}\rangle}

\title{Measurements of hadron production in 90 GeV/\emph{c} proton-carbon interactions}




\affiliation{National Nuclear Research Center, Baku, Azerbaijan}
\affiliation{Faculty of Physics, University of Sofia, Sofia, Bulgaria}
\affiliation{LPNHE, Sorbonne University, CNRS/IN2P3, Paris, France}
\affiliation{Karlsruhe Institute of Technology, Karlsruhe, Germany}
\affiliation{HUN-REN Wigner Research Centre for Physics, Budapest, Hungary}
\affiliation{E\"otv\"os Lor\'and University, Budapest, Hungary}
\affiliation{Institute for Particle and Nuclear Studies, Tsukuba, Japan}
\affiliation{Okayama University, Okayama, Japan}
\affiliation{University of Bergen, Bergen, Norway}
\affiliation{University of Oslo, Oslo, Norway}
\affiliation{Jan Kochanowski University, Kielce, Poland}
\affiliation{Institute of Nuclear Physics, Polish Academy of Sciences, Cracow, Poland}
\affiliation{National Centre for Nuclear Research, Warsaw, Poland}
\affiliation{Jagiellonian University, Cracow, Poland}
\affiliation{AGH - University of Krakow, Krakow, Poland}
\affiliation{University of Silesia, Katowice, Poland}
\affiliation{University of Warsaw, Warsaw, Poland}
\affiliation{University of Wroc{\l}aw,  Wroc{\l}aw, Poland}
\affiliation{Warsaw University of Technology, Warsaw, Poland}
\affiliation{Affiliated with an institution covered by a cooperation
agreement with CERN}
\affiliation{University of Belgrade, Belgrade, Serbia}
\affiliation{Fermilab, Batavia, USA}
\affiliation{University of Notre Dame, Notre Dame, USA}
\affiliation{University of Colorado, Boulder, USA}
\affiliation{University of Hawaii at Manoa, Honolulu, USA}
\affiliation{University of Pittsburgh, Pittsburgh, USA}

\author{H.~\surname{Adhikary}}
\affiliation{Jan Kochanowski University, Kielce, Poland}
\author{P.~\surname{Adrich}}
\affiliation{National Centre for Nuclear Research, Warsaw, Poland}
\author{K.K.~\surname{Allison}}
\affiliation{University of Colorado, Boulder, USA}
\author{N.~\surname{Amin}}
\affiliation{Karlsruhe Institute of Technology, Karlsruhe, Germany}
\author{E.V.~\surname{Andronov}}
\affiliation{Affiliated with an institution covered by a cooperation agreement with CERN}
\author{I.-C.~\surname{Arsene}}
\affiliation{University of Oslo, Oslo, Norway}
\author{M.~\surname{Bajda}}
\affiliation{Jagiellonian University, Cracow, Poland}
\author{Y.~\surname{Balkova}}
\affiliation{University of Silesia, Katowice, Poland}
\author{D.~\surname{Battaglia}}
\affiliation{University of Notre Dame, Notre Dame, USA}
\author{A.~\surname{Bazgir}}
\affiliation{Jan Kochanowski University, Kielce, Poland}
% ...

\collaboration{NA61/SHINE Collaboration}
\noaffiliation

\begin{abstract}
This paper presents the multiplicity of neutral and charged hadrons produced in 90 GeV$/c$ proton-carbon interactions from a dataset taken by the NA61/SHINE experiment in 2017. Particle identification via \dedx was performed for the charged hadrons
$\pi^\pm$, $K^\pm$, and $p / \bar{p}$; the neutral hadrons $K^0_S$, 
$\Lambda$, and $\bar{\Lambda}$ were identified via an invariant mass analysis of their decays to charged hadrons.
Double-differential multiplicity results as a function of laboratory momentum and polar angle are presented for each particle species; these results provide vital constraints on the predicted neutrino beam flux for current and future long-baseline neutrino oscillation experiments.
\end{abstract} 

\maketitle

\section{Introduction}
\label{sec:introduction}

During the creation of a neutrino beam at long-baseline neutrino oscillation experiments, hadronic interactions between primary beam protons and atomic nuclei create neutral and charged hadrons in the beamline. These hadrons can then decay into neutrinos or re-interact with beamline material into neutrino-producing particles.

The Neutrinos at the Main Injector (NuMI) facility creates a neutrino beam by striking a carbon target with a 120 GeV$/c$ proton beam \cite{numibeamline}, and the Long-Baseline Neutrino Facility (LBNF), which will provide the neutrino beam for the Deep Underground Neutrino Experiment (DUNE)~\cite{dune_facility}, will most likely use the same primary interaction as NuMI \cite{dune_physics}. For experiments like NuMI Off-axis $\nu_{e}$ Appearance (NOvA) \cite{patterson2013nova} and DUNE, understanding the initial hadron production in the creation of their neutrino beam is a critical component of estimating the neutrino beam flux; an accurate estimate of the neutrino beam flux is necessary for precisely measuring neutrino flavor oscillation, cross sections, and any other results from these long-baseline neutrino oscillation experiments.

While the primary interaction for long-baseline neutrino oscillation experiments starting at Fermilab is between a 120 GeV$/c$ proton and a carbon nucleus, secondary and tertiary interactions occurring inside and outside the target volume contribute significantly to the neutrino beam flux \cite{aliaga2016neutrino}. Measuring hadronic production for these secondary and tertiary interactions, like 90 GeV$/c$ proton-carbon interactions, will enable more accurate predictions of the neutrino beam flux.

The NA61/SPS Heavy Ion and Neutrino Experiment (NA61/SHINE)\cite{abgrall2014na61} is a fixed-target experiment located at the North Area of the CERN Super Proton Synchrotron (SPS). NA61/SHINE has a dedicated program of hadron production measurements relevant to long-baseline neutrino oscillation experiments, and previous measurements at NA61/SHINE have significantly improved neutrino beam flux estimations for the T2K experiment \cite{abgrall2011measurements, abgrall2012measurement, abgrall2014measurements, abgrall2015measurements, abgrall2013pion, abgrall2016measurements, abgrall2019measurements, acharya2021measurement}, which uses a primary proton beam with a momentum of 31 GeV$/c$. NA61/SHINE has also published several papers on hadron production reactions at higher energies relevant to Fermilab neutrino experiments \cite{aduszkiewicz2018measurements, aduszkiewicz2019measurementscrosssection, aduszkiewicz2019measurements, adhikary2023measurementsneutral, adhikary2023measurementscharged}.

In 2017, NA61/SHINE recorded the dataset being analyzed in this manuscript, a 90 GeV$/c$ proton beam on a thin carbon target. The measured differential multiplicities include the important $\nu$- and $\bar{\nu}$-producing reactions $p + \text{C} \rightarrow \pi^{\pm} + X$
and $p + \text{C} \rightarrow K^{\pm} + X$. To constrain the re-interaction of outgoing protons and anti-protons, which can lead to additional neutrinos and anti-neutrinos, the reactions $p + \text{C} \rightarrow p + X$ and $p + \text{C} \rightarrow \bar{p} + X$ are also measured.

A significant number of charged hadrons can originate from the decay of neutral hadrons produced from the initial proton-nucleus interaction. This analysis measures the differential multiplicity of $K^0_S$, $\Lambda$, and $\bar{\Lambda}$ to constrain these decay contributions.

This publication details the process of identifying and measuring the number of neutral and charged hadrons resulting from the interaction of a 90 GeV/$c$ proton with a carbon atom. For each particle species, the number of produced particles of that species are reported in bins of the particle's momentum and angle with respect to the incident proton.  The reported multiplicities have been corrected for detector acceptance.  Particle identification for charged hadrons is performed via energy loss, denoted \dedx, and the main decay mode of each neutral species is used to identify the neutral hadrons: $K^0_S \rightarrow \pi^+ \pi^-$ (69.2\%), $\Lambda \rightarrow p \pi^-$ (64.1\%), and $\bar{\Lambda} \rightarrow \bar{p} \pi^+$ (64.1\%) \cite{pdg2024}. 

This paper is organized as follows. First, Section \ref{sec:na61Detector} describes the experimental setup of the NA61/SHINE detector, and then Section \ref{sec:reconstructionAndSimulation} briefly explains the software needed for track reconstruction and detector simulation. After that, Sections \ref{sec:neutralAnalysis} and \ref{sec:chargedAnalysis} walk through the neutral- and charged-hadron analyses, respectively. Section \ref{sec:systematicUncertainties} explains the calculation of each systematic uncertainty for the analyses, and finally Section \ref{sec:hadronMultiplicityMeasurements} shows sample multiplicity results before Section \ref{sec:summary} summarizes the paper.


\section{The NA61/SHINE Detector}
\label{sec:na61Detector}

\begin{figure*}[t!]
\centering
\includegraphics[width=.9\textwidth]{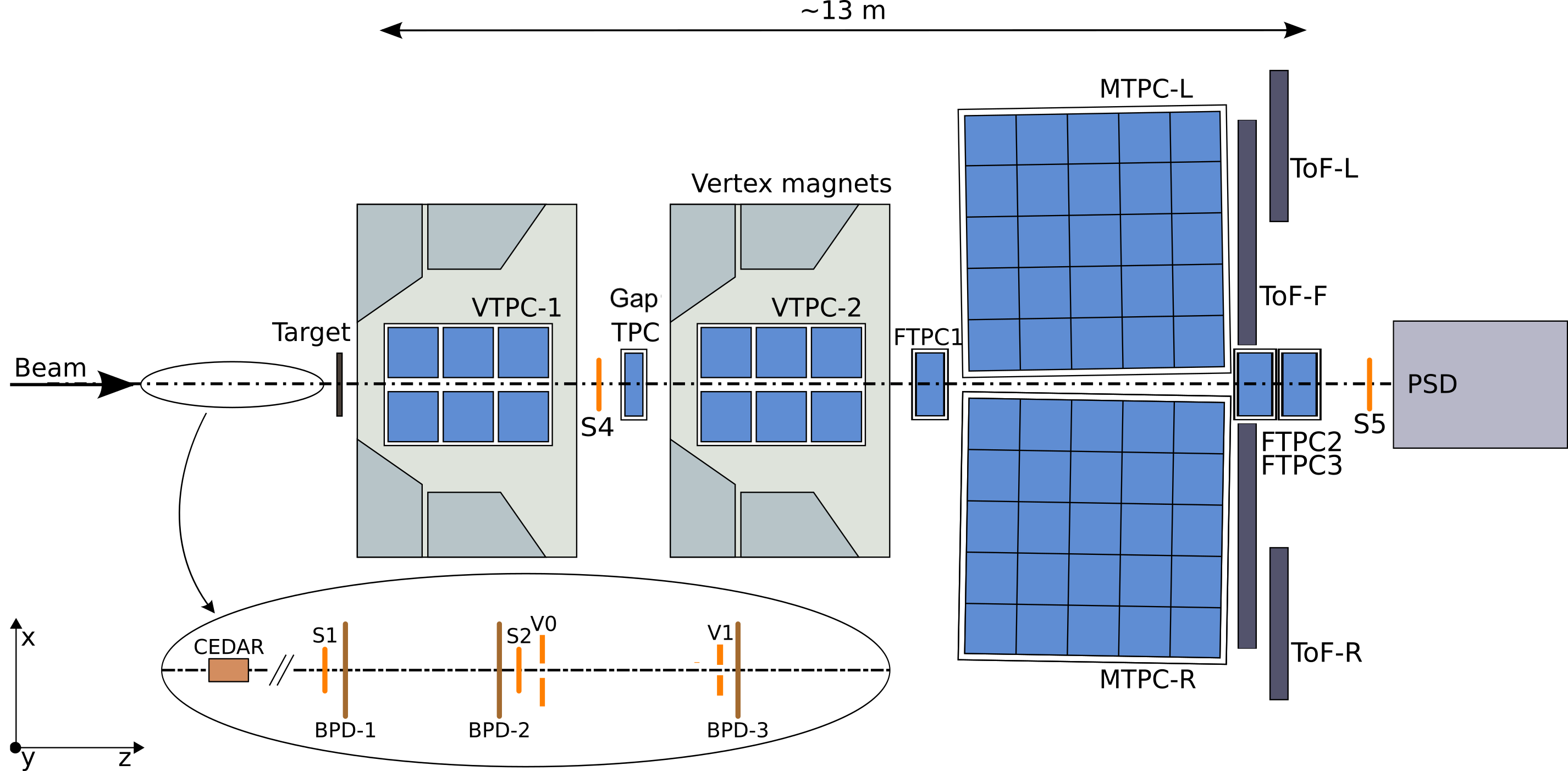}

\caption{Top-view schematic layout of the NA61/SHINE detector at the CERN SPS \cite{abgrall2014na61} showing the components present in the 2017 90 GeV$/c$ proton-carbon data taking. The detector configuration upstream of the target is shown in the inset. The alignment of the chosen coordinate system is shown on the plot; its origin $(x=y=z=0)$ lies in the middle of VTPC-2, on the beam axis. The nominal beam direction is along the $z$-axis.  The target is placed at $z=-590.20$ cm. The magnetic field bends charged particle trajectories in the $x$-$z$ (horizontal) plane. The drift direction in the TPCs is along the (vertical) $y$-axis.}
\label{fig:na61schematic}
\end{figure*}

NA61/SHINE is a large-acceptance hadron spectrometer \cite{abgrall2014na61}, shown in Figure \ref{fig:na61schematic}, located on the H2 beamline in Experimental Hall North 1 in CERN's North Area complex; the detector coordinate system is visible
in Figure \ref{fig:na61schematic}. Primary 400 GeV$/c$ proton beams are available from the SPS,
as well as ions with momenta in the range [13$A$ -- 158$A$] GeV$/c$; additionally, directing the primary protons into a production target provides secondary hadron beams with momenta 13--350 GeV$/c$. As the secondary hadron beams contain a mixture of hadrons and leptons, beam particle identification at the event level is necessary. 

The NA61/SHINE triggering system uses the Cerenkov Differential Counter with Achromatic Ring Focus (CEDAR) \cite{bovet1978cedar, bovet1982cedar}, located upstream of the NA61/SHINE spectrometer, for beam particle identification. In addition to the CEDAR detector, two scintillator counters, S1 and S2, and two veto scintillators, V0 and V1, select identified beam particles with acceptable trajectories; the veto scintillators have cylindrical holes centered on the beam. If a signal is detected in a veto counter the beam particle is rejected. The final part of the triggering system is the S4 scintillator, which is placed downstream of the target. With a radius of 1 cm, the S4 scintillator provides information on the angular scatter of the beam particle from possible interactions inside the target. For the 2017 90 GeV$/c$ proton-carbon dataset, there were four main trigger labels:

\renewcommand{\theenumi}{\roman{enumi}}%
\begin{enumerate}
\item T1 (identified beam particle): CEDAR $\cdot$ S1 $\cdot$ S2 $\cdot$ $\overline{\text{V0}}$ $\cdot$ $\overline{\text{V1}}$,
\item T2 (identified beam particle interaction):
  CEDAR $\cdot$ S1 $\cdot$ S2 $\cdot$ $\overline{\text{V0}}$ $\cdot$
  $\overline{\text{V1}}$ $\cdot$ $\overline{\text{S4}}$,
\item T3 (unidentified beam particle): S1 $\cdot$ S2 $\cdot$ $\overline{\text{V0}}$ $\cdot$ $\overline{\text{V1}}$,
\item T4 (unidentified beam particle interaction): S1 $\cdot$ S2 $\cdot$ $\overline{\text{V0}}$ $\cdot$ $\overline{\text{V1}}$ $\cdot$ $\overline{\text{S4}}$.
\end{enumerate}

Before the target, three Beam Position Detectors (BPDs), gaseous strip detectors, measure incoming beam particle trajectories. Placed 29.5 m (BPD-1), 8.2 m (BPD-2), and 0.7 m (BPD-3) upstream of the target, a straight line fit to the BPD measurements represents the incoming beam trajectory.

After the triggering system, eight Time Projection Chambers (TPCs) provide tracking of charged particles as well as energy loss (\dedx) measurements; the \dedx information is used to identify charged hadrons on a track-by-track basis. Two of the TPCs, Vertex TPC-1 and Vertex TPC-2 (VTPC-1 and VTPC-2), are placed inside the vertex magnets, which provide bending power up to 9 T$\cdot$m and enable track momentum measurements. The Gap TPC (GTPC) and the three Forward TPCs (FTPCs) \cite{rumberger2020forward} measure forward-going tracks passing through the gap between the VTPCs and the Main TPCs (MTPCs).

A Time-of-Flight (ToF) system provides particle identification via mass determination in select regions of phase space; the ToF-Forward (ToF-F) was not used in the analysis as it was recently installed and was still in its commissioning phase during the taking of the 90 GeV$/c$ proton-carbon data;
the S5 counter and the Projectile-Spectator-Detector (PSD) were
also not used in the analysis.

For the study of 90 GeV$/c$ proton-carbon interactions, a thin carbon target with dimensions of 25 mm (W) $\times$ 25 mm (H) $\times$ 14.8 mm (L) (3.1\% interaction length) and density $\rho = 1.80 \pm 0.01$ g/cm$^3$ was used.
Data were also collected with the target removed from the beamline, in order to subtract background from interactions occurring outside the target volume. See Table \ref{tb:selectedEventCounts} for the number of T2 triggers recorded.

\section{Data Reconstruction and Simulation}
\label{sec:reconstructionAndSimulation}

\begin{figure*}[t!]
\centering
\includegraphics[width=.9\textwidth]{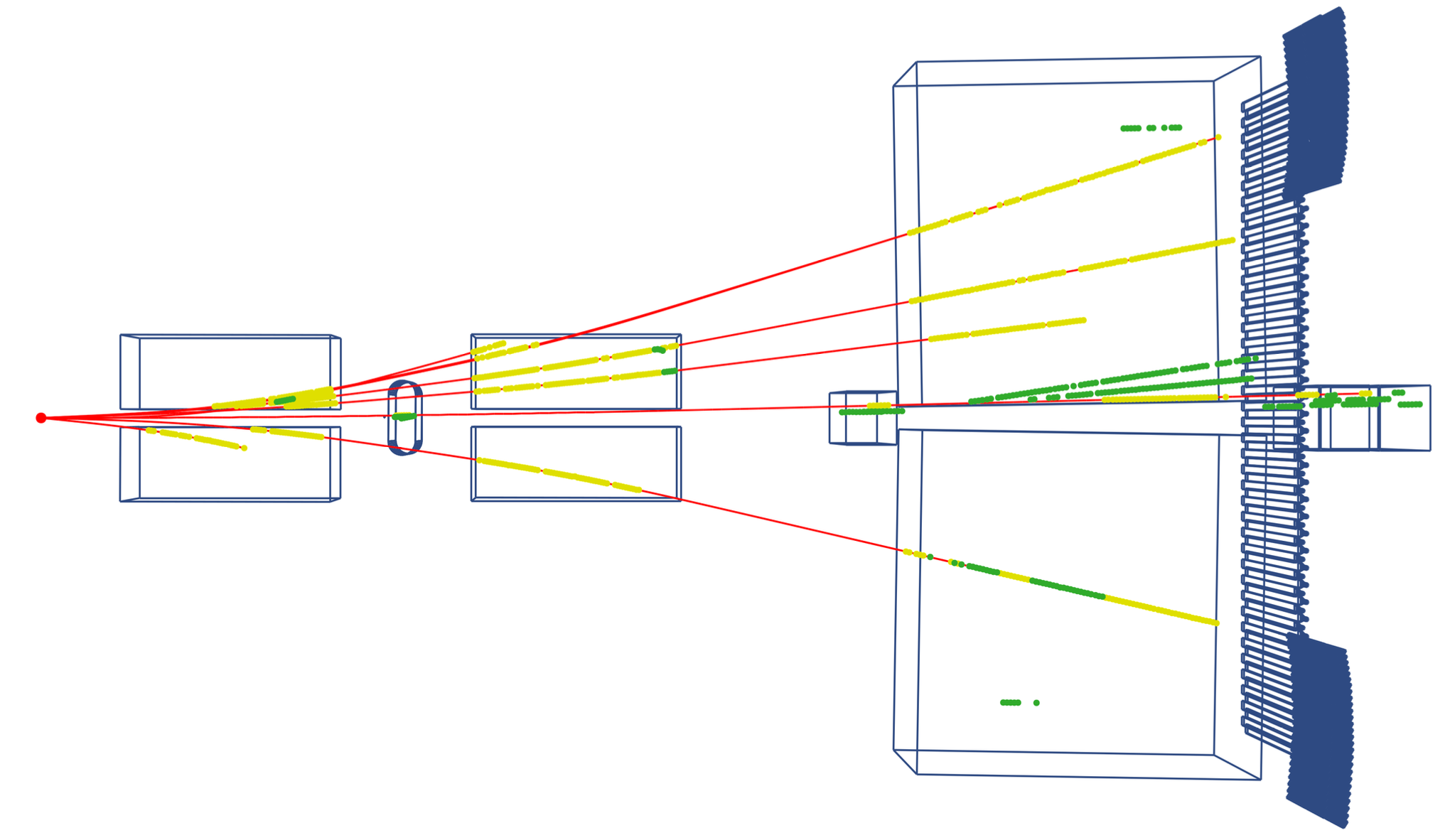}
\caption{Top-view event display of a 90 GeV$/c$ proton-carbon interaction in data in the NA61/SHINE detector.  The green points are hits that are reconstructed in local tracks in a single TPC and the yellow points are hits that are reconstructed in global tracks.  The red lines show the trajectories of global tracks that reconstruct at the main vertex in the thin target.}
\label{fig:na61evdisplay}
\end{figure*}

In each interaction, individual charged particle tracks are reconstructed in three-dimensional space in the NA61/SHINE detector.  An example of an event display showing a single reconstructed 90 GeV$/c$ proton-carbon interaction is shown in Fig.~\ref{fig:na61evdisplay}.  The neutral-hadron analysis relies on the NA61/SHINE Legacy Reconstruction Chain \cite{sipos2012offline}, which includes a $V^0$ finder and a Minuit-based $V^0$ fitter; the inclusion of the newly installed FTPCs for the charged-hadron analysis necessitated new track reconstruction software. The reconstruction with the FTPCs uses a Cellular-Automaton-based track seeding algorithm and a Kalman Filter track fitter \cite{adhikary2023measurementscharged}.

A \GeantFour-based \cite{agostinelli2003geant4, allison2006geant4, allison2016recent} package simulates the passage of particles through the NA61/SHINE detectors and the detector response, and it is also used to evaluate reconstruction efficiency and detector acceptance. Reconstructing simulated events forms the basis for the Monte Carlo (MC) corrections, as described in Section \ref{sec:neutralMonteCarloCorrections}; the nominal MC corrections were calculated with \GeantFour version 10.7.0, using the FTFP\_BERT physics list.

\section{Neutral Hadron Analysis}
\label{sec:neutralAnalysis}

\subsection{Event Selection}
\label{sec:neutralEventSelection}

There are three event-level cuts used in the selection of neutral and charged tracks; the number of T2 events passing all of the event-level cuts are shown in Table \ref{tb:selectedEventCounts}.

\begin{table*}[htbp]
\centering
\begin{tabular}{ccc}
& Target-Inserted & Target-Removed \\
\hline
Recorded & 2.2 M & 0.16 M  \\
Selected & 1.5 M & 0.08 M
\end{tabular}
\caption{The number of recorded and selected target-inserted and target-removed T2 events.}
\label{tb:selectedEventCounts}
\end{table*}

\begin{description}
\item[Beam Divergence Cut (BPD Cut)]

The first cut applied removes events with a beam particle projected to miss the target or the S4 scintillator. Beam tracks with a significant angle will miss the S4 scintillator and cause a false interaction trigger in the absence of any significant proton-target interactions. The BPD cut ensures the trajectory of the beam track is within 0.95 cm of the center of the S4 scintillator.

\item[Well-Measured Beam Trajectory (BPD Status) Cut]

In order to properly measure the beam track trajectory, a cut on the BPD reconstruction is required. The BPD status cut requires a reconstructed cluster in BPD-3 and a convergent straight-line fit with data from at least one more BPD. A cluster in BPD-3 is always required to ensure there was no significant scatter of the beam particle upstream of BPD-3.

\item[Off-Time Beam Particle Cut]

As a second beam particle near the arrival of the triggering beam particle can falsely trigger a non-interaction and produce fake main vertex tracks, events with a secondary beam particle within $\pm 2.5$ \textmu{}s of the primary beam particle are excluded from the analysis. To apply the cut, the Waveform Analyzer (WFA) records the signals from the triggering scintillators near the time of the triggering beam particle.

\end{description}
 
After the event-level cuts, the cuts differ between the neutral- and the charged-hadron analyses. Section \ref{sec:neutralTrackSelection} describes the cuts used to select neutral particles, and Section \ref{sec:chargedTrackSelection} describes the charged-track selection cuts.

\subsection{Selection of Neutral Particle Candidates}
\label{sec:neutralTrackSelection}

In the neutral-hadron analysis, topological and purity selection criteria are applied to improve the sample purity and remove false $V^0$s, where $V^0$ refers
to a neutral-particle candidate.

\begin{description}
\item[$V^0$ Topological Cuts]  Cuts are applied to remove backgrounds to true $V^0$ hadron events.
\begin{enumerate}
\item To remove fake $V^0$ contributions from the primary interaction, selected $V^0$s are required to be separated from the main vertex by at least 3.5 cm, and the distance between the extrapolated $V^0$ track's position at the main vertex and the beam particle's position at the main vertex must be less than 4 cm in $x$ and less than 2 cm in $y$.    
\item For the charged decay tracks to be reconstructed accurately, they each need at least 12 total measurements, known as clusters, in the VTPCs.
\item To reject converted photons, the transverse momentum of the decay in the $V^0$ rest frame is required to be larger than 30 MeV/$c$: $p^+_T + p^-_T > 30$ MeV$/c$. This cut does not significantly affect the $\Lambda$ and $\bar{\Lambda}$ samples.
\item As mentioned earlier, the decay modes $K^0_S \rightarrow \pi^+ \pi^-$ (69.2\%), $\Lambda \rightarrow p \pi^-$ (64.1\%), and $\bar{\Lambda} \rightarrow \bar{p} \pi^+$ (64.1\%) \cite{pdg2024} are used for the selection of the neutral species using \dedx information.
\end{enumerate}

\item[$V^0$ Purity Cuts] After the topological cuts, the remaining cuts are designed to increase the $V^0$ sample purity, and they are specific to each neutral hadron species.  

\renewcommand{\theenumi}{\roman{enumi}}%
\begin{enumerate}
\item The first purity cut restricts the angle formed by the child tracks in the decay frame with respect to the $V^0$ direction of travel ($\theta^{+ *}$ for the positively charged decay product and $\theta^{- *}$ for the negatively charged one). For $K^0_S$ the cuts are $-0.9 < \cos(\theta^{+ *}) < 0.7$ and $-0.7 < \cos(\theta^{- *}) < 0.9$. For $\Lambda$ the allowed ranges are $-0.7 < \cos(\theta^{+ *}) < 0.9$ and $-0.9 < \cos(\theta^{- *}) < 0.7$; the ranges for $\bar{\Lambda}$ are $-0.9 < \cos(\theta^{+ *}) < 0.7$ and $-0.7 < \cos(\theta^{- *}) < 0.9$.

\item The invariant mass of the reconstructed particle, $M = \sqrt{ m_+^2 + m_-^2 + 2 E_+^* E_-^* -2 \vec{p}_+^{\ *} \cdot \vec{p}_-^{\ *}}$, is calculated using the selected decay channel and is restricted for each neutral species. Here $m_{+}$ is the mass of the positively charged particle, and $p_{+}^{\ *}$ and $E_{+}^{\ *}$ are the momentum and energy of the positively charged particle in the decay frame. Likewise $m_{-}$ is the mass of the negatively charged particle, and $p_{-}^{\ *}$ and $E_{-}^{\ *}$ are the momentum and energy of the negatively charged particle in the decay frame. For $K^0_S$ the allowed range is $0.40 < M <  0.65$ GeV$/c^2$, while the range for $\Lambda$ and $\bar{\Lambda}$ is $ 1.09 < M < 1.215 $ GeV$/c^2$.

\item Following the procedure from the analysis of the 2016 and 2017 $p$ + C 120 GeV$/c$ data \cite{adhikary2023measurementscharged}, a cut on the decay products' energy loss requires the energy loss of each child track to be within 15\% of the expected \dedx for the specific child species; this restriction reduces the pion contamination in the $\Lambda$ and $\bar{\Lambda}$ selections without significantly affecting the signal distribution. The uncertainty associated with applying this cut is discussed in Section \ref{sec:systematicUncertainties}.

\item The last purity cut requires the proper lifetime $\tau$ of each particle species to be greater than 0.25 $c \tau_{PDG}$ \cite{pdg2024}.
\end{enumerate}

\end{description}

\begin{figure*}[t]
  \centering
    \includegraphics[width=0.45\textwidth]{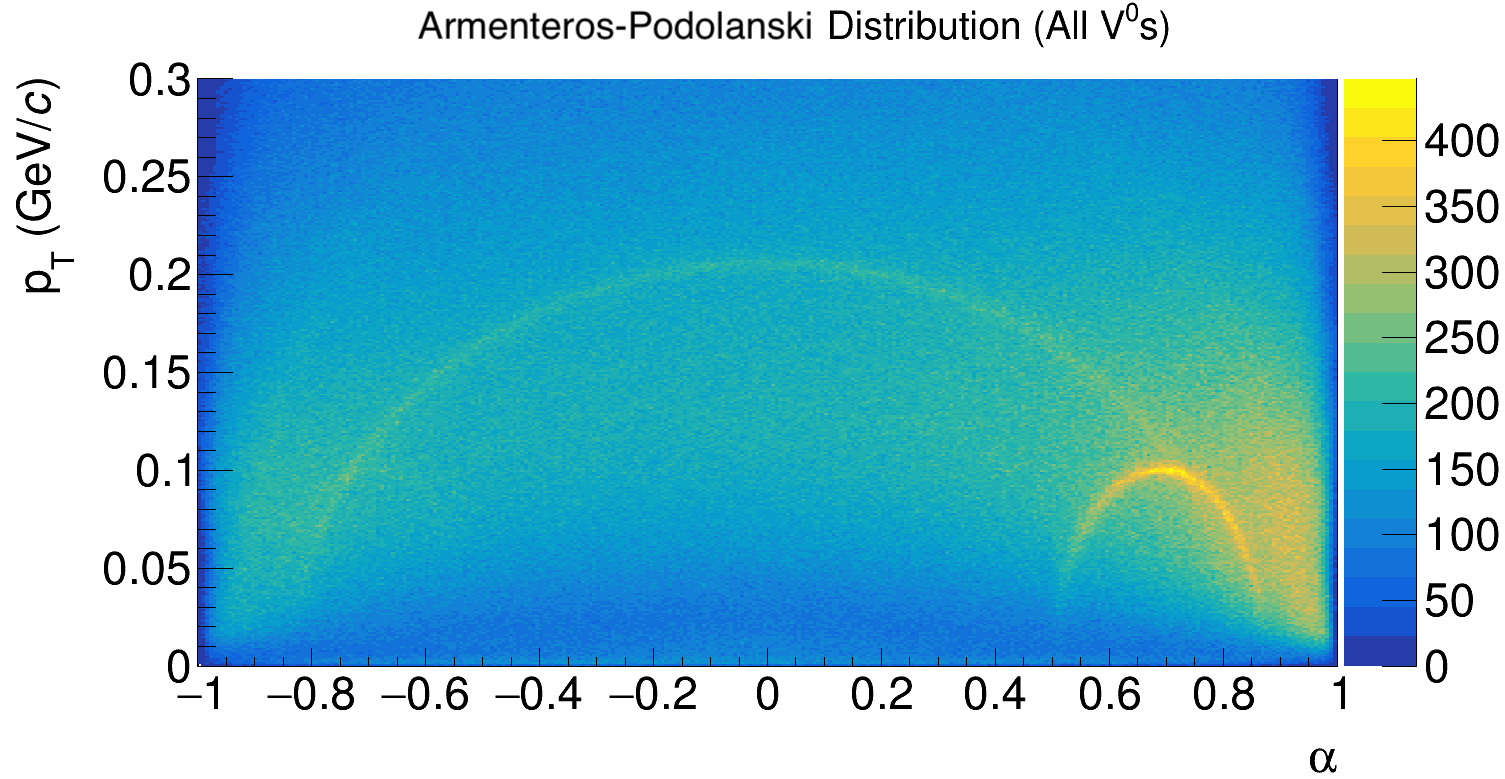}
    \hspace{2em}
    \includegraphics[width=0.45\textwidth]{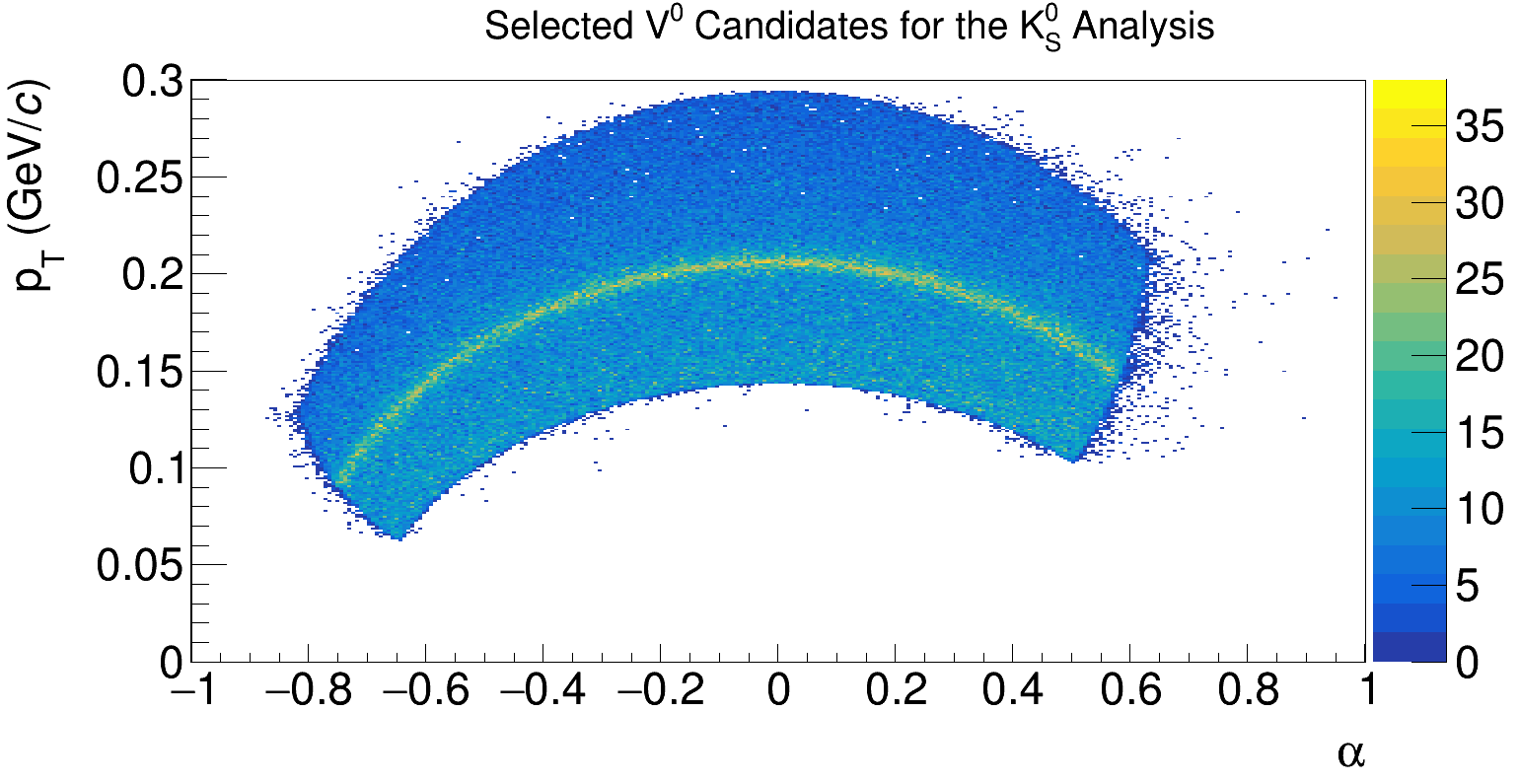} \\
    \includegraphics[width=0.45\textwidth]{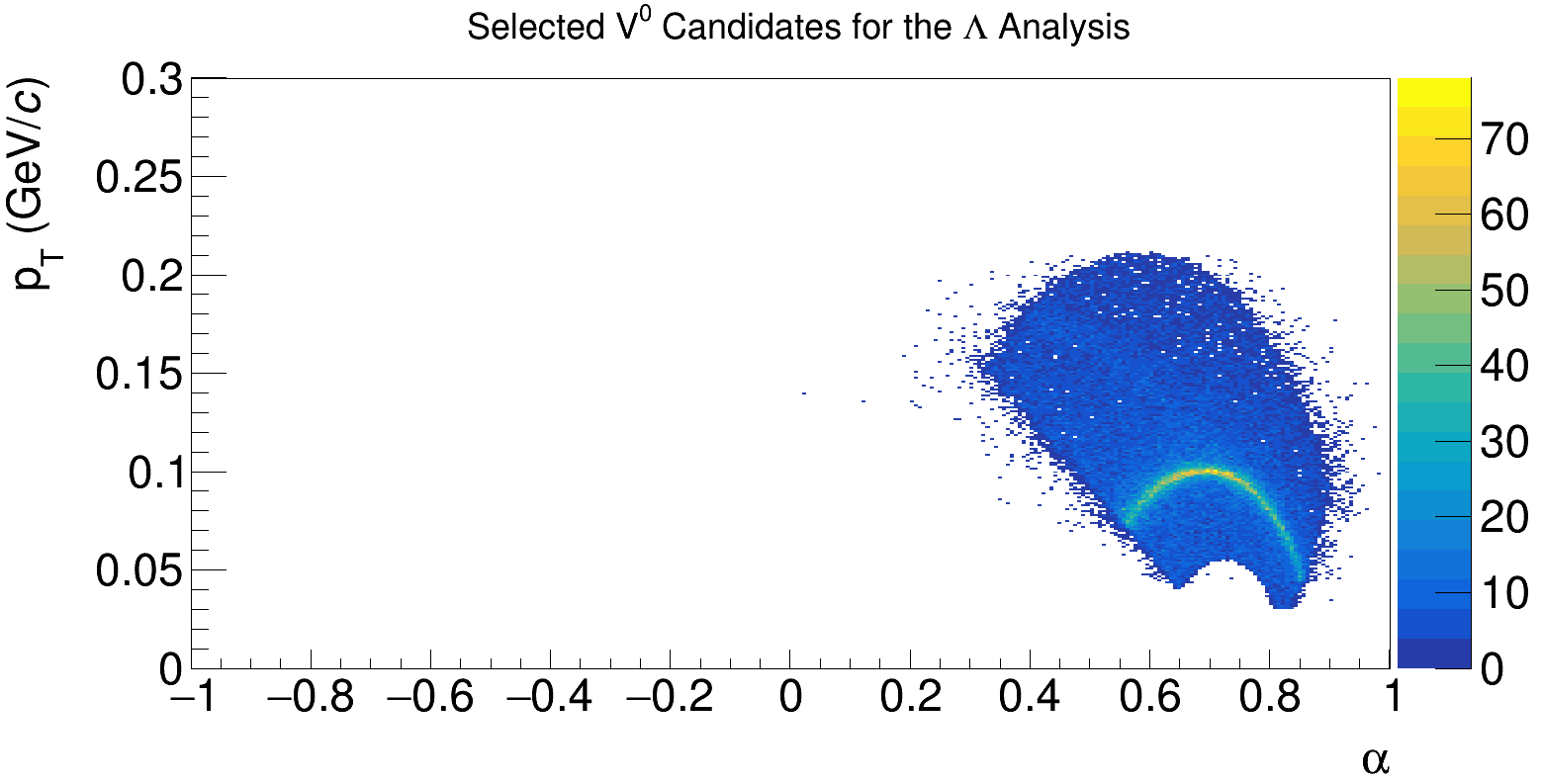}
    \hspace{2em}
    \includegraphics[width=0.45\textwidth]{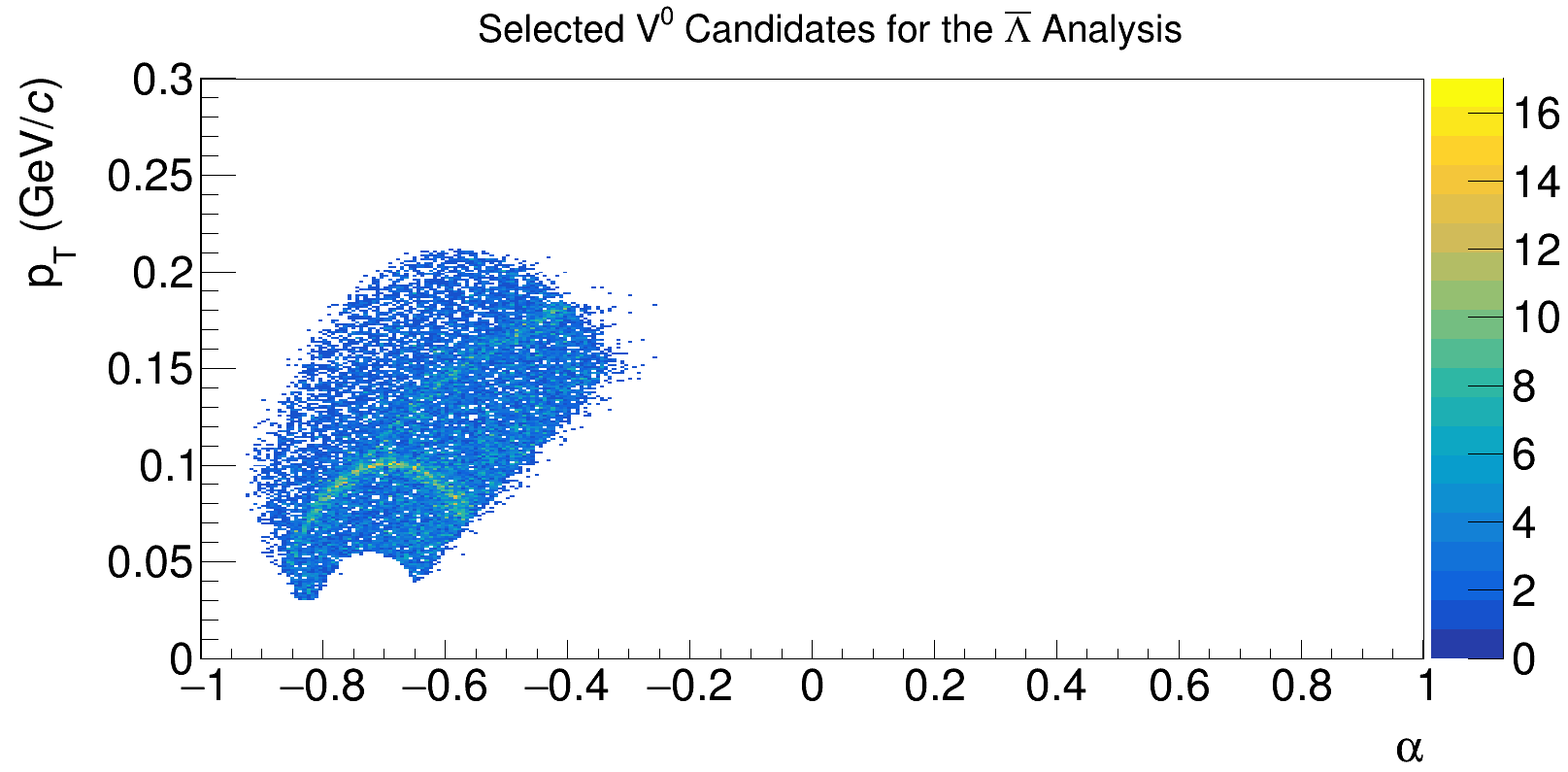}
  \caption{
    Armenteros--Podolanski distributions before (top left) and after applying all of the cuts in the neutral-hadron analyses to select $K^0_S$ (top-right), $\Lambda$ (bottom-left), and $\bar{\Lambda}$ (bottom-right). $\alpha$ is the longitudinal momentum asymmetry of the decay products, as defined in Equation \ref{eq:apAlpha}.
  }
  \label{fig:apPlotCutFlow}
\end{figure*}

Figure \ref{fig:apPlotCutFlow} shows the effects of the cuts applied to the Armenteros--Podolanski distributions \cite{podolanski1954iii}, which plot the transverse momentum $p_T$ of the $V^0$ versus the longitudinal momentum asymmetry in the decay frame:
\begin{equation}  
  \begin{aligned}
    \alpha = \frac{ p^{+, *}_L - p^{-, *}_L}{p^{+, *}_L + p^{-, *}_L} \raisebox{0.6ex}{.}
  \end{aligned}
  \label{eq:apAlpha}
\end{equation}
Here $p^{+, *}_L$ is the longitudinal momentum of the positively charged
decay product in the $V^0$ decay frame,
and $p^{-, *}_L$ is for the negatively charged particle.

Table \ref{tb:neutralTrackCounts} shows the number of selected $V^0$ candidates for each particle species in the neutral-hadron analysis after applying the selection criteria.

\begin{table*}[htbp]
\centering
\begin{tabular}{cccc}
& $K^0_S$ \vspace{1mm} &  $\Lambda$& $\bar{\Lambda}$ \\
\hline
Target-Inserted & 263 k & 67 k & 9 k \\
Target-Removed & 22 k & 4 k & 0.5 k
\end{tabular}
\caption[Neutral Particles Passing Selection Cuts]{The number of target-inserted and target-removed neutral particle candidates passing all of the selection cuts.}
\label{tb:neutralTrackCounts}
\end{table*}

\subsection{Invariant Mass Distribution Fits}

Once all of the cuts have been applied, the remaining $V^0$ candidates are placed into kinematic bins, defined in ranges of total momentum $p$ and angle $\theta$, and an invariant mass spectrum fit extracts the number of signal $V^0$s in each bin. The signal shape is described by the Cauchy distribution, also known as a Lorentz distribution, given by
\begin{equation}  
  \begin{aligned}
    f_s(m; m_0, \gamma) = \frac{1}{\pi \gamma} \left[ \frac{1}{ (m - m_{PDG} - m_0)^2 + \gamma^2}
      \right].
  \end{aligned}
  \label{eq:cauchyDistribution}
\end{equation}
The parameter $\gamma$ describes the distribution width, $m_{PDG}$ is the Particle Data Group mass \cite{pdg2024}, and $m_0$ is a mass offset, as the fit mass is allowed to float.
A second-degree polynomial,
$f_{\text{bg}}$,
is used to fit the background.

A continuous log-likelihood function is constructed, with the parameter $c_s$ controlling the signal to background ratio:
\begin{equation}  
  \begin{aligned}
    \ln(L) = \sum_{\text{All V0}} \ln[ c_s f_s(m; m_0, \gamma) + (1 - c_s) f_{\text{bg}}].
  \end{aligned}
  \label{eq:neutralLogLikelihood}
\end{equation}

After the fit is performed, the raw signal yield $y^{\text{raw}}$ is extracted from the total number of $V^0$ candidates $N_{\text{V0}}$ in the kinematic bin, where
\begin{equation}  
  \begin{aligned}
    y^{\text{raw}} = c_s N_{\text{V0}}.
  \end{aligned}
  \label{eq:neutralRawYield}
\end{equation}

\begin{figure*}[t]
  \centering
    \includegraphics[width=0.32\textwidth]{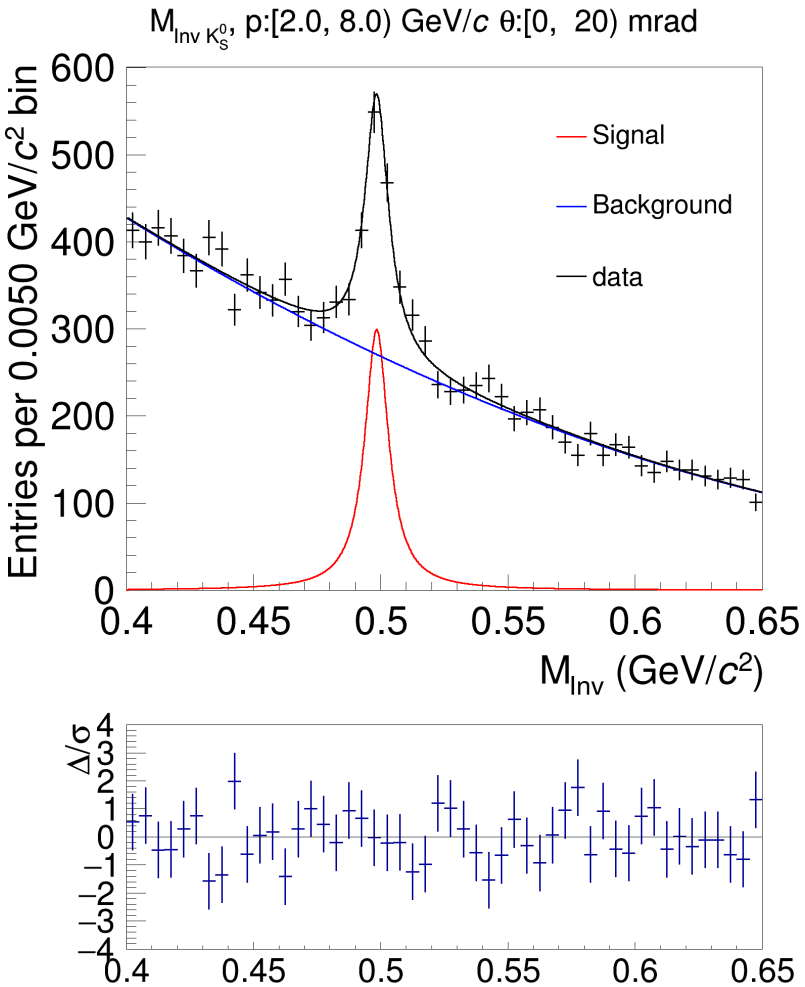}
    \hspace{1em}
    \includegraphics[width=0.30\textwidth]{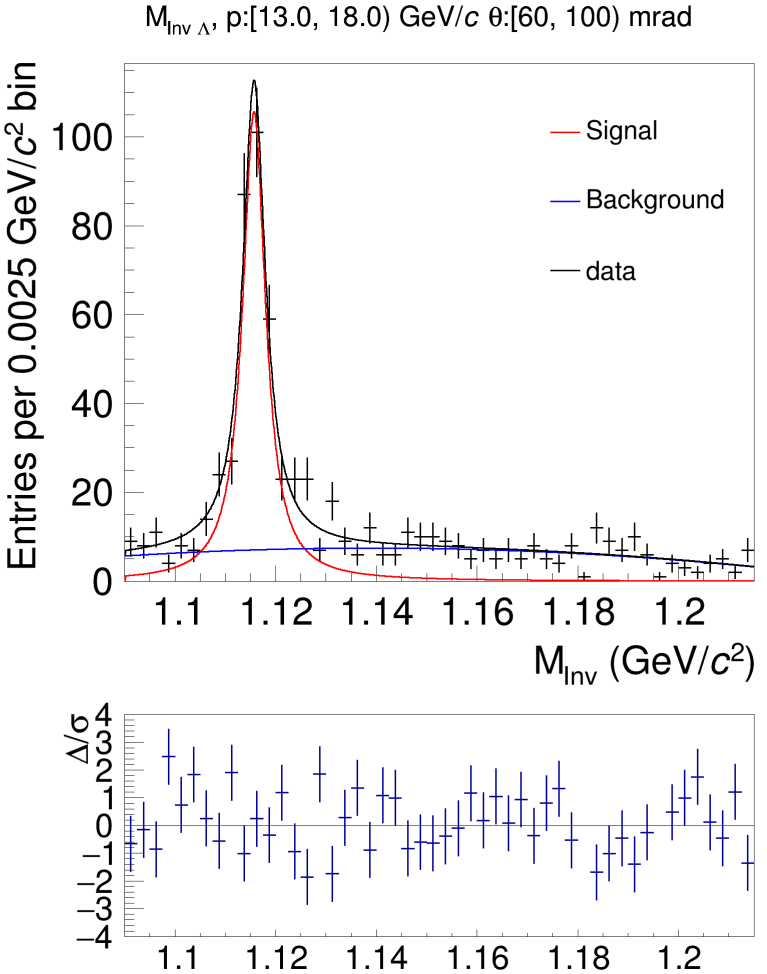}
    \hspace{1em}
    \includegraphics[width=0.30\textwidth]{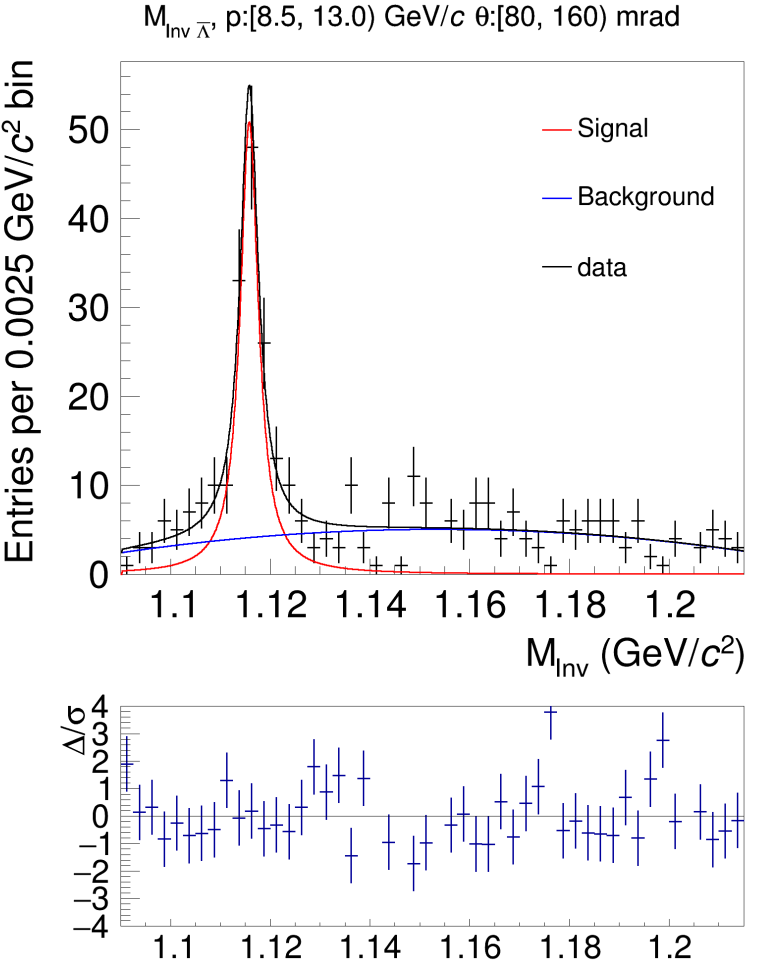}
  \caption{
    Example invariant mass fits for $K^0_S$, $\Lambda$, and $\bar{\Lambda}$. The signal shape is shown alone in red, the background is shown in blue, and the sum of the signal and background is shown in black. At the bottom of each fit is a plot of the fit residual over the statistical uncertainty on the number of entries for each bin; the fit residual is the fit result minus the number of entries.
  }
  \label{fig:sampleMassFit}
\end{figure*}

Figure \ref{fig:sampleMassFit} shows an example mass fit for each of the neutral-hadron species. 

\subsection{Monte Carlo Corrections}
\label{sec:neutralMonteCarloCorrections}

Monte Carlo corrections are applied to correct for tracks removed by cuts, and to correct for detector acceptance, background contributions, and reconstruction inefficiencies. For each kinematic bin $i$ in the neutral-hadron analysis, the total correction factor can be written:
\begin{multline}
  c_i = \frac{N(\text{Sim. neutral particles from prod. evts})_i}
  {N(\text{Selected recons. neutral particles})_i} = \\
  c_{\text{acc.}} \times c_{\text{sel.}} \times c_{\text{rec. eff.}} \times c_{\text{f.d.}} \times c_{\text{br.}}.
  \label{eq:chargedMonteCarloCorrections}
\end{multline}
Here $c_{\text{acc.}}$ corrects for particles removed by acceptance cuts, $c_{\text{sel.}}$ corrects for tracks removed by track quality cuts, $c_{\text{rec. eff.}}$ is the correction factor associated with any reconstruction efficiencies, and $c_{\text{f.d.}}$ is the correction associated for feed-down particles; in the neutral-hadron analysis, feed-down particles originate from weakly decaying $\Xi$ and $\Omega$ baryons. As only one decay channel is selected, $c_{\text{br.}}$ corrects for the missing decays.
The Monte Carlo corrections are calculated by repeating the analysis on Monte Carlo samples. The number of simulated particles in each kinematic bin is counted, and then divided by the number of selected reconstructed particles.

\section{Charged Hadron Analysis}
\label{sec:chargedAnalysis}

\subsection{Selection of Charged Tracks}
\label{sec:chargedTrackSelection}

The event-level cuts in the charged-hadron analysis are the same as those used in the neutral-hadron analysis, described in Section \ref{sec:neutralEventSelection}.  Additional cuts on the event topology, track reconstruction, detector acceptance, and energy loss are described below.

\begin{description}
\item[Topological Cuts]  This charged-hadron analysis classifies tracks as Right-Side Tracks (RSTs) and Wrong-Side Tracks (WSTs), according to the track's charge $q$ and the orientation of the track's momentum $\Vec{p}$ with respect to the magnetic field.  RSTs are bent by the magnets in the same direction as the initial $x$-component of their momentum  $p_x$.  Since tracks curving in that direction align better with the geometry of the TPC readout pads, they typically exhibit a narrower \dedx distribution and momentum range than WSTs. The RST/WST designation is only applied for tracks with polar angle $\theta \geq 10$ mrad, as the azimuthal angle $\phi$ is difficult to measure at small $\theta$, and the distinction is defined by
\begin{equation}
\begin{cases}
p_x/q > 0 & \text{RST}, \\
p_x/q < 0 & \text{WST}.
\end{cases}
\label{eq:rstWSTDesignation}
\end{equation}
In this analysis, final identified hadron spectra is calculated with RSTs, and WSTs are used as a consistency check.

\item[Track Quality and Vertex Cuts]  The track quality cuts ensure proper energy loss and track reconstruction.
\begin{enumerate}
    \item  Well-measured momentum and sufficient \dedx samples requires at least 20 total clusters in VTPC-1 + VTPC-2, or three clusters in the GTPC and 20 additional clusters in the MTPCs, or three clusters in the GTPC and six additional clusters in the FTPCs. (GTPC clusters are used for tracking only, and not for \dedx.)

\item In addition to the cluster requirements, the reconstructed main vertex needs to be within $\pm 5$ cm of the target center in $z$, and tracks must be reconstructed within 2 cm [total $(x,y)$ distance] of the beam particle's position at the main vertex;
these vertex cuts ensure measured tracks are produced from the primary proton-carbon interaction.

\end{enumerate}

\item[Acceptance Cuts]  The acceptance of the NA61/SHINE detector in $\phi$ varies significantly with changes in $\theta$. For each angular bin $\theta$, only regions of uniform $(p, \phi)$ acceptance are allowed. Selecting only uniform acceptance regions allows for the extrapolation of track multiplicity into unmeasured regions, as particle production is independent of $\phi$. 

\item[\dedx Cuts]  Energy loss, shown in Figure \ref{fig:energyLossvsMomentum}, is used for particle identification.
\begin{enumerate} 
\item In the vicinity of Bethe--Bloch crossings, \dedx cannot be used to separate out the charged particle species. The Bethe--Bloch crossings are defined as momenta regions where two species' Bethe--Bloch expectations are within 5\% of one another. For $\pi^\pm$, the proton cross-over region $p \in [1.64,2.02]$ GeV$/c$ is omitted. For $p / \bar{p}$, both the pion and kaon Bethe--Bloch regions are omitted, as well as the small region between the two crossings, giving an exclusion range of $p \in [1.64, 4.32]$ GeV$/c$. For $K^\pm$, the total exclusion region is $p \in [0.95, 4.32]$ GeV$/c$.

\begin{figure*}[t]
  \centering
  \includegraphics[width=0.48\textwidth]{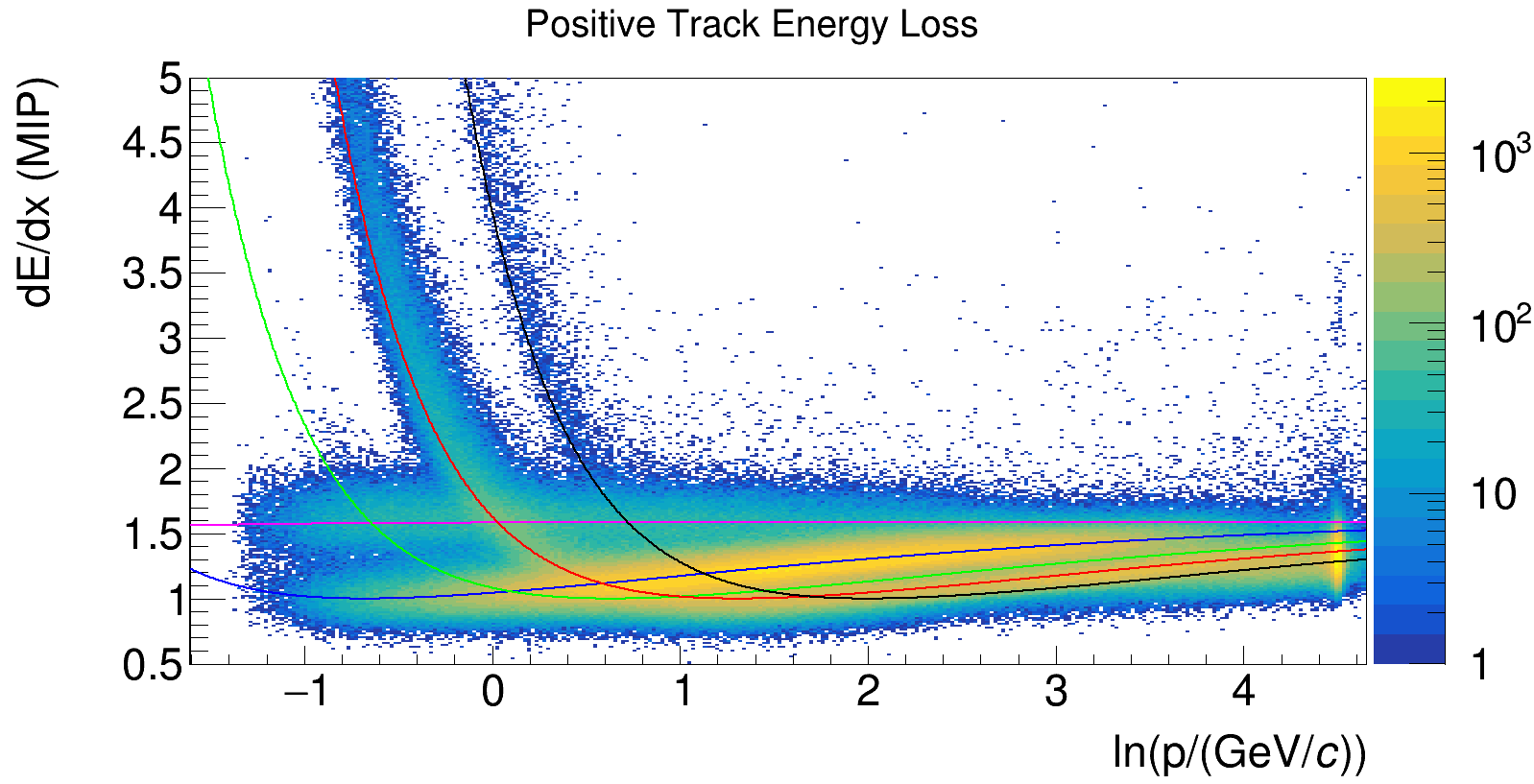}
  \hspace{1em}
  \includegraphics[width=0.48\textwidth]{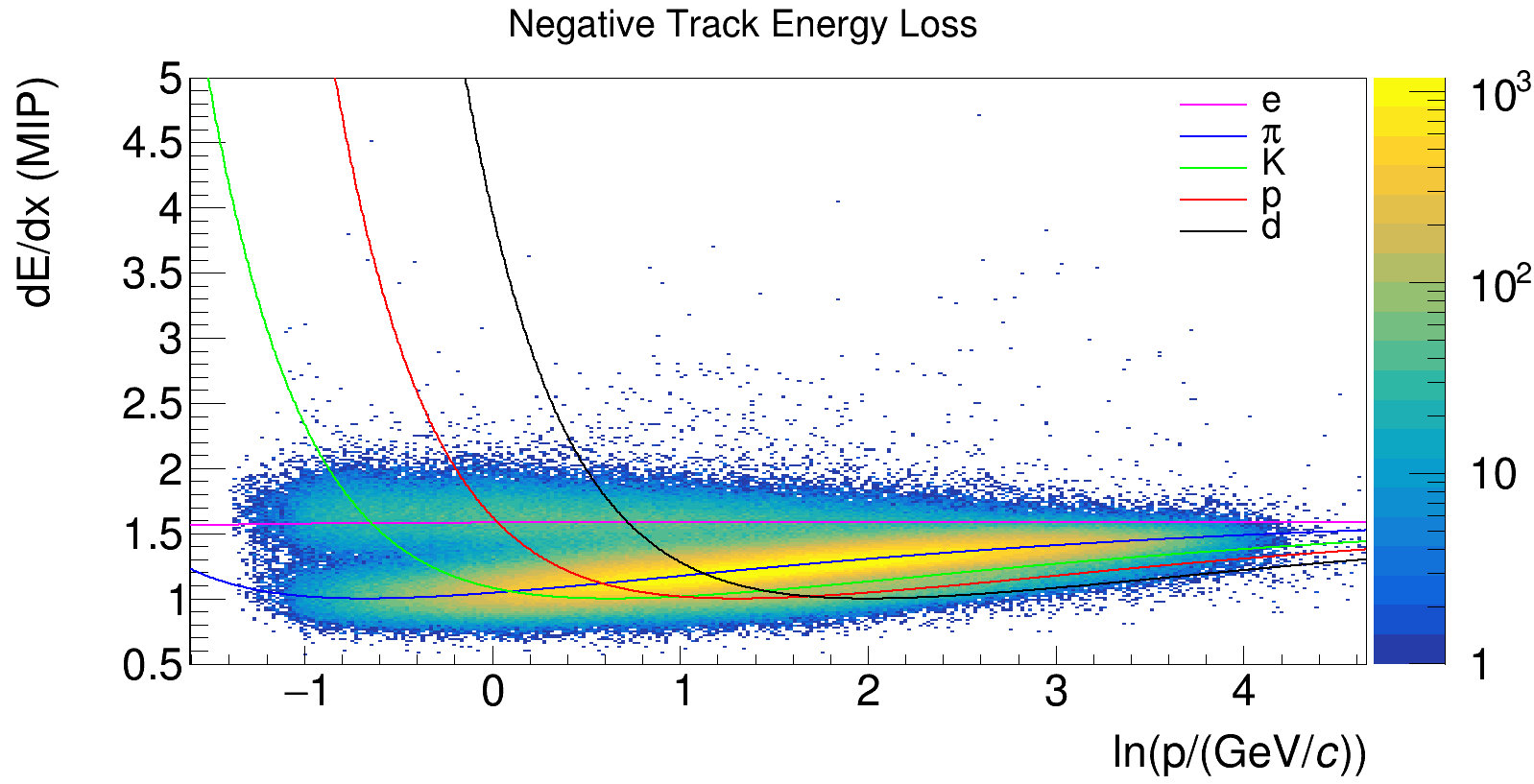}
\caption{Two-dimensional distributions of charged track \dedx vs. $\ln(p)$ after applying track quality cuts. The lines on the plot represent the Bethe--Bloch predictions for each particle species in the energy loss fit. A prominent peak in the positively charged track \dedx distribution is visible at the beam momentum of 90 GeV/$c$ (ln($p$ / [GeV/$c$]) = 4.50) in the left plot; the right plot shows the negatively charged track \dedx distribution.}
\label{fig:energyLossvsMomentum}
\end{figure*}

\item A second \dedx quality cut is applied to remove doubly charged tracks and tracks with large energy loss distortions; this cut removes tracks with $p > 2.2$ GeV$/c$ and \dedx $> 2.0$ times that of a minimum-ionizing particle (MIP).
\end{enumerate}
\end{description}

Table \ref{tb:chargedTrackCounts} shows the number of selected tracks for each particle species in the charged-hadron analysis after applying all of the selection criteria.

\begin{table*}[htbp]
\centering
\begin{tabular}{cccc}
& $\pi^\pm$ & $p$ / $\bar{p}$ & $K^\pm$\\
\hline
Target-Inserted & 1.1 M & 0.9 M & 0.7 M \\
Target-Removed & 10 k & 11 k & 8 k
\end{tabular}
\caption[Charged Tracks Passing Selection Cuts]{The number of target-inserted and target-removed charged tracks passing all of the selection cuts.}
\label{tb:chargedTrackCounts}
\end{table*}

\subsection{Energy Loss Fits}

The average energy loss $\langle \epsilon \rangle$ for a charged particle traversing a medium depends on the particle velocity $\beta$, allowing for separation of particle masses in given momentum ranges. Charged tracks passing the selection criteria are separated by charge and sorted into kinematic analysis bins, and then a likelihood-based fit is performed in each analysis bin to estimate the fractional content of each particle species; the considered species are $e$, $\pi$, $K$, $p$, and $d$.

The fit function used in this analysis is identical to the one used in the analysis of the 2016 and 2017 120 GeV$/c$ proton-carbon data \cite{adhikary2023measurementscharged}.  For a given momentum range and particle species, the \dedx distribution will resemble a straggling function \cite{bichsel2006approximation}, with a long tail toward large energy deposit. Truncating the distribution at the $[0,50]$ percentiles, meaning the largest 50\% of samples are removed, allows the distribution to be described by an asymmetric Gaussian function: 
\begin{equation}
  f(\epsilon,\sigma) =
  \frac{1}{\sqrt{2\pi\sigma}}e^{-\frac{1}{2}\left(\frac{ \epsilon -\mu}{\delta\sigma}\right)^2},
  \qquad \delta =
  \begin{cases}
    1-d, & \epsilon \leq \mu \\
    1+d, & \epsilon > \mu
  \end{cases}.
  \label{eq:asymmetricGaussian}
\end{equation}
Here $\epsilon$ is the track energy loss, $d$ describes the asymmetry of the distribution, $\sigma$ is the base distribution width, and $\mu$ is the distribution peak, given by $\mu = \langle \epsilon \rangle - \frac{4 d \sigma}{\sqrt{2 \pi}}$.

The width of the distribution $\sigma$ depends on the number of energy loss samples in each detector, the mean \dedx $\langle \epsilon \rangle$, and a scaling parameter $\alpha$:
\begin{equation} 
  \sigma = \frac{ \langle\epsilon\rangle^\alpha }{ \sqrt{ \frac{N_\text{cl\;Up}}{\sigma^2_\text{0\;Up}} +
      \frac{N_\text{cl\;V}}{\sigma^2_\text{0\;V}} +
      \frac{N_\text{cl\;M}}{\sigma^2_\text{0\;M}} +
      \frac{N_\text{cl\;F}}{\sigma^2_\text{0\;F}}} }.
      \label{eq:dedxWidth}
\end{equation}
Here $N_{\text{cl, Up}}$ denotes the number of \dedx samples in the two upstream sectors of VTPC-1, $N_{\text{cl, V}}$ denotes samples in the remainder of the VTPCs, $N_{\text{cl, M}}$ denotes the MTPCs, and $N_{\text{cl, F}}$ denotes the FTPCs; the variations in the TPC base widths arise from differing pad geometries in the various TPC regions.  Experimentally, NA61/SHINE needs four different base widths to accurately describe the energy loss of tracks: $\sigma^2_\text{0\;Up}$ for the upstream VTPC1 sectors one and four, $\sigma^2_\text{0\;V}$ for the rest of the VTPC sectors, $\sigma^2_\text{0\;M}$ for the MTPC sectors, and $\sigma^2_\text{0\;F}$ for the FTPC sectors.

The constructed log-likelihood function is a sum over all tracks:
\begin{multline}
LL(\epsilon,Y_e^\pm,Y_\pi^\pm,Y_K^\pm,Y_P^\pm,Y_d^\pm) = \\
  \sum_{i}^{i \; \in                                                                                        
    \; + \; \text{tracks}} \left( \sum_j
  \frac{Y_j^+}{\sqrt{2\pi\sigma_i}}e^{-\frac{1}{2}\left(\frac{\epsilon_i-\mu_j}{\delta\sigma_i}\right)^2}
  \right) + \\
  \sum_{k}^{k \; \in                                                                                        
    \; - \; \text{tracks}} \left( \sum_l
  \frac{Y_l^-}{\sqrt{2\pi\sigma_k}}e^{-\frac{1}{2}\left(\frac{\epsilon_k-\mu_l}{\delta\sigma_k}\right)^2}
  \right), \\
  \qquad
  \begin{cases}
    j \in e^+, \pi^+, K^+, p^+, d^+ \\
    l \in e^-, \pi^-, K^-, p^-, d^-
  \end{cases}.
  \label{eq:dedxFitFunction}
\end{multline}
Here $Y_j$ is the fractional yield for each particle species.

The only constraints applied enforce the sum of the particle yields to be unity and the ordering of particle species \dedx for a given momentum range. Once the raw fractional yields $Y_j$ are obtained, the raw yield is obtained for particle species $j$ in kinematic bin $i$ from the total number of tracks $N_i$:
\begin{equation}
    y_{j,\;i}^\text{raw} = N_i Y_{j,\;i}.
    \label{eq:rawChargedYield}
\end{equation}

\begin{figure*}[t]
  \centering
  \includegraphics[width=0.495\textwidth]{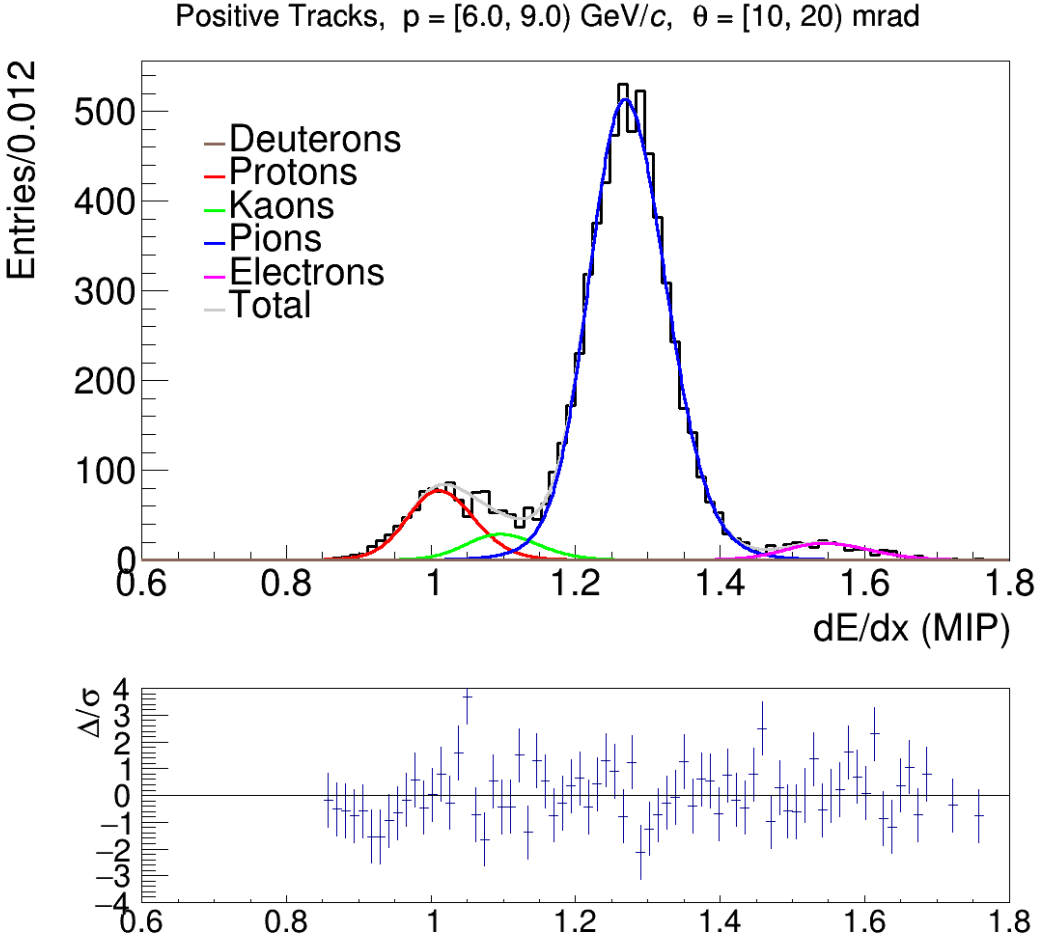}
  \includegraphics[width=0.495\textwidth]{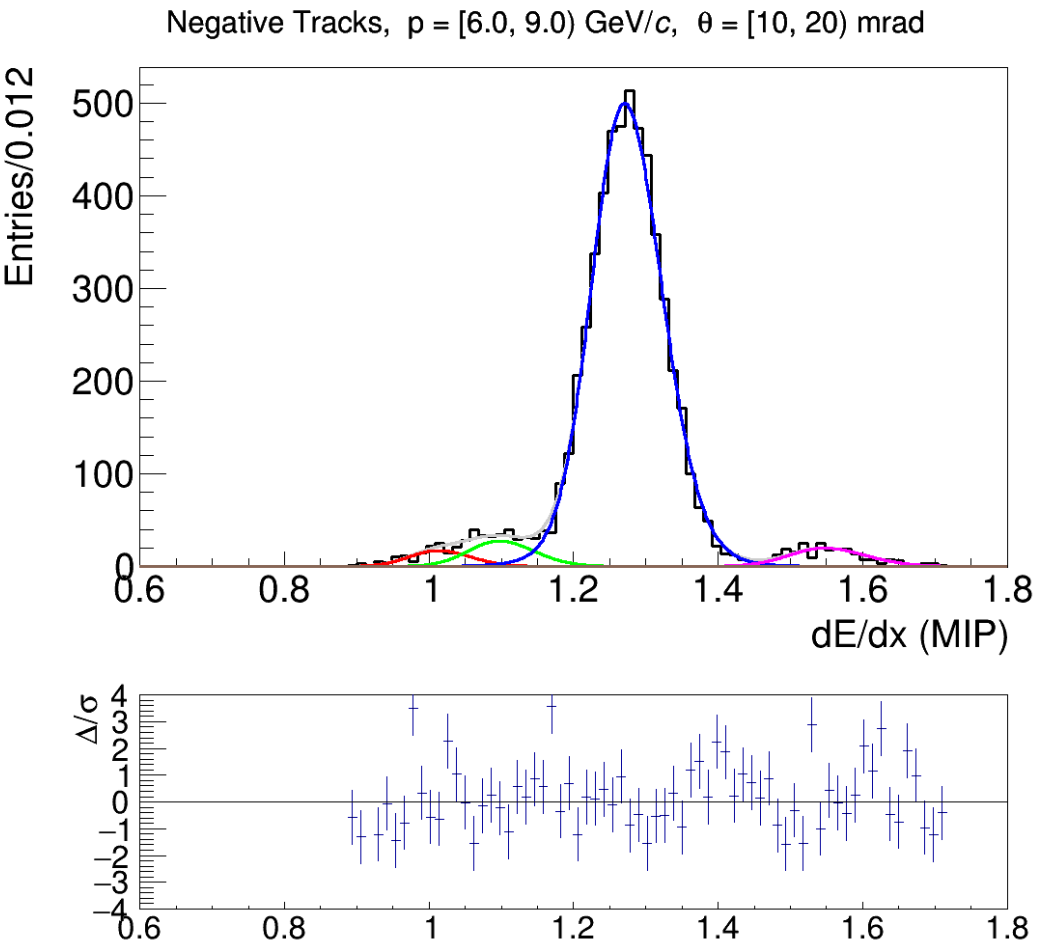}
\caption{Example of a \dedx distribution fit for one kinematic bin. Both the positively and negatively charged track fits are shown, and both fits show a clear abundance of pions. At the bottom of each fit is a plot of the fit residual over the statistical uncertainty on the number of entries for each bin; the fit residual is the fit result minus the number of entries.}
\label{fig:sampleDEDXFit}
\end{figure*}

It is difficult to extract accurate $K^{\pm}$ and $\bar{p}$ yields in kinematic bins with low statistics, so bins with fewer than 16 track counts at the $K^{-}$ and $\bar{p}$ peaks are excluded from the charged-hadron analysis. An example of an energy loss fit is shown in Figure \ref{fig:sampleDEDXFit}.

\subsection{Monte Carlo Corrections}
\label{sec:chargedMonteCarloCorrections}

The Monte Carlo corrections in the charged-hadron analysis are calculated in the same manner as described in Section \ref{sec:neutralMonteCarloCorrections} for the neutral-hadron analysis, except that in the charged-hadron analysis, there is no decay channel selection correction.

\subsection{Energy Loss Fit Bias Corrections}
\label{sec:dedxFitBiasCorrections}

An additional correction is applied to remove any biases from the energy loss fitting procedure. Using the variations in the fit parameters observed across different kinematic bins $i$, the fit parameters are randomly shifted according to their distributions, and the individual track \dedx is re-simulated with the shifted parameters. After performing 50 trials in each bin, the explicit correction is given by $1/(1 + c_i^{\text{Fit}})$, where
\begin{equation}
  c^\text{Fit}_i = \frac{1}{N_\text{trials}}
  \sum^{N_\text{trials}}_{n = 1}\left( \frac{y_n^\text{fit} -
    y_n^\text{true}}{y_n^\text{true}} \right).
    \label{eq:chargedFitBiasCorrection}
\end{equation}
Here $N_\text{trials} = 50$ and $y_{fit}$ and $y_{true}$ are the fit and true yields, respectively. Typical fit bias corrections are less than 4\%.

\subsection{Feed-Down Re-Weighting}
\label{sec:chargedFeedDownReWeighting}

The feed-down correction factor in Equation \ref{eq:chargedMonteCarloCorrections} for charged hadrons resulting from the decays of neutral hadrons is estimated using Monte Carlo models. As these models often do not accurately predict weakly decaying neutral-hadron multiplicities, the feed-down corrections can be constrained with the results of the neutral-hadron analysis, described in Section \ref{sec:neutralAnalysis}; the neutral-hadron analysis measured the multiplicity of $K^0_S$, $\Lambda$, and $\bar{\Lambda}$. Each kinematic bin $i$ gets a re-weighting factor, given by
\begin{equation}
    w_i = \frac{m^\textrm{Data}_i}{m^\textrm{MC}_i},
    \label{eq:feedDownReweighting}
\end{equation}
where ${m^\textrm{Data}_i}$ is the measured multiplicity of a particular neutral hadron, and ${m^\textrm{MC}_i}$ is the Monte Carlo multiplicity. This re-weighting is applied to pions and protons originating from the decay of $K^0_S$, $\Lambda$, or $\bar{\Lambda}$; regions not covered by the neutral-hadron analysis are not re-weighted.The values are typically a few percent for pions and less than 10\% for protons.

Re-weighting the feed-down corrections can significantly constrain the uncertainty resulting from the application of the Monte Carlo feed-down corrections, as shown in the previous analysis of the 2016 and 2017 $p$ + C 120 GeV$/c$ data \cite{adhikary2023measurementscharged}.

\section{Systematic Uncertainties}
\label{sec:systematicUncertainties}

The systematic uncertainties are handled in the same manner for the neutral- and charged-hadron analyses. Any kinematic bin with a total uncertainty greater than 50\% is excluded from the analysis. This section will discuss the considered sources of systematic uncertainties for both analyses. 

\begin{figure*}[t]
  \centering
    \raisebox{0.25\height}{\includegraphics[width=0.25\textwidth]{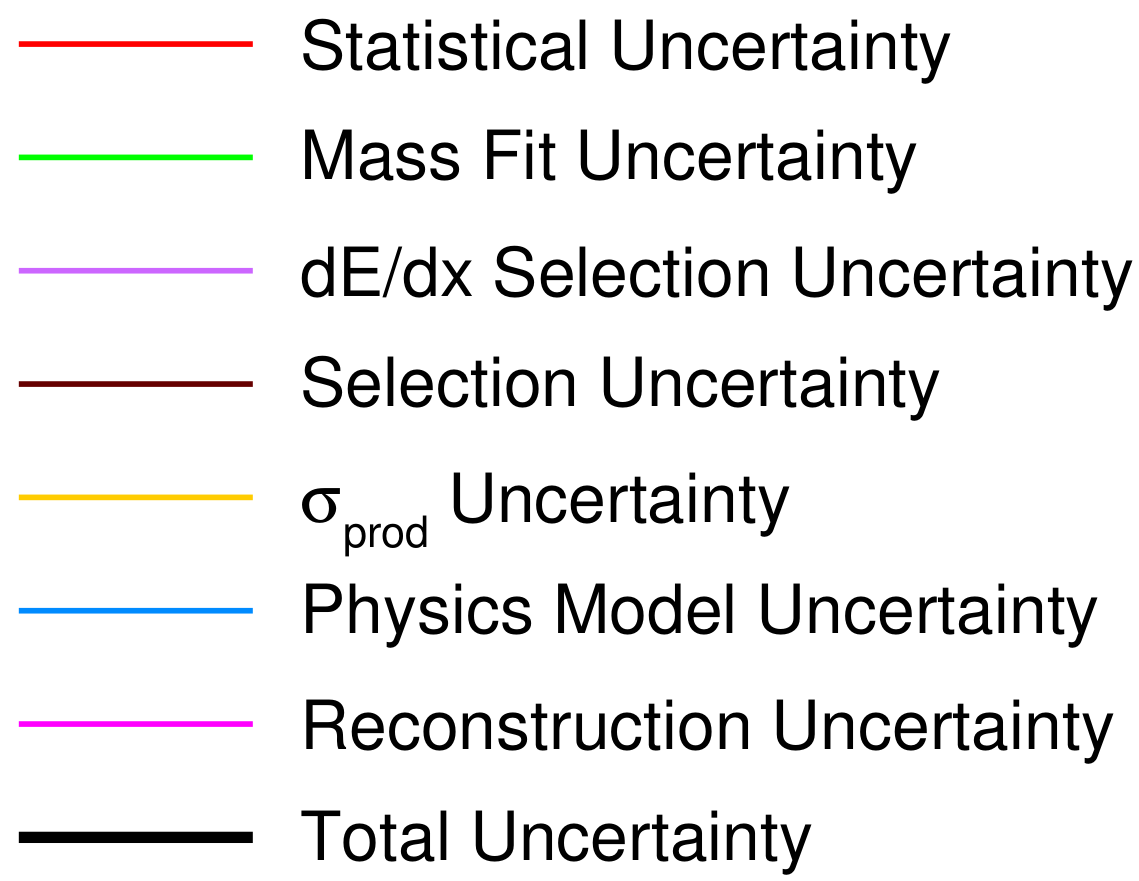}}
    \hspace{0.2cm}
    \includegraphics[width=0.33\textwidth]{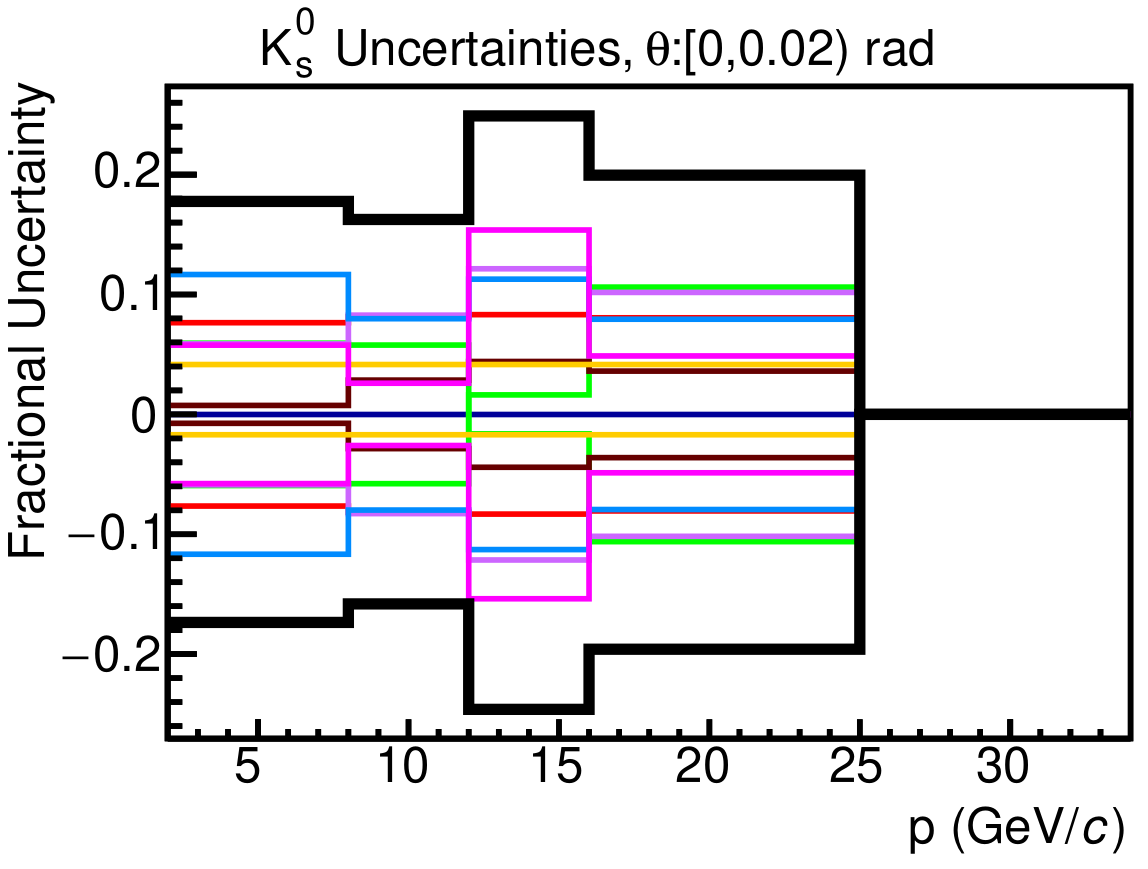}
    \hspace{0.2cm}
    \includegraphics[width=0.33\textwidth]{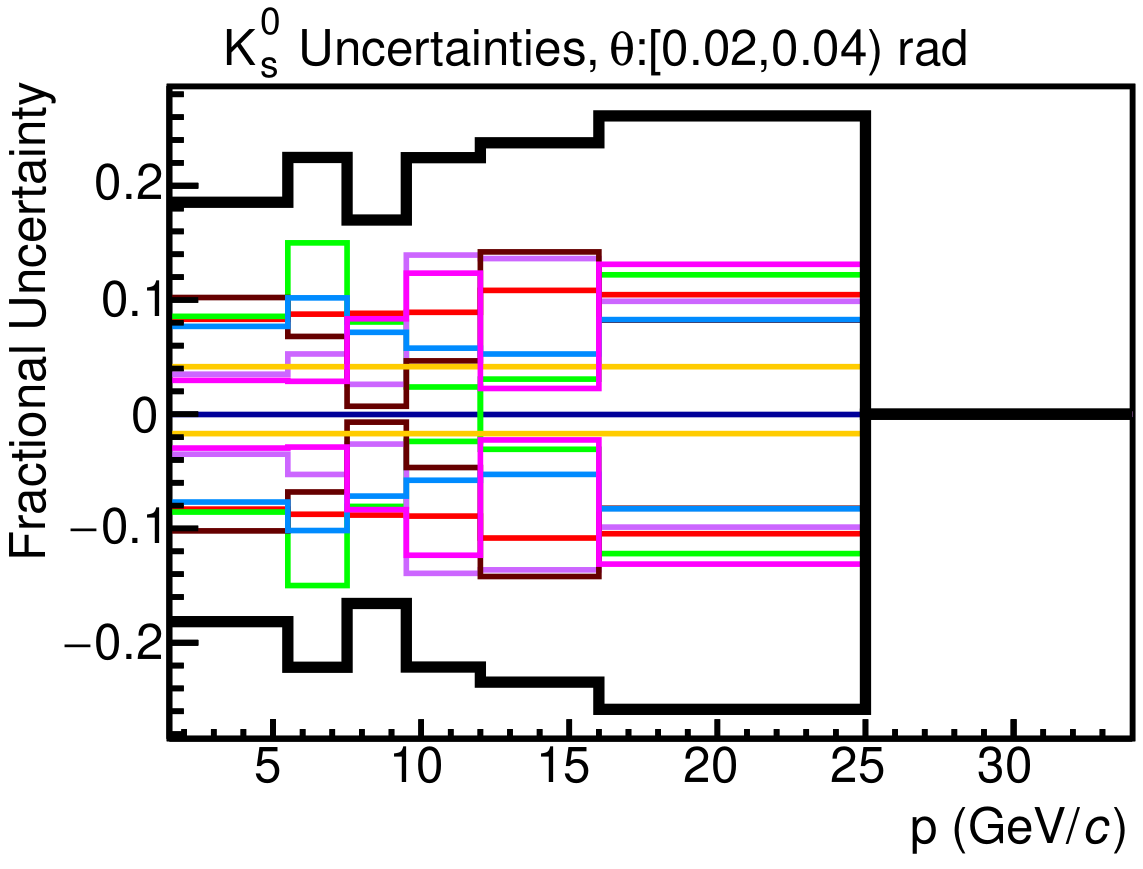}
\caption{Example $K^0_S$ systematic uncertainty breakdown for the angular bins [0, 0.02) rad and [0.02, 0.04) rad. For each kinematic bin shown in the plot, the total uncertainty is shown in black, and all of the constituent uncertainties are shown as well. These uncertainties correspond to the multiplicity measurements shown in Figure \ref{fig:sampleK0SMultiplicities}.}
\label{fig:sampleK0SUncertainties}
\end{figure*}

\begin{figure*}[t]
  \centering
    \raisebox{0.23\height}{ \includegraphics[width=0.27\textwidth]{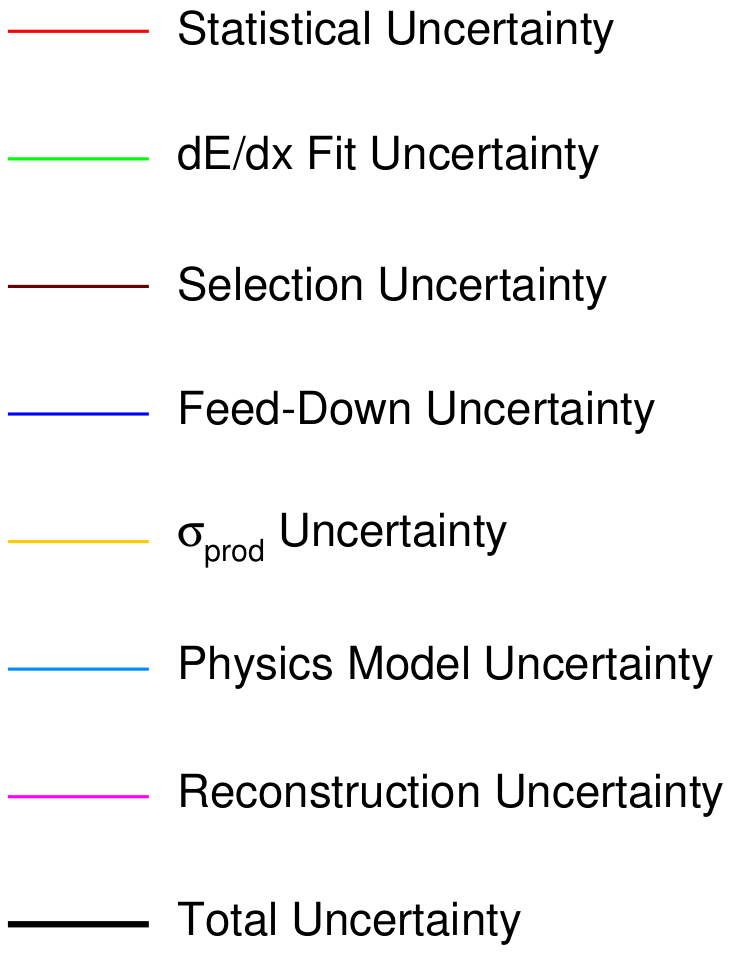}}
    \hspace{0.2cm}
    \includegraphics[width=0.33\textwidth]{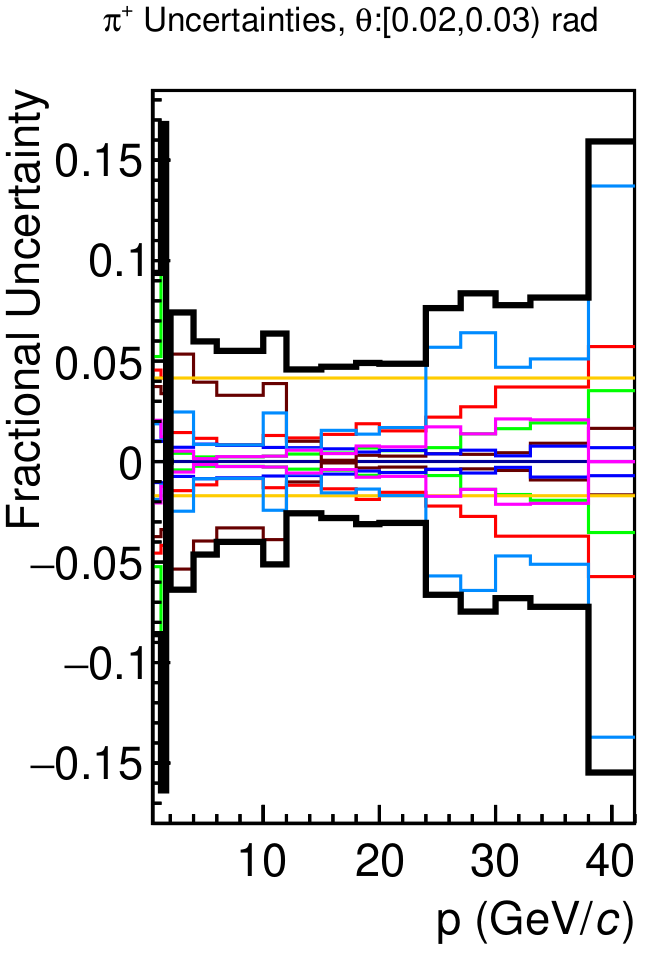}
    \hspace{0.2cm} 
    \includegraphics[width=0.33\textwidth]{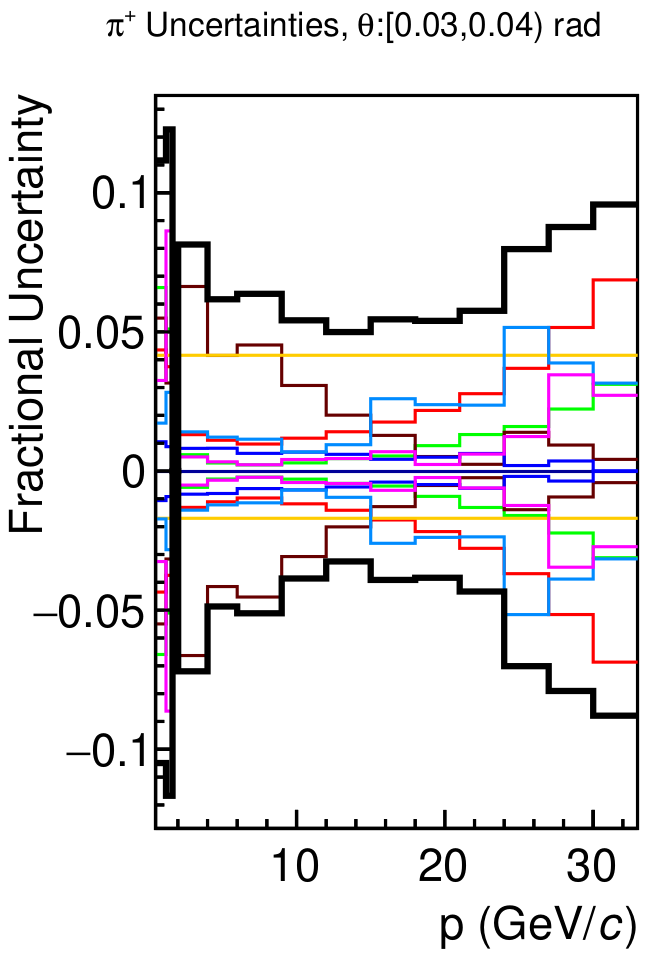} \\
    \hspace{0.3cm} \hspace{0.27\textwidth}
    \includegraphics[width=0.33\textwidth]{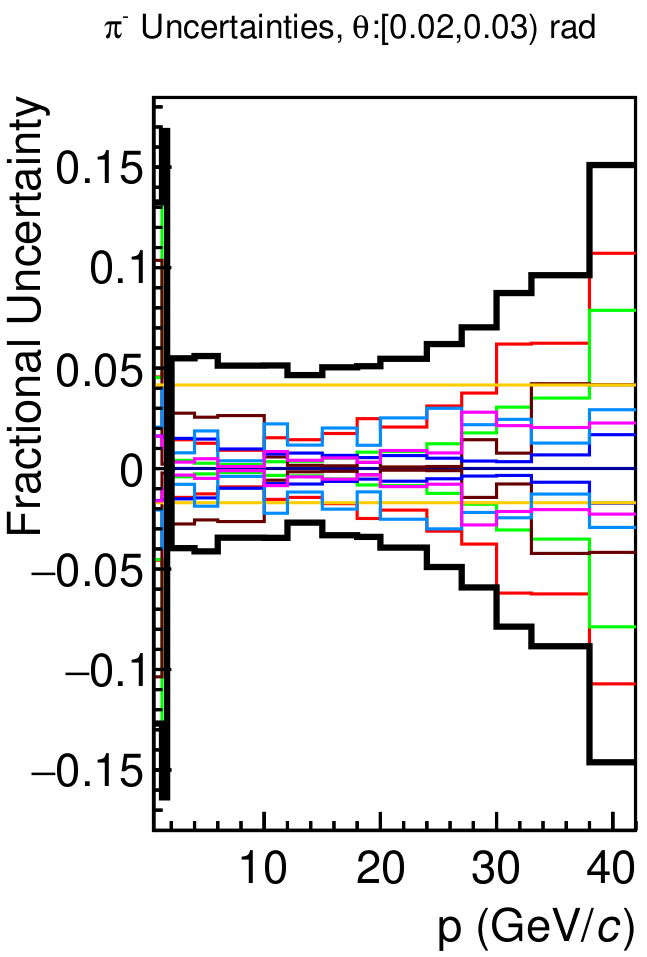}
    \hspace{0.2cm}
    \includegraphics[width=0.33\textwidth]{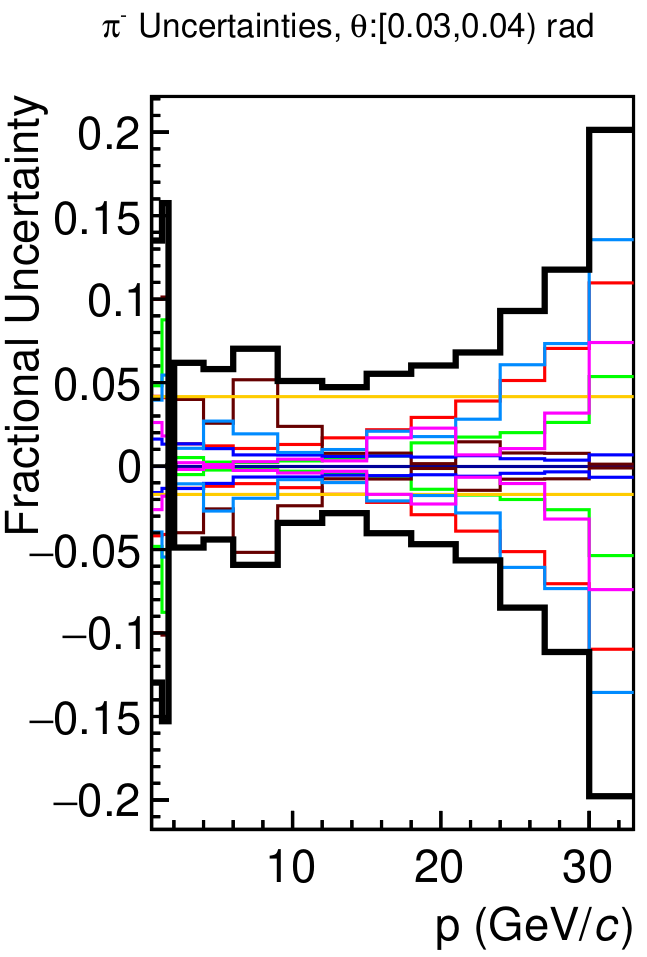}
\caption{Example $\pi^{\pm}$ systematic uncertainty breakdown for the angular bins [0.02, 0.03) rad and [0.03, 0.04) rad. For each kinematic bin shown in the plot, the total uncertainty is shown in black, and all of the constituent uncertainties are shown as well. These uncertainties correspond to the multiplicity measurements shown in Figure~\ref{fig:samplePionMultiplicities}.
The bins before the Bethe--Bloch overlap region generally have higher uncertainty than the bins after the overlap region, as can be seen here.}
\label{fig:samplePionUncertainties}
\end{figure*}

Example breakdowns of the individual uncertainties for $K^0_S$ and $\pi^{\pm}$ can be seen in Figures \ref{fig:sampleK0SUncertainties} and \ref{fig:samplePionUncertainties}, respectively. For the neutral-hadron analysis, the total uncertainty on the multiplicity measurements is typically 10-25\%, and for the charged-hadron analysis the uncertainty is typically 5-15\%.

\begin{description}
\item[Reconstruction]

Differences between the true detector positions during data taking and the positions used in reconstruction and Monte Carlo can affect final calculated multiplicities. To estimate the contribution of detector alignment, residual distributions between track and point measurements along the track are used to estimate any potential detector  misalignment for TPCs used in the momentum reconstruction of tracks. From the width of these distributions, for the charged-hadron analysis the VTPCs are independently shifted by $\pm 100$ \textmu{}m in the $x$-dimension, and FTPC-1 and the GTPC are shifted by $\pm 50$ \textmu{}m in $x$. 

For the neutral hadron analysis, to account for each neutral particle requiring two charged tracks, the VTPCs are conservatively shifted by an additional 100 \textmu{}m. The differences between the resulting systematic uncertainty multiplicity measurements and the standard multiplicity measurement are added in quadrature to obtain the final reconstruction uncertainty.

\item[Selection]

Likely due to dead channels in the front-end electronics and unsimulated periodic detector noise resulting in cluster loss, simulated tracks contain 5-10\% more clusters than tracks from data, on average.
In order to account for this effect, the Monte Carlo corrections are re-calculated with the number of clusters reduced by a conservative 15\%. The resulting difference in the calculated multiplicities is taken as a systematic uncertainty.

\item[Physics Model]

Different underlying physics models will lead to slightly different Monte Carlo corrections, which in turn will give different final calculated multiplicities. The standard Monte Carlo corrections are calculated with the  \GeantFour FTFP\_BERT physics list, and the largest difference in the calculated multiplicity between physics list FTFP\_BERT and the lists FTF\_BIC, QBBC, and QGSP\_BERT is taken as a systematic uncertainty for each kinematic bin. It should be noted that while the final Monte Carlo calculated multiplicities for two different physics lists can be quite similar, the Monte Carlo corrections can still differ significantly, which results in different calculated final differential multiplicities for the data depending on the physics list used. The upper and lower uncertainties are added in quadrature to make this uncertainty symmetric.

\item[Production Cross Section]

The upper and lower uncertainties on the production cross section \cite{allison2024hadron} are propagated through the multiplicity analysis to obtain the associated uncertainties. The result is a uniform fractional uncertainty of $(+4.0, -1.6)$\%.

\item[Decay Product \dedx Selection]

To estimate the uncertainty associated with the decay product \dedx selection, the decay product Bethe--Bloch cut is relaxed by 5\%, and the data and Monte Carlo samples are reprocessed.
The differences between the resulting multiplicities and the standard multiplicities are taken as a systematic uncertainty.

\item[Momentum]

As the invariant fit mass in Equation \ref{eq:cauchyDistribution} is allowed to float, any uncertainty related to the momentum reconstruction can be studied by looking at the difference in the fit masses for a fit performed with the aggregation of all 
kinematic bins; the differences between the fit masses and currently accepted values \cite{pdg2024} would arise from momentum mis-reconstruction. The mass shift values for $K^0_S$, $\Lambda$, and $\bar{\Lambda}$ were 0.0067\%, 0.0057\%, and 0.016\% respectively; these shifts are on the order of magnitude expected from the fit uncertainty. As this systematic uncertainty is negligible compared to the other uncertainties, it is not included in either the neutral-hadron or charged-hadron analyses.

\item[Feed-down]

The feed-down uncertainty originates from particles produced in the primary interaction weakly decaying into the analyzed particle species. For the neutral-hadron analysis, weak decays of $\Xi$ and $\Omega$ baryons which can decay to $\Lambda$ and $\bar{\Lambda}$ are considered, and the feed-down corrections come entirely from Monte Carlo, as there is a lack of data measuring the production of $\Xi$ and $\Omega$ baryons in 90 GeV$/c$ proton-carbon interactions. As production rates of these baryons vary up to 50\% among different physics models, to estimate the feed-down uncertainty, the number of feed-down tracks is varied by $\pm 50$\%, and the changes in the final calculated multiplicities are taken as a systematic uncertainty.

For the charged-hadron analysis, the process is the same, except $\pi^{\pm}$, $K^{\pm}$, and
$p / \bar{p}$ particles originating from the decay of $K^0_S$, $\Lambda$, or $\bar{\Lambda}$ in a region of phase space covered by the neutral-hadron analysis can be re-weighted, as discussed in Section \ref{sec:chargedFeedDownReWeighting}. The uncertainty on the neutral hadron measurement is then applied; if the kinematics are not covered by the neutral-hadron analysis, an uncertainty of 50\% is used. The collected uncertainties are then averaged for each kinematic bin to assign a total feed-down uncertainty, and the resulting changes in the final calculated multiplicities are taken as a systematic uncertainty. For covered regions, the uncertainty is typically much smaller than 50\% \cite{adhikary2023measurementscharged}.

\item[Fit]

For the neutral-hadron analysis, the uncertainty associated with the invariant mass fit is estimated by looking at the number of true versus fit fractions for each Monte Carlo invariant mass fit with the \GeantFour physics lists FTFP\_BERT, QGSP\_BERT, QBBC, and FTF\_BIC.
The fractional differences are averaged, and the average difference is taken as a systematic uncertainty. This is generally a more conservative estimate of the fit uncertainty than taking the errors on the fit parameters.

For the charged-hadron analysis, the standard deviation of fit biases from the \dedx fit bias corrections, as discussed in Section \ref{sec:dedxFitBiasCorrections}, is calculated as
\begin{equation}
  \sigma^\textrm{Fit}_i =  \sqrt{\frac{1}{N_\textrm{trl}} \sum^{N_\textrm{trl}}_{i = 1}\left( \frac{y_i^\textrm{fit} -
      y_i^\textrm{true}}{y_i^\textrm{true}} - \Big \langle
    \frac{y^\textrm{fit} - y^\textrm{true}}{y^\textrm{true}} \Big
    \rangle \right)^2}.
    \label{eq:chargedFitBiasUncertainty}
\end{equation}
The standard deviation of the fractional multiplicity given by the $N_\textrm{trl}=$50 Monte Carlo trials is taken as the fit uncertainty for the charged-hadron analysis, and is propagated through as a systematic uncertainty.

\end{description}

After obtaining the systematic uncertainties, the total systematic error for each multiplicity measurement is calculated by adding all of the uncertainty sources together in quadrature; the feed-down, physics model, and production cross section
uncertainties are uncorrelated, and the rest of the systematic uncertainties are correlated.

\section{Neutral- and Charged-Hadron Multiplicity Results}
\label{sec:hadronMultiplicityMeasurements}

Once the raw yields are obtained along with all of the correction factors, the differential production multiplicities, defined as the number of produced hadrons per production interaction in each kinematic bin, can be calculated. (A production interaction is defined as an interaction resulting in the production of new hadrons, and it excludes quasi-elastic interactions.) The double-differential production multiplicities are given by
\begin{equation} 
  \frac{d^2 n_i}{dp d\theta} = \frac{ c^\textrm{total}_i
    \sigma_\textrm{trig}}{(1-\epsilon) \sigma_\textrm{prod} \Delta p
    \Delta \theta} \left(
  \frac{y^{\textrm{I}}_i}{N_{\textrm{trig}}^{\textrm{I}}} -
  \frac{\epsilon y^{\textrm{R}}_i}{N_{\textrm{trig}}^{\textrm{R}}}
  \right).
  \label{eq:neutralDifferentialMultiplicity}
\end{equation}
Here $n_i$ is the number of produced hadrons in kinematic bin $i$ with production angle $\theta$ and total momentum $p$. Also $y_i^\textrm{I}$ is the raw yield with the target inserted, $y^\textrm{R}_i$ is the raw yield with the target removed, $N^\text{I}_\text{trig}$ and $N^\text{R}_\text{trig}$ the number of recorded triggers with the target inserted and removed, $c^{\text{total}}_i$ is the total correction factor, and $\epsilon = P_\textrm{trig}^\textrm{R}/ P_\textrm{trig}^\textrm{I}$ is the removed-to-inserted trigger probability ratio. The trigger and production cross sections are $\sigma_\textrm{trig}$ and $\sigma_\textrm{prod}$, respectively, and $\Delta p \Delta \theta$ is the size of the kinematic bin.

The statistical uncertainty is calculated from the number of raw fit particles in each kinematic bin, and it is added in quadrature with the total systematic uncertainty to obtain the total uncertainty for each multiplicity measurement.

Following previously established procedures for measuring the trigger, production, and inelastic cross sections \cite{aduszkiewicz2019measurementscrosssection}, $\sigma_\textrm{trig} = 234.5 \pm 1.2$ mb, $\sigma_\textrm{prod} = 222.2 \pm 1.2$ (stat.) $\pm \ ^{0.2}_{8.0}$ (model)$ \pm 3.3$ (syst) mb, and $\sigma_\textrm{inel} = 240.8 \pm 1.2$ (stat.) $\pm \ ^{10.8}_{9.3}$ (model)$ \pm \ ^{3.4}_{3.6}$ (syst) mb are obtained for 90 GeV/$c$ proton-carbon interactions~\cite{allison2024hadron}.  

\begin{figure*}[t]
  \centering
    \includegraphics[width=0.49\textwidth]{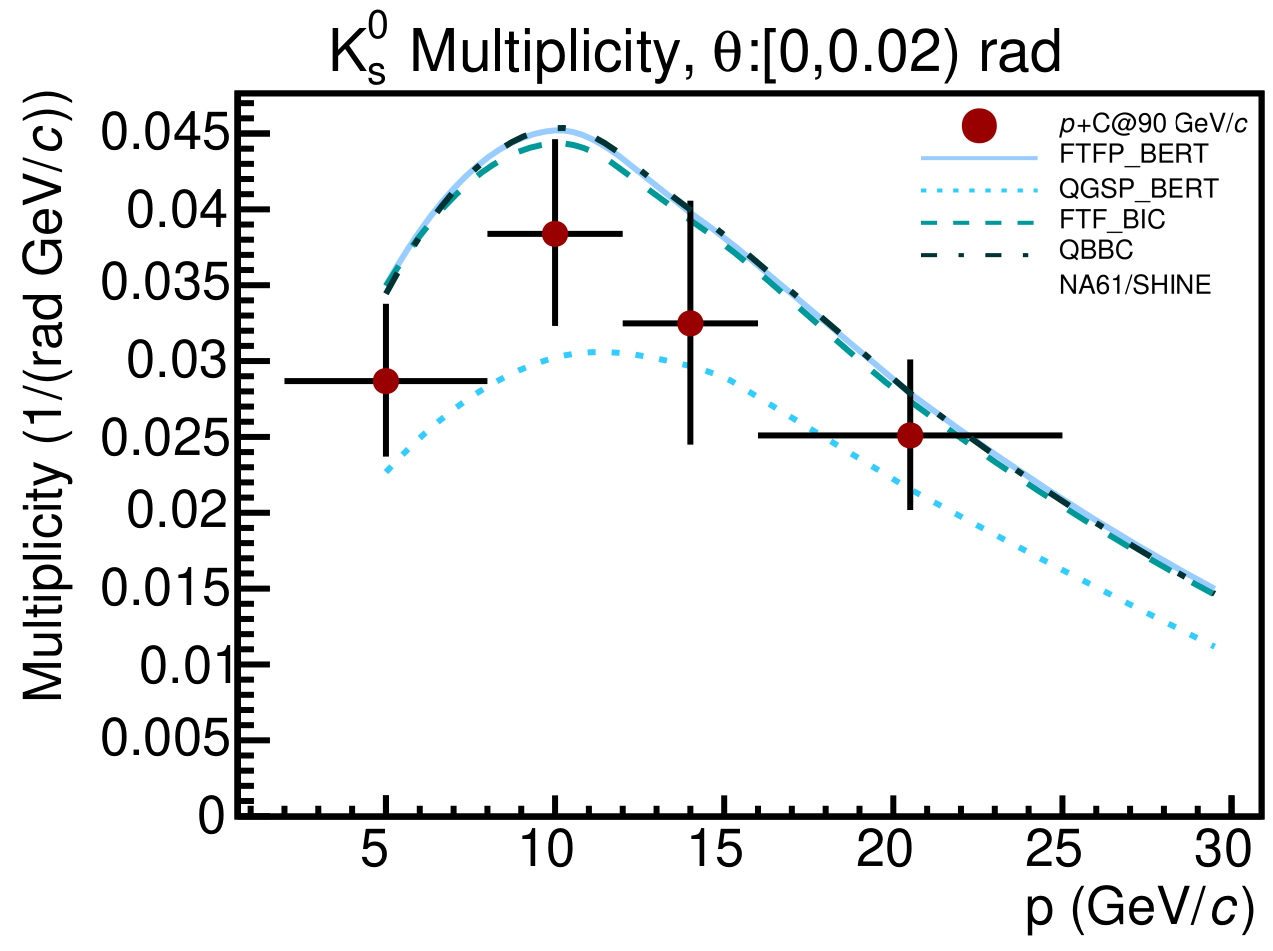}
    \includegraphics[width=0.49\textwidth]{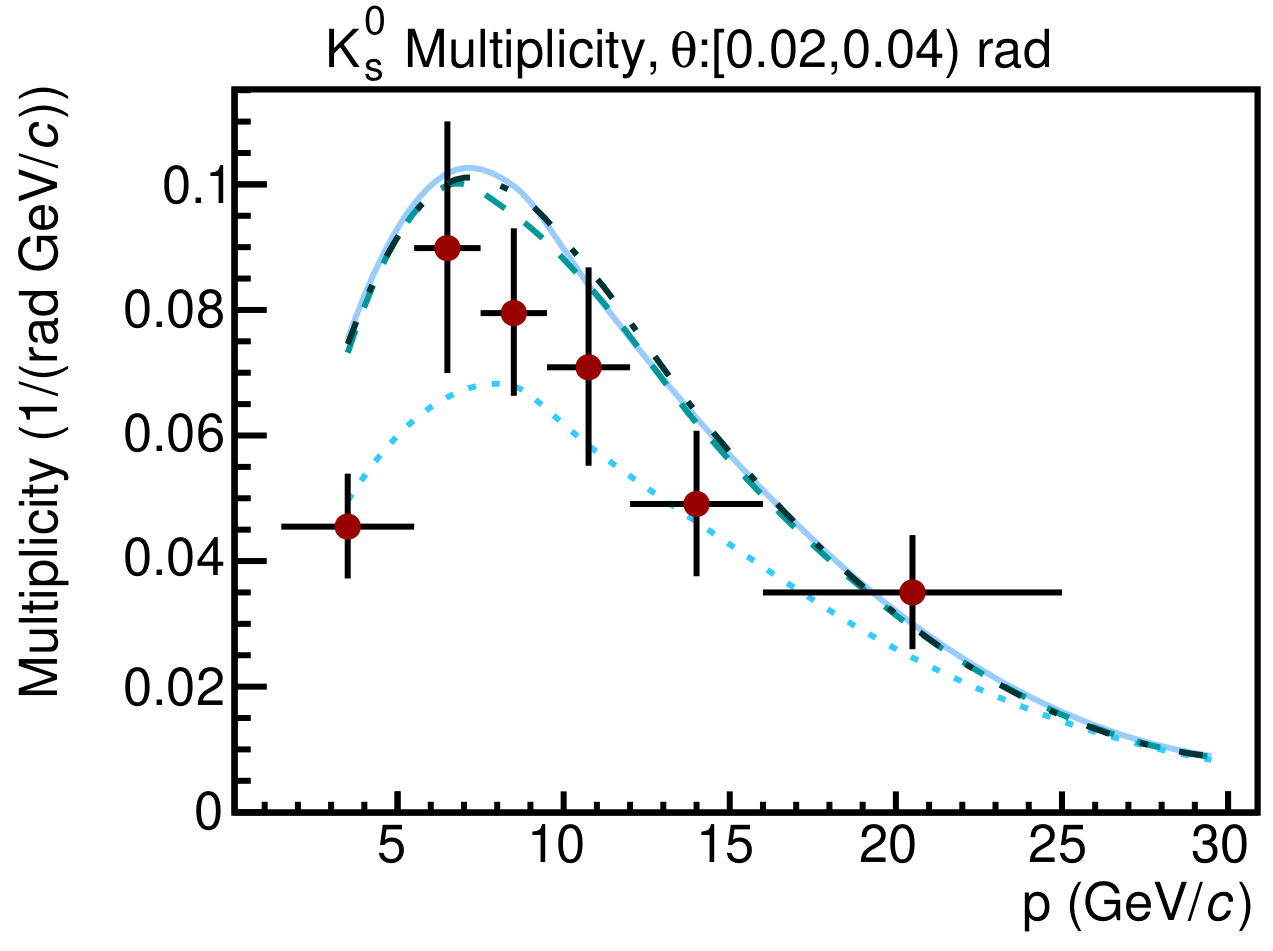}
\caption{Example $K^0_S$ multiplicity measurements for the angular bins [0, 0.02) rad and [0.02, 0.04) rad. The uncertainties shown are the total ones. The data points, shown in red, are compared to four physics lists from \GeantFour version 10.7.0.}
\label{fig:sampleK0SMultiplicities}
\end{figure*}

\begin{figure*}[t]
  \centering
    \includegraphics[width=0.49\textwidth]{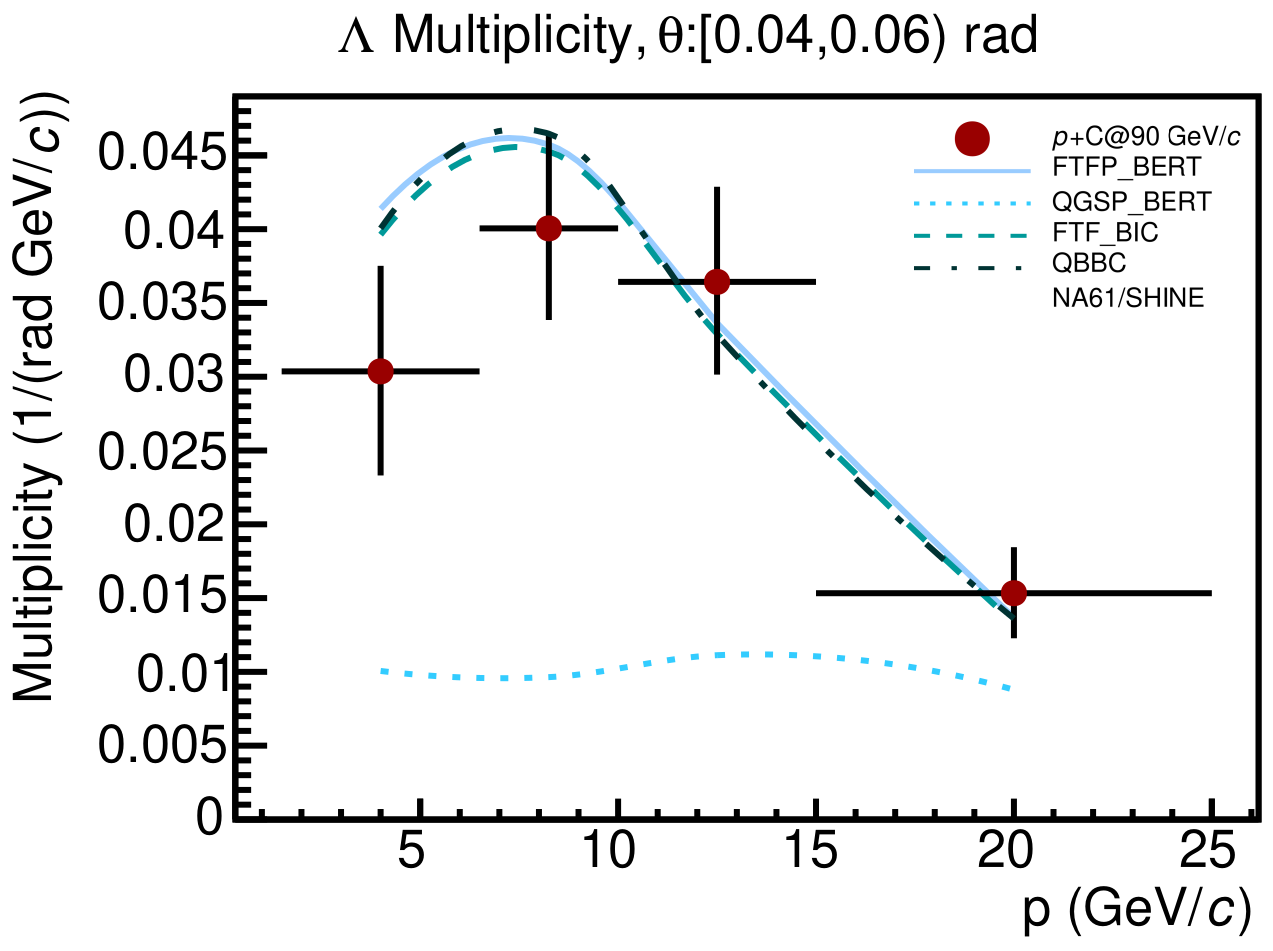}
    \includegraphics[width=0.49\textwidth]{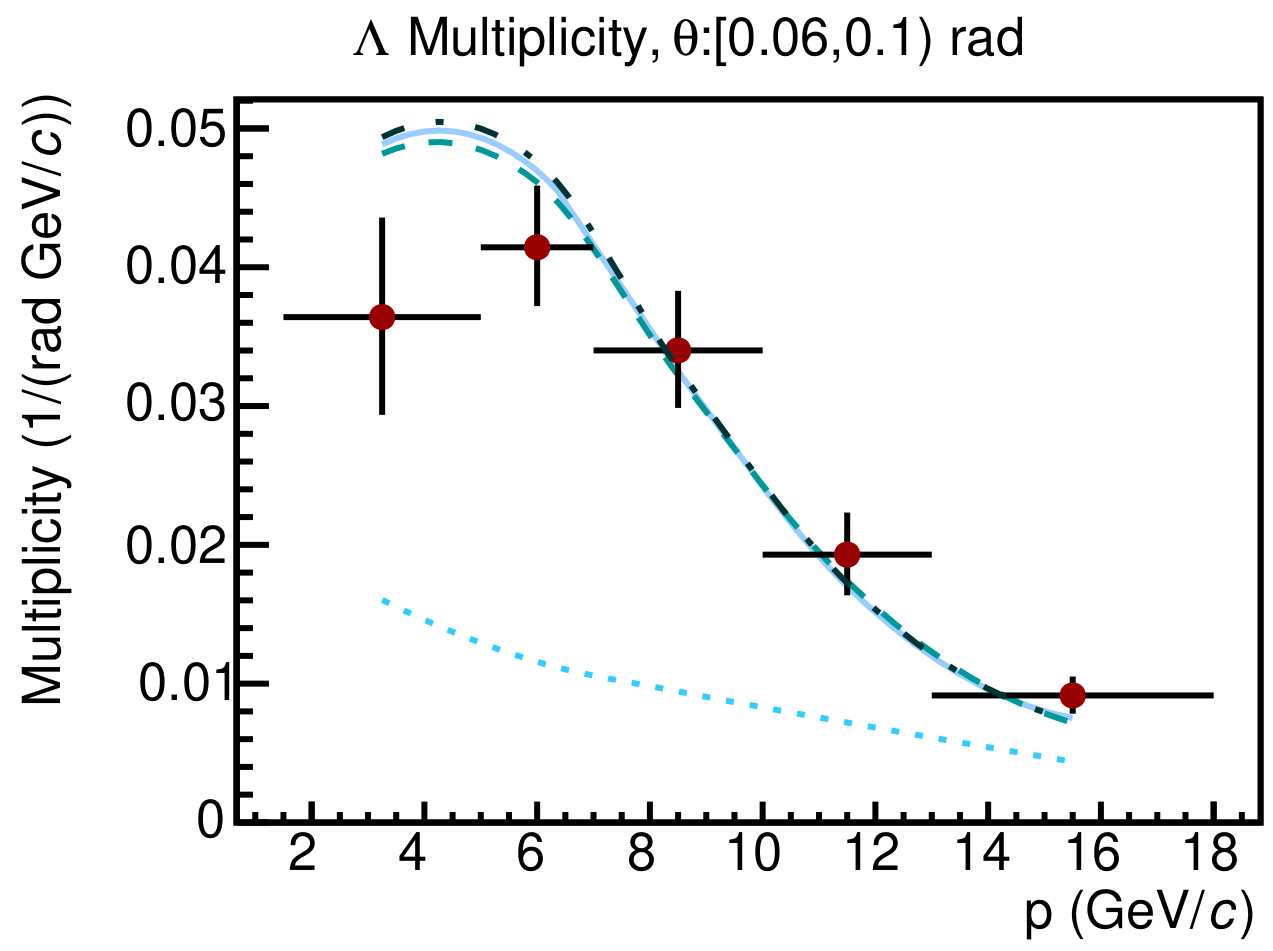}
\caption{Example $\Lambda$ multiplicity measurements for the angular bins [0.04, 0.06) rad and [0.06, 0.10) rad. The uncertainties shown are the total ones. The data points, shown in red, are compared to four physics lists from \GeantFour version 10.7.0.}
\label{fig:sampleLambdaMultiplicities}
\end{figure*}

\begin{figure*}[t]
  \centering
    \includegraphics[width=0.49\textwidth]{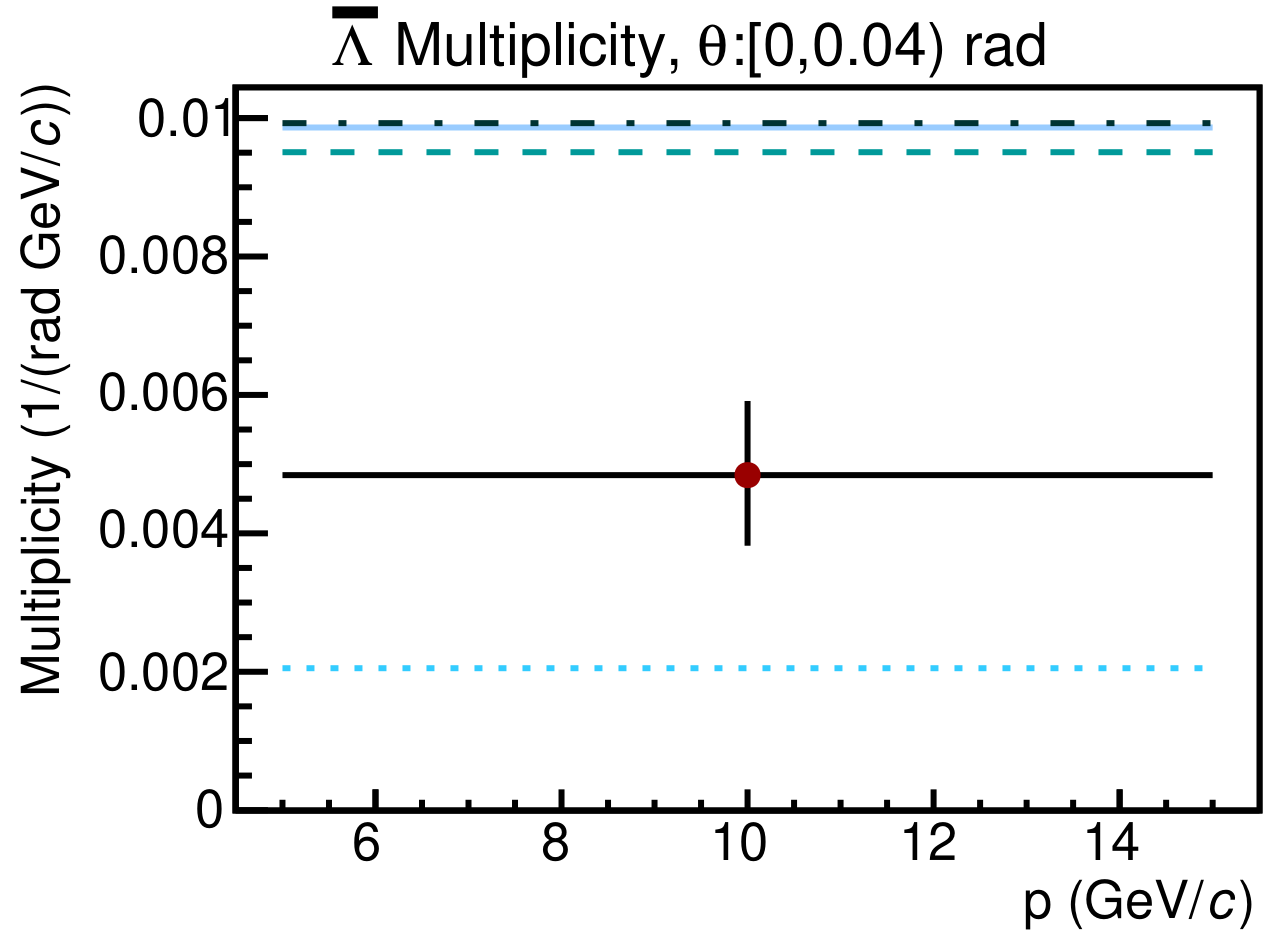}
    \includegraphics[width=0.49\textwidth]{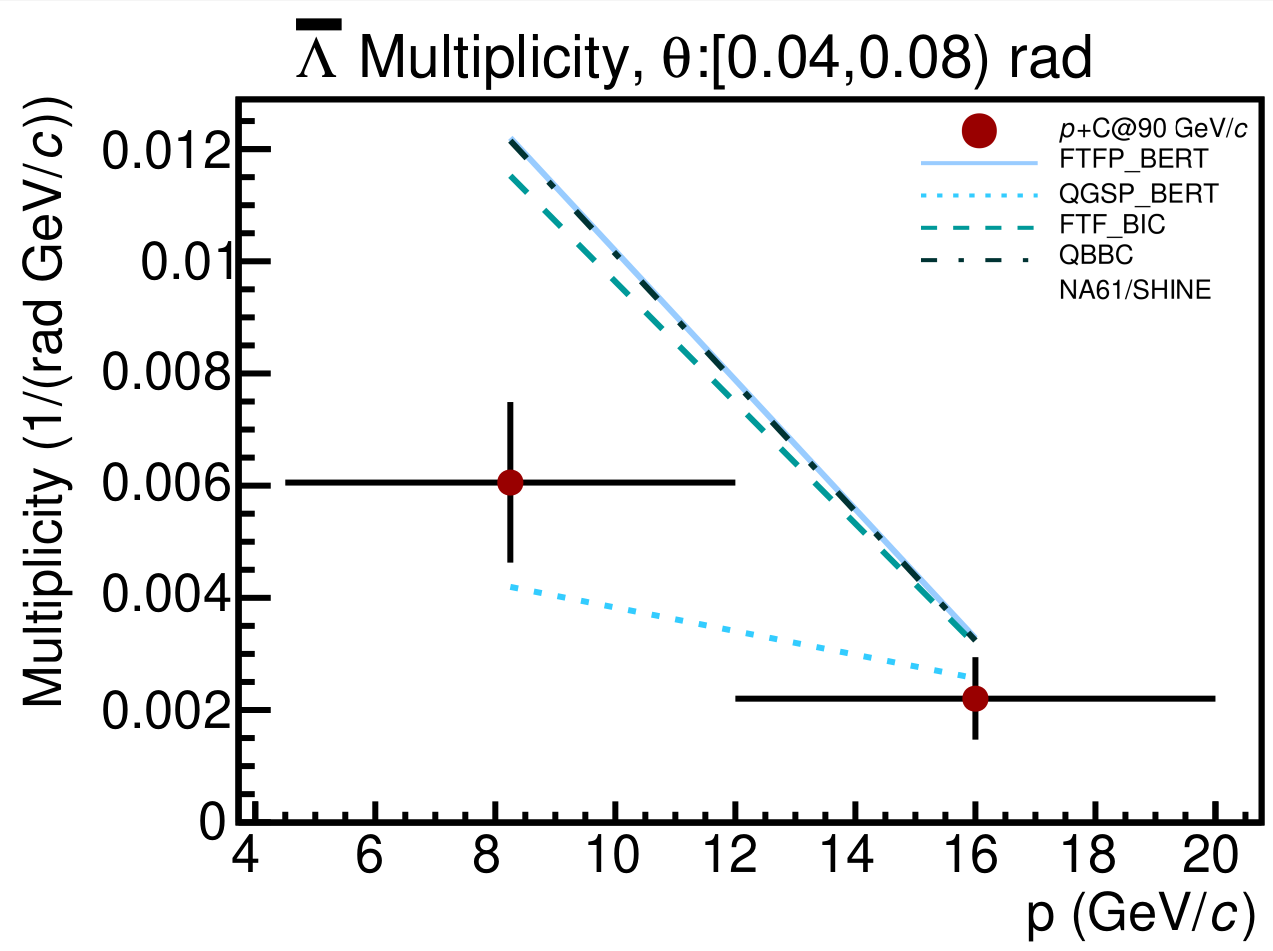}
\caption{Example $\bar{\Lambda}$ multiplicity measurements for the angular bins [0, 0.04) rad and [0.04, 0.08) rad. The uncertainties shown are the total ones. The data points, shown in red, are compared to four physics lists from \GeantFour version 10.7.0.}
\label{fig:sampleALamMultiplicities}
\end{figure*}

\begin{figure*}[t]
  \centering
    \includegraphics[width=0.49\textwidth]{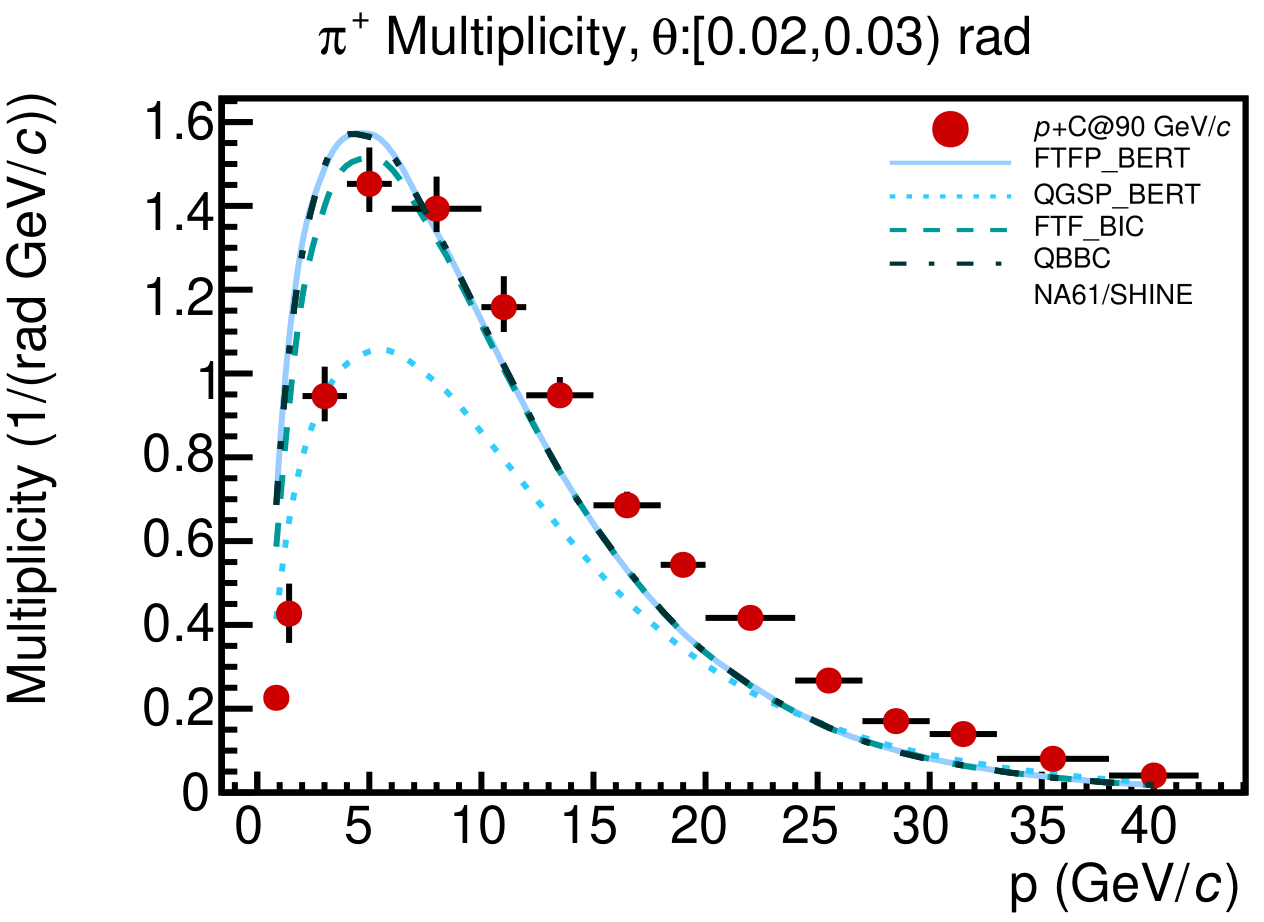}
    \includegraphics[width=0.49\textwidth]{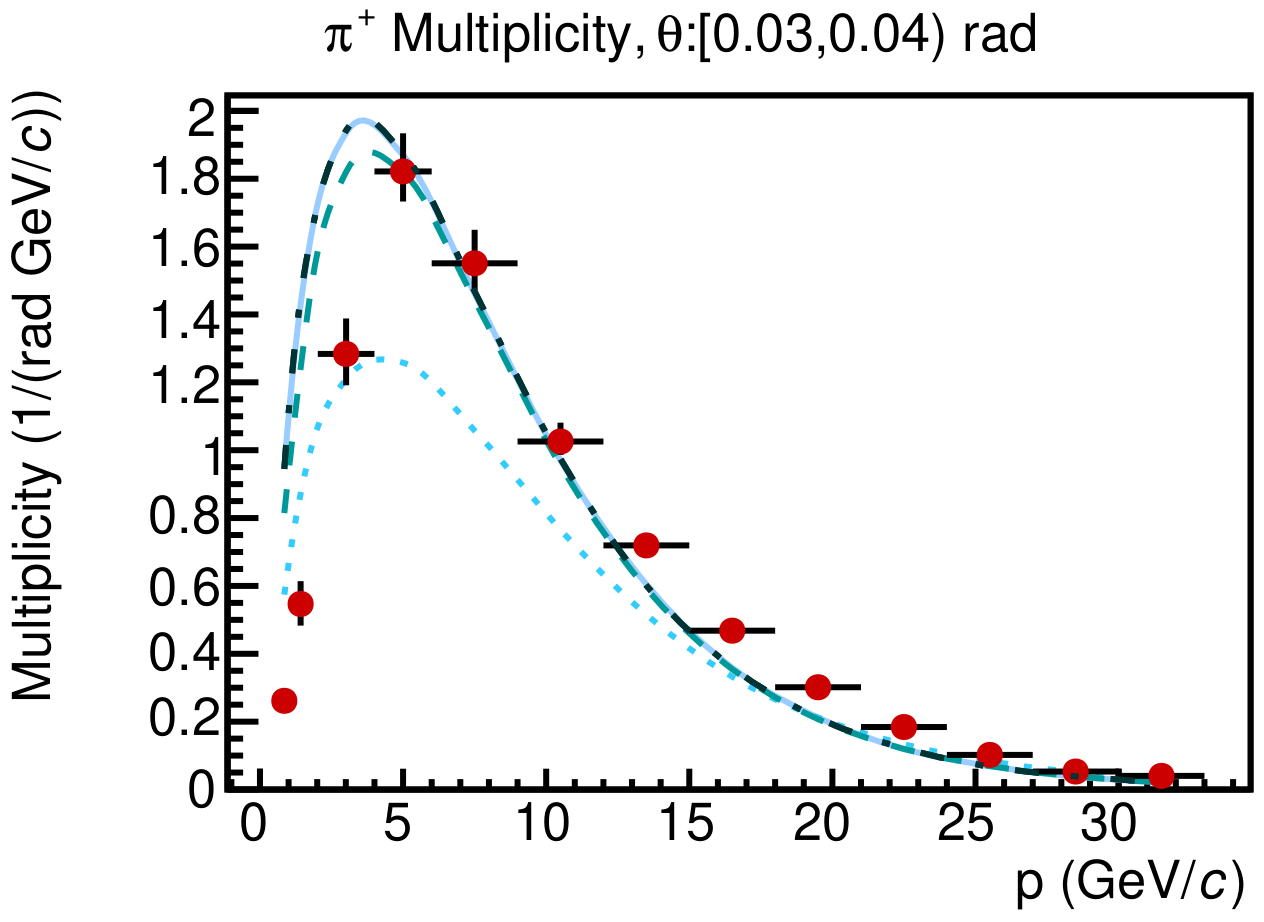} \\
    \includegraphics[width=0.49\textwidth]{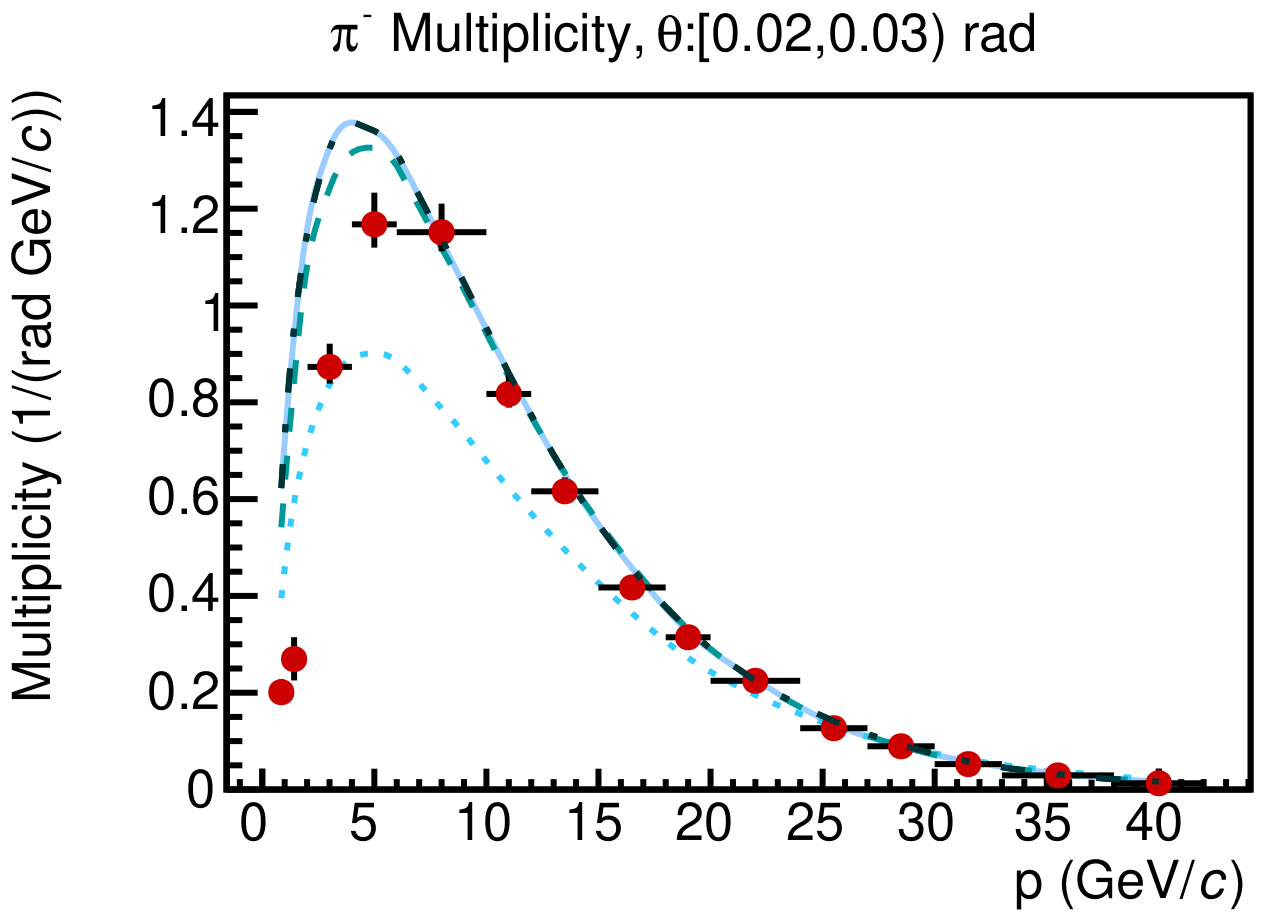}
    \includegraphics[width=0.49\textwidth]{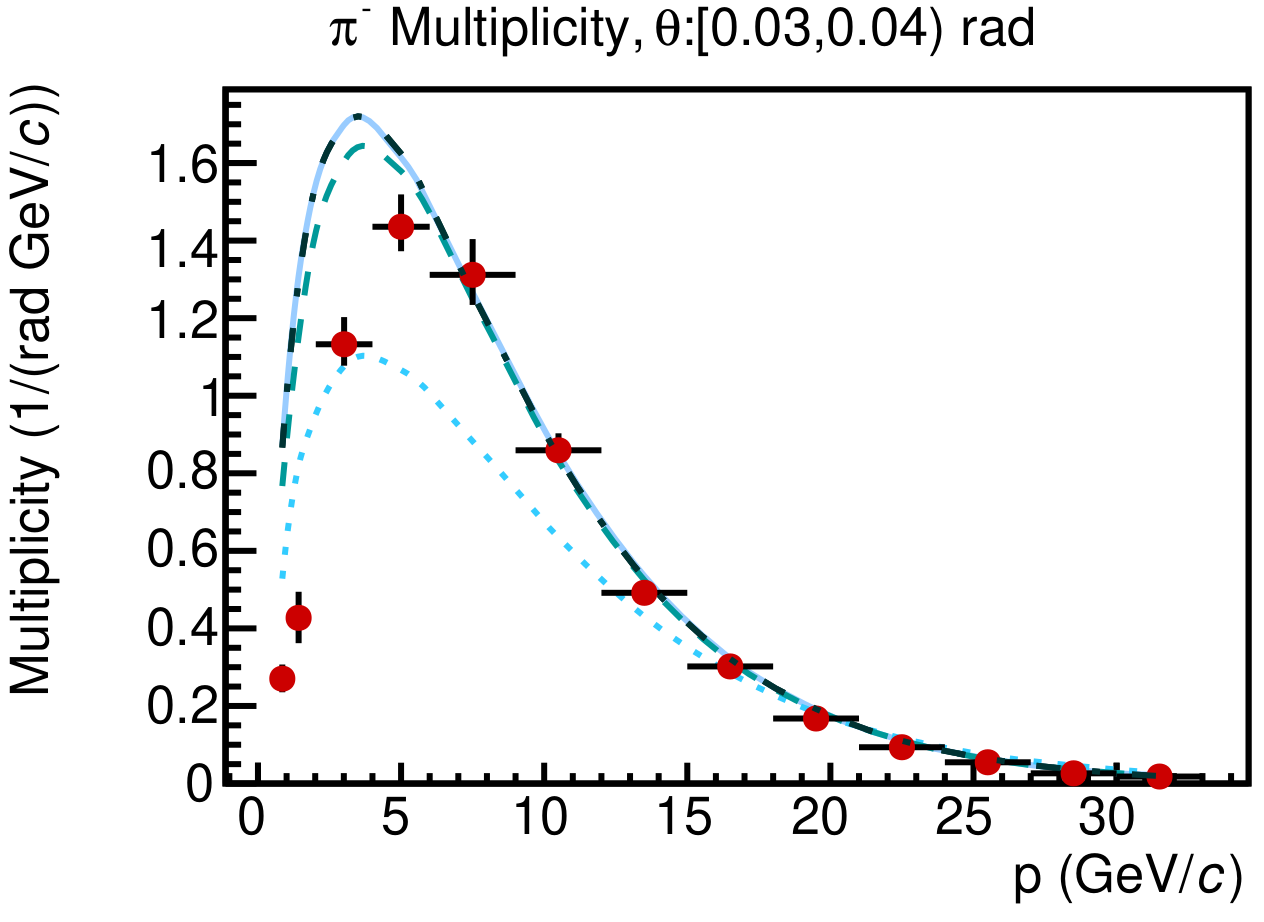}
\caption{Example $\pi^{\pm}$ multiplicity measurements for the angular bins [0.02, 0.03) rad and [0.03, 0.04) rad. The uncertainties shown are the total ones. The data points, shown in red, are compared to four physics lists from \GeantFour version 10.7.0.}
\label{fig:samplePionMultiplicities}
\end{figure*}

\begin{figure*}[t]
  \centering
    \includegraphics[width=0.49\textwidth]{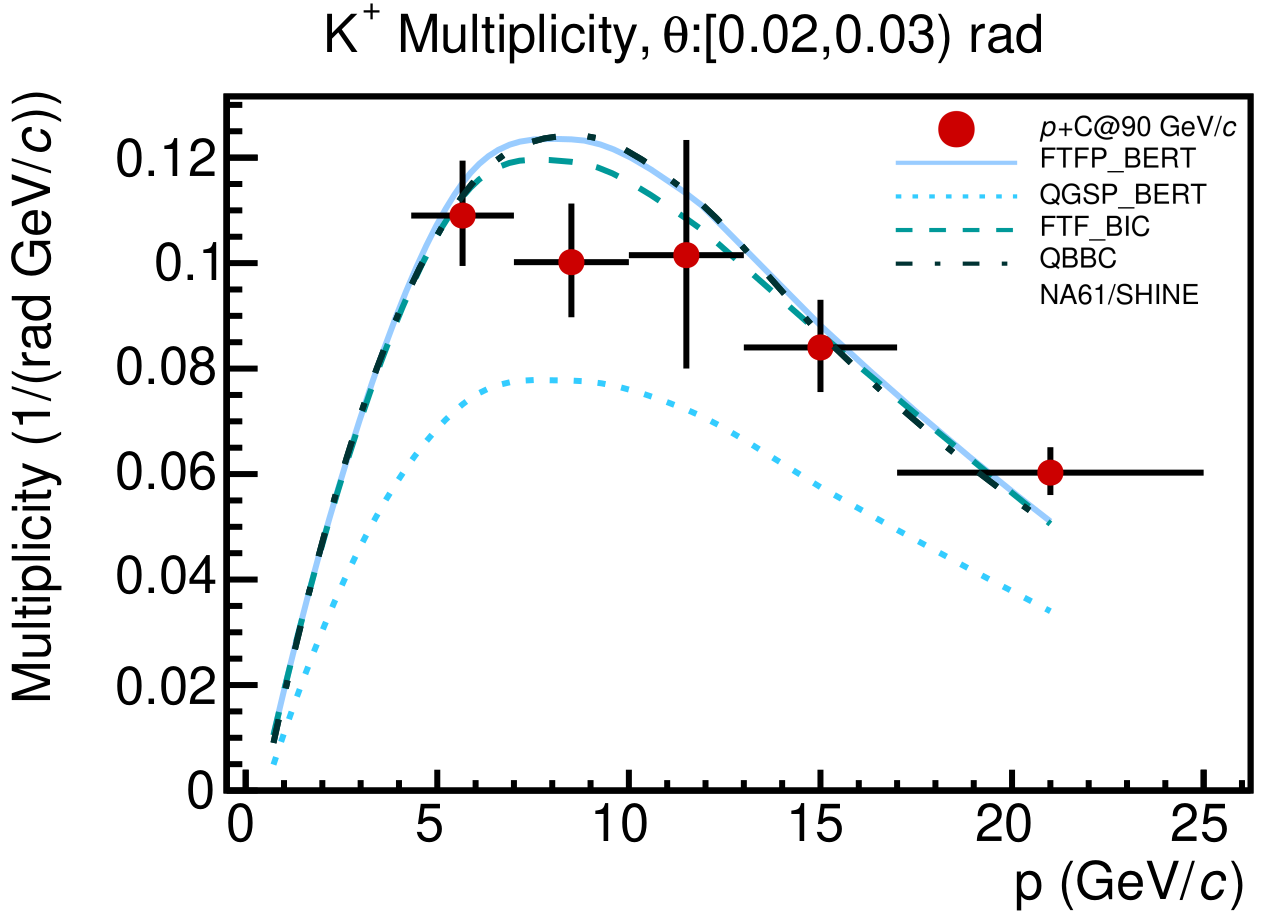}
    \includegraphics[width=0.49\textwidth]{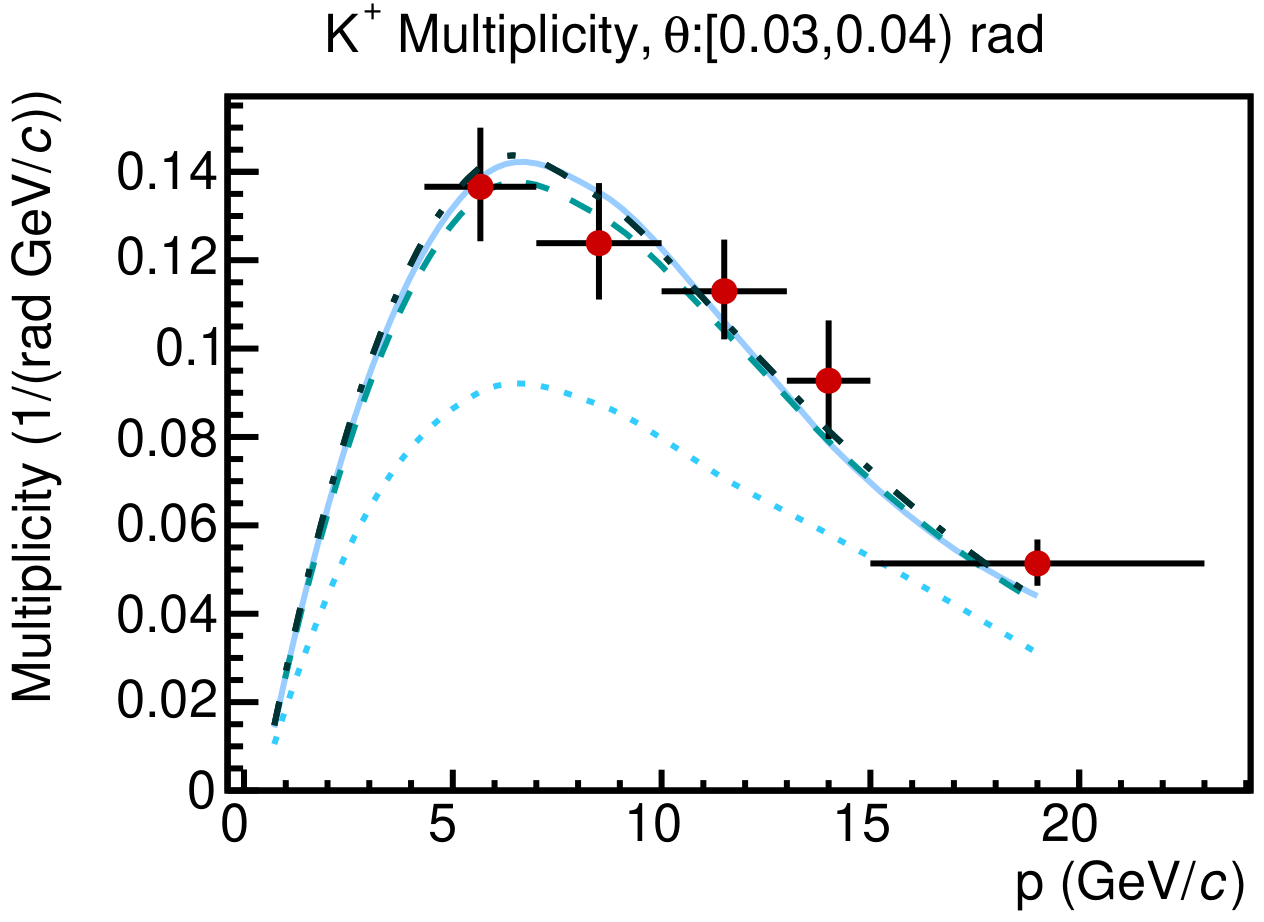} \\
    \includegraphics[width=0.49\textwidth]{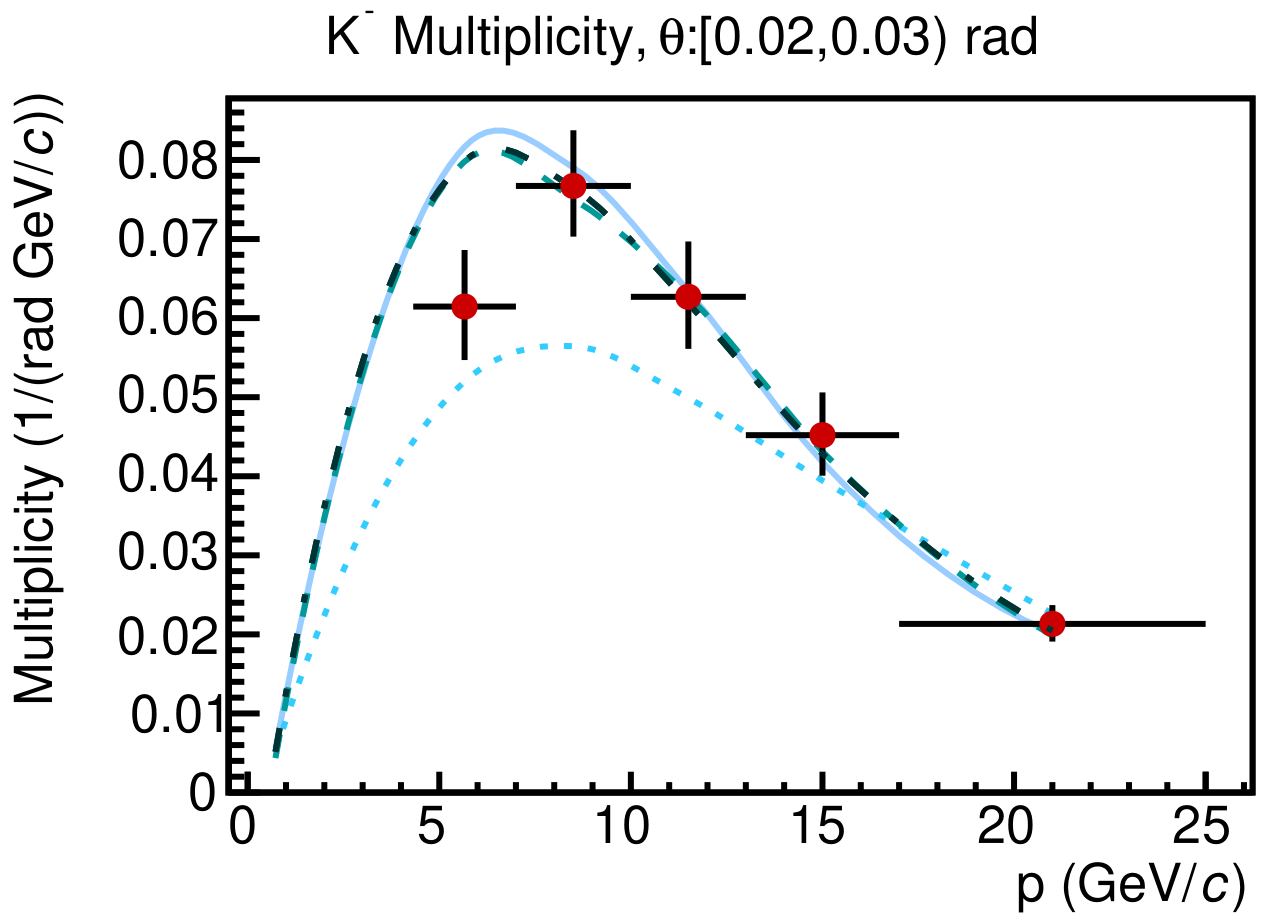}
    \includegraphics[width=0.49\textwidth]{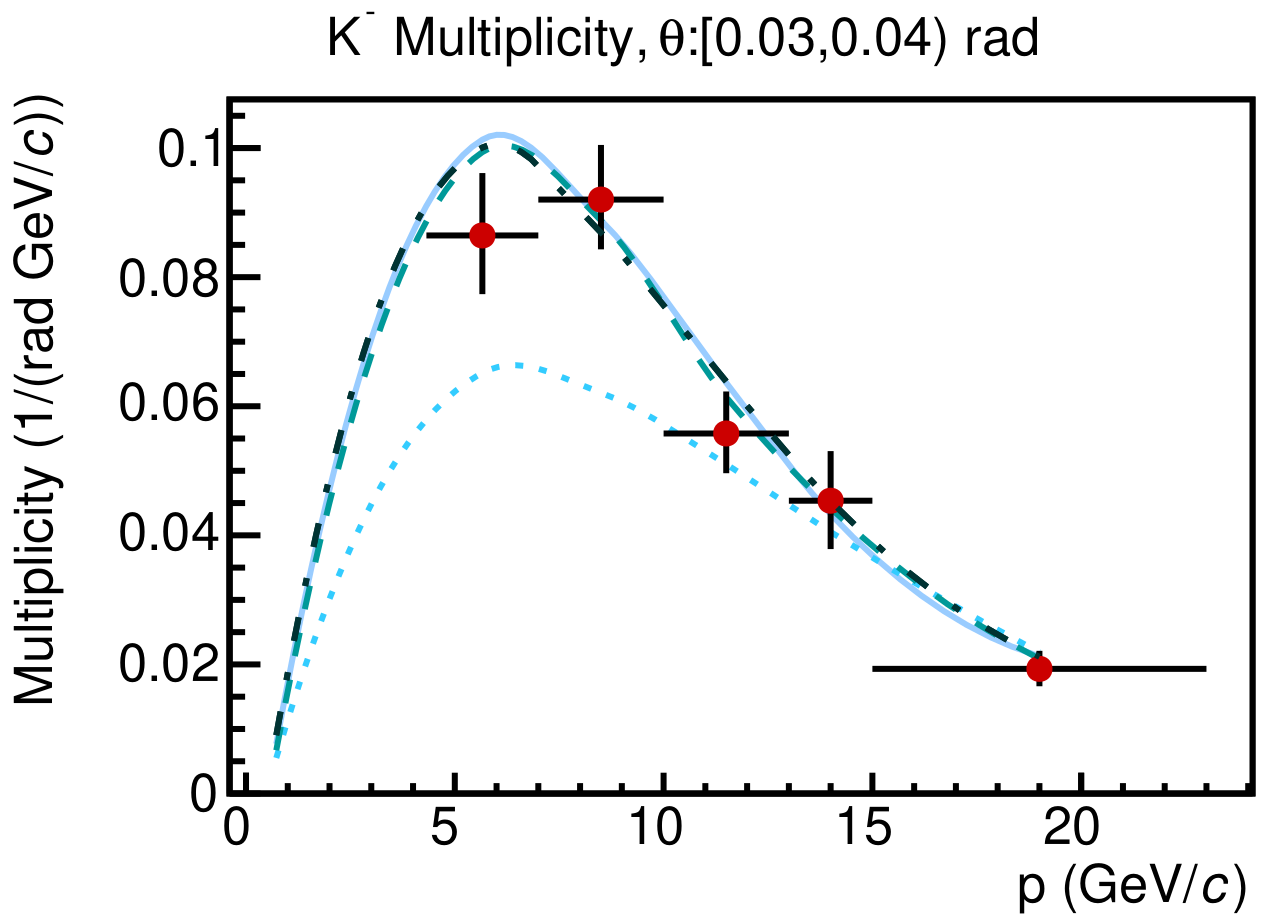}
\caption{Example $K^{\pm}$ multiplicity measurements for the angular bins [0.02, 0.03) rad and [0.03, 0.04) rad. The uncertainties shown are the total ones. The data points, shown in red, are compared to four physics lists from \GeantFour version 10.7.0.}
\label{fig:sampleKaonMultiplicities}
\end{figure*}

\begin{figure*}[t]
  \centering
    \includegraphics[width=0.49\textwidth]{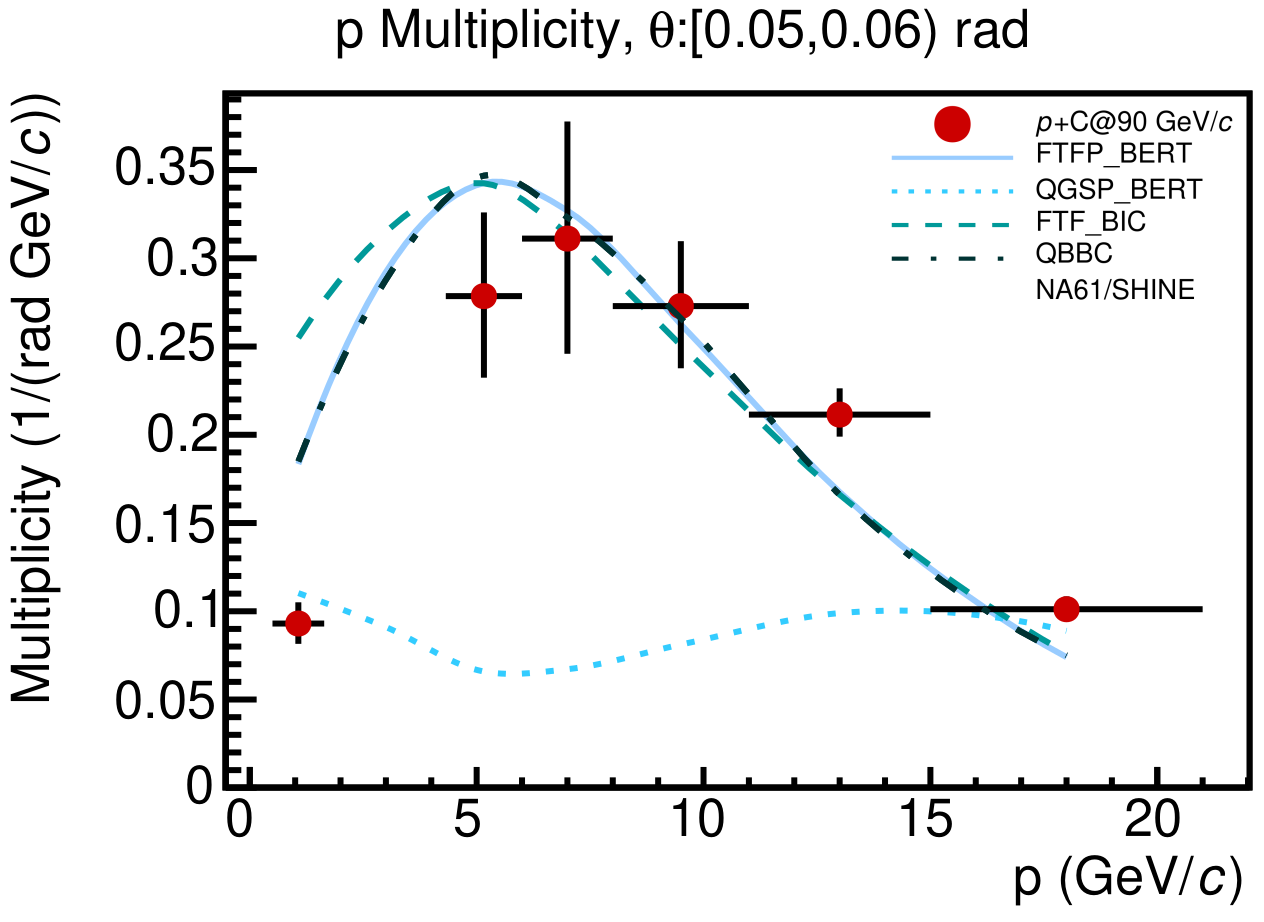}
    \includegraphics[width=0.49\textwidth]{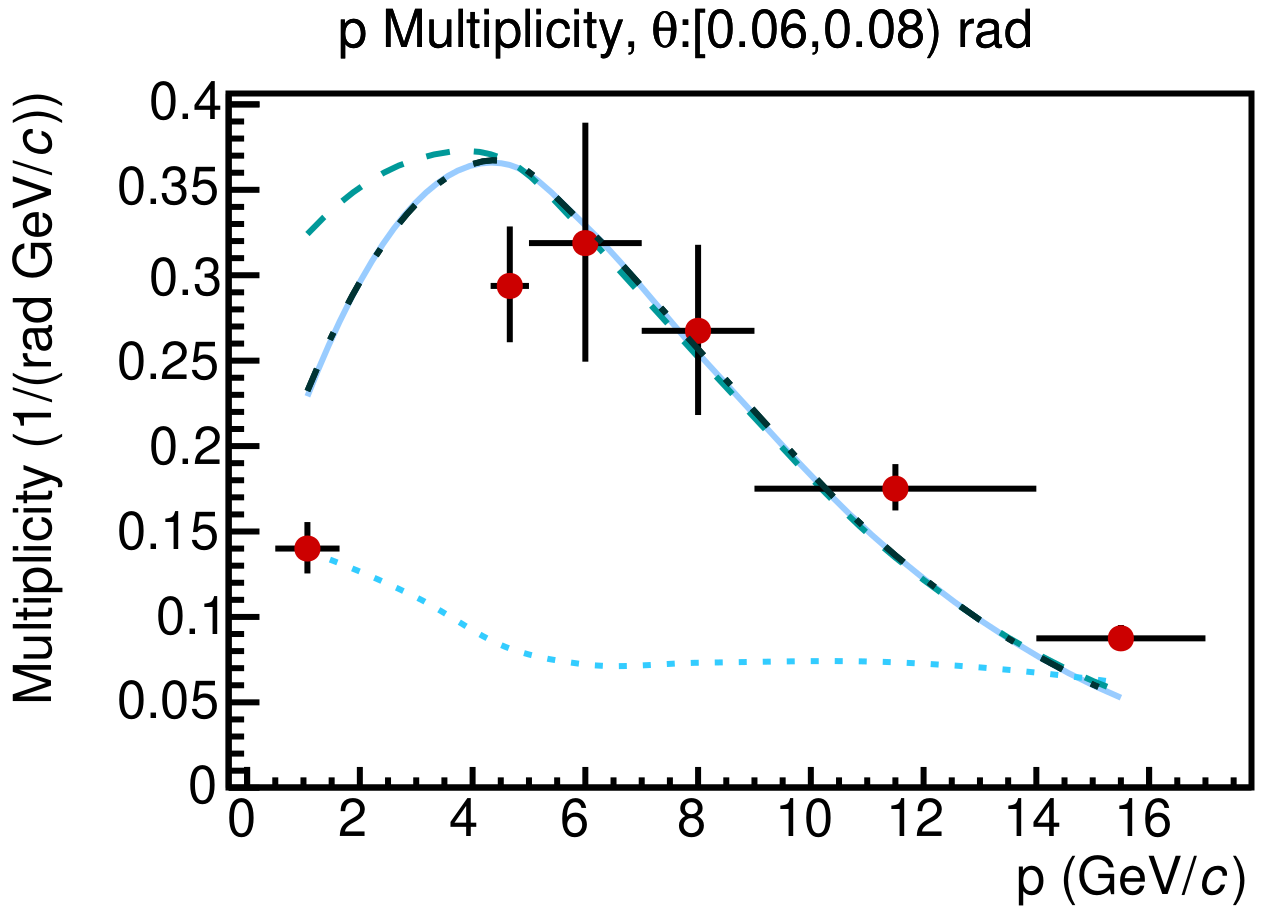} \\
    \includegraphics[width=0.49\textwidth]{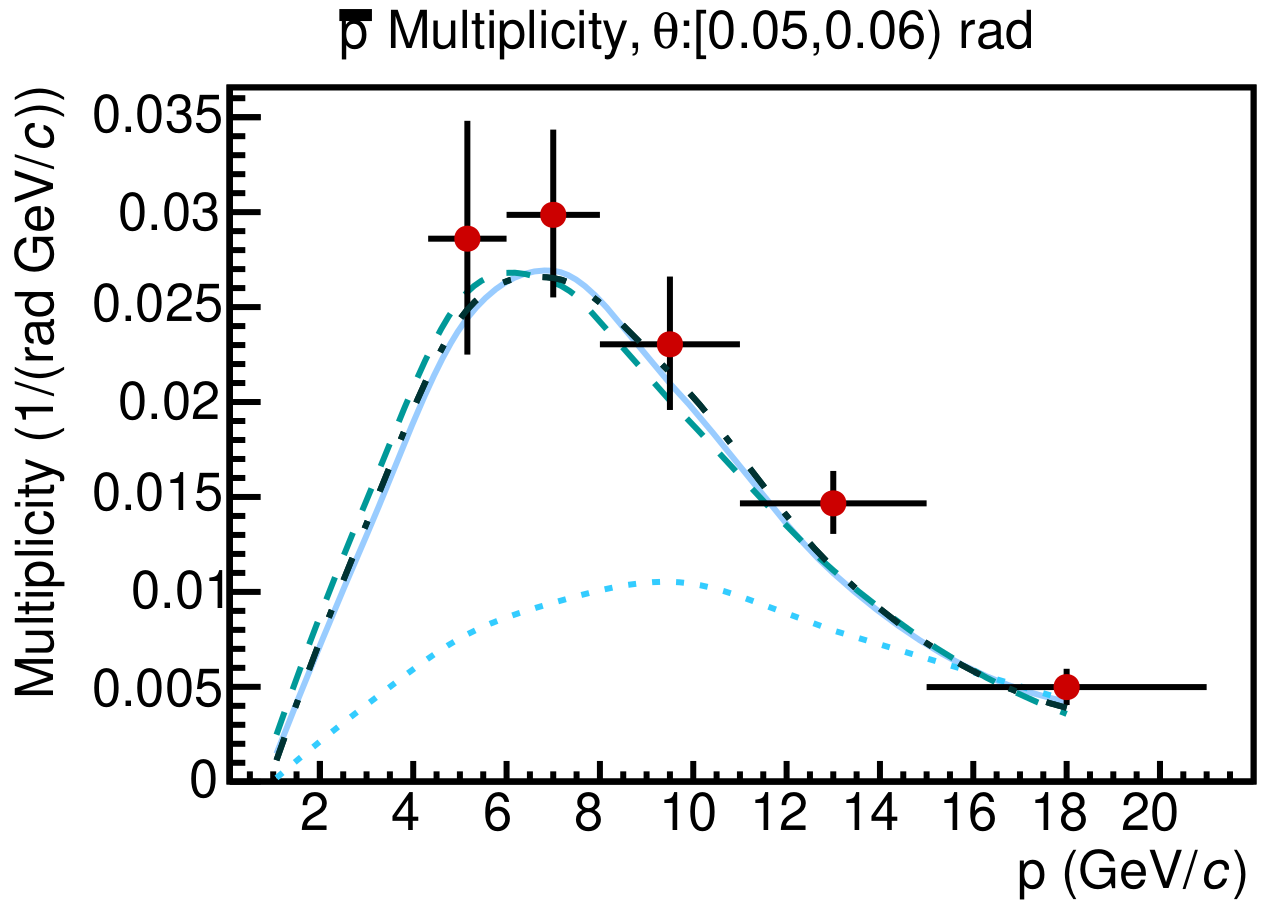}
    \includegraphics[width=0.49\textwidth]{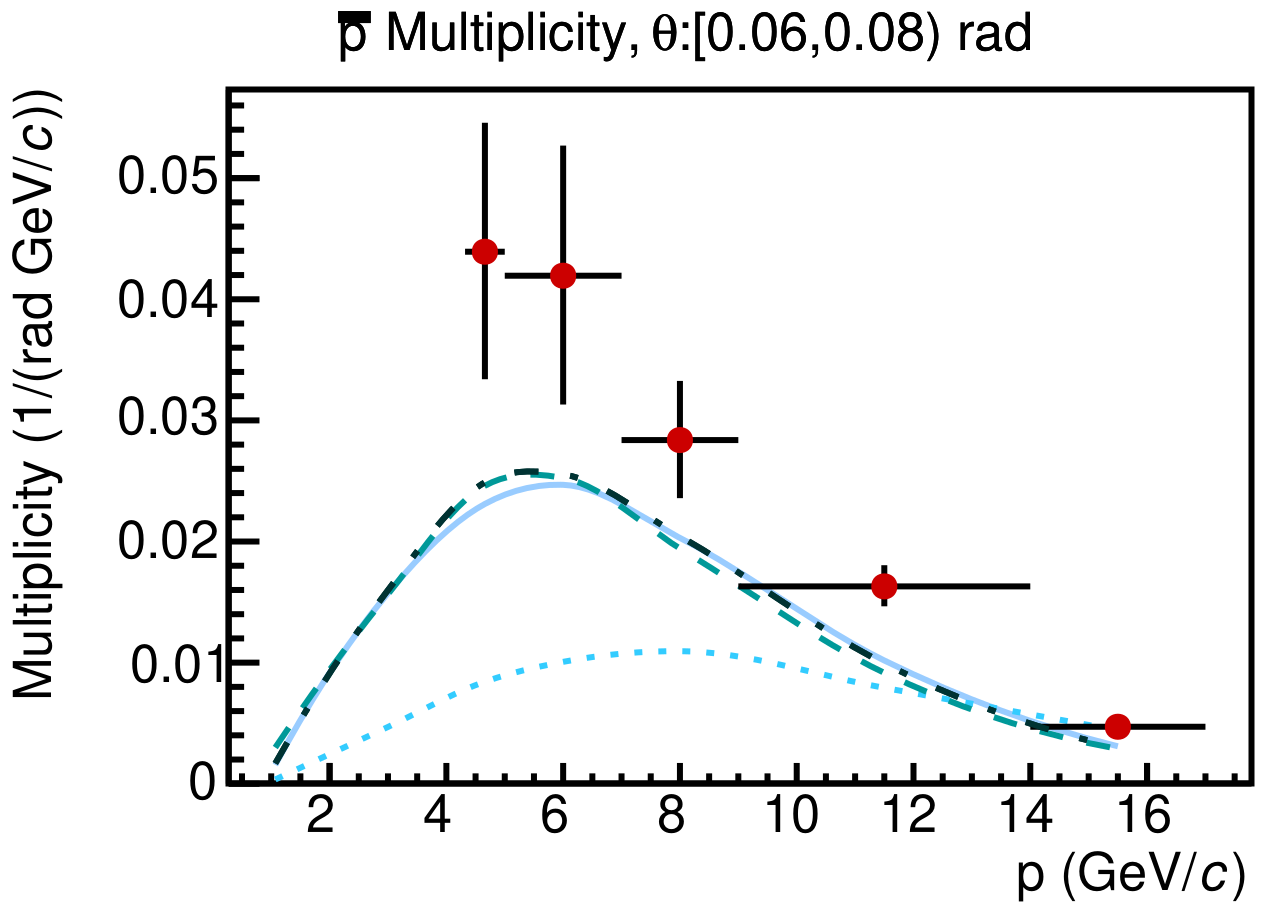}
\caption{Example $p / \bar{p}$ multiplicity measurements for the angular bins [0.05, 0.06) rad and [0.06, 0.08) rad. The uncertainties shown are the total ones. The data points, shown in red, are compared to four physics lists from \GeantFour version 10.7.0.}
\label{fig:sampleProtonMultiplicities}
\end{figure*}

Figures \ref{fig:sampleK0SMultiplicities}-\ref{fig:sampleALamMultiplicities} show example differential multiplicities for $K^0_S$, $\Lambda$, and $\bar{\Lambda}$ from the neutral analysis, and Figures \ref{fig:samplePionMultiplicities}-\ref{fig:sampleProtonMultiplicities} show sample $\pi^\pm$, $K^\pm$, and $p / \bar{p}$ differential multiplicities. A lifetime cross check, which bins the neutral species in bins of proper lifetime to ensure the reconstructed lifetime is in agreement with the PDG lifetime~\cite{pdg2024}, was performed in~\cite{allison2024hadron}.

\section{Summary}
\label{sec:summary}

\begin{figure*}[t]
  \centering
    \includegraphics[width=0.50\textwidth]{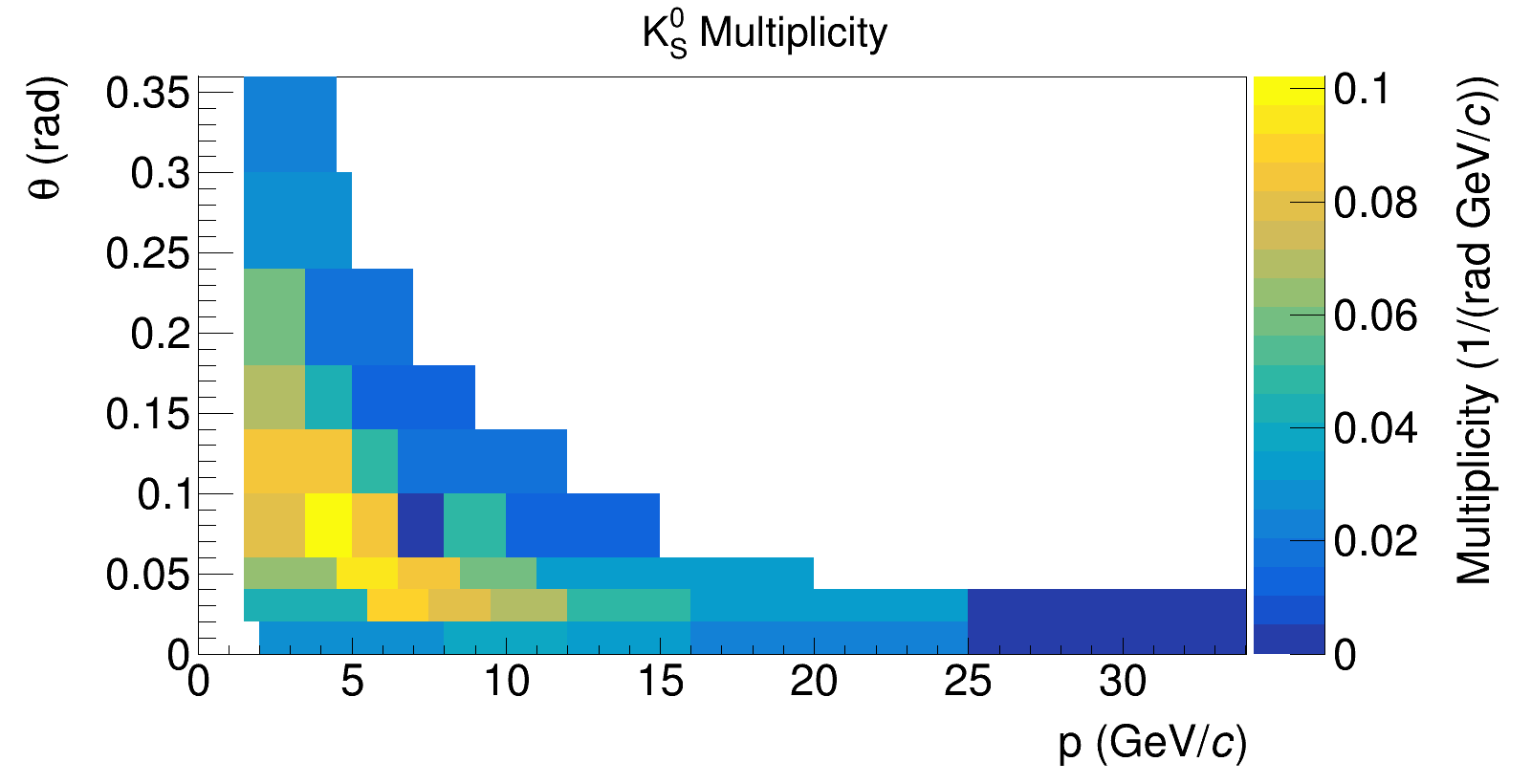}
\caption{Multiplicity measurements for the $K^0_S$ analysis. Numerical results can be found in \cite{pC90EDMS}.}
\label{fig:2DMultiplicitiesK0S}
\end{figure*}

\begin{figure*}[t]
  \centering
    \includegraphics[width=0.50\textwidth]{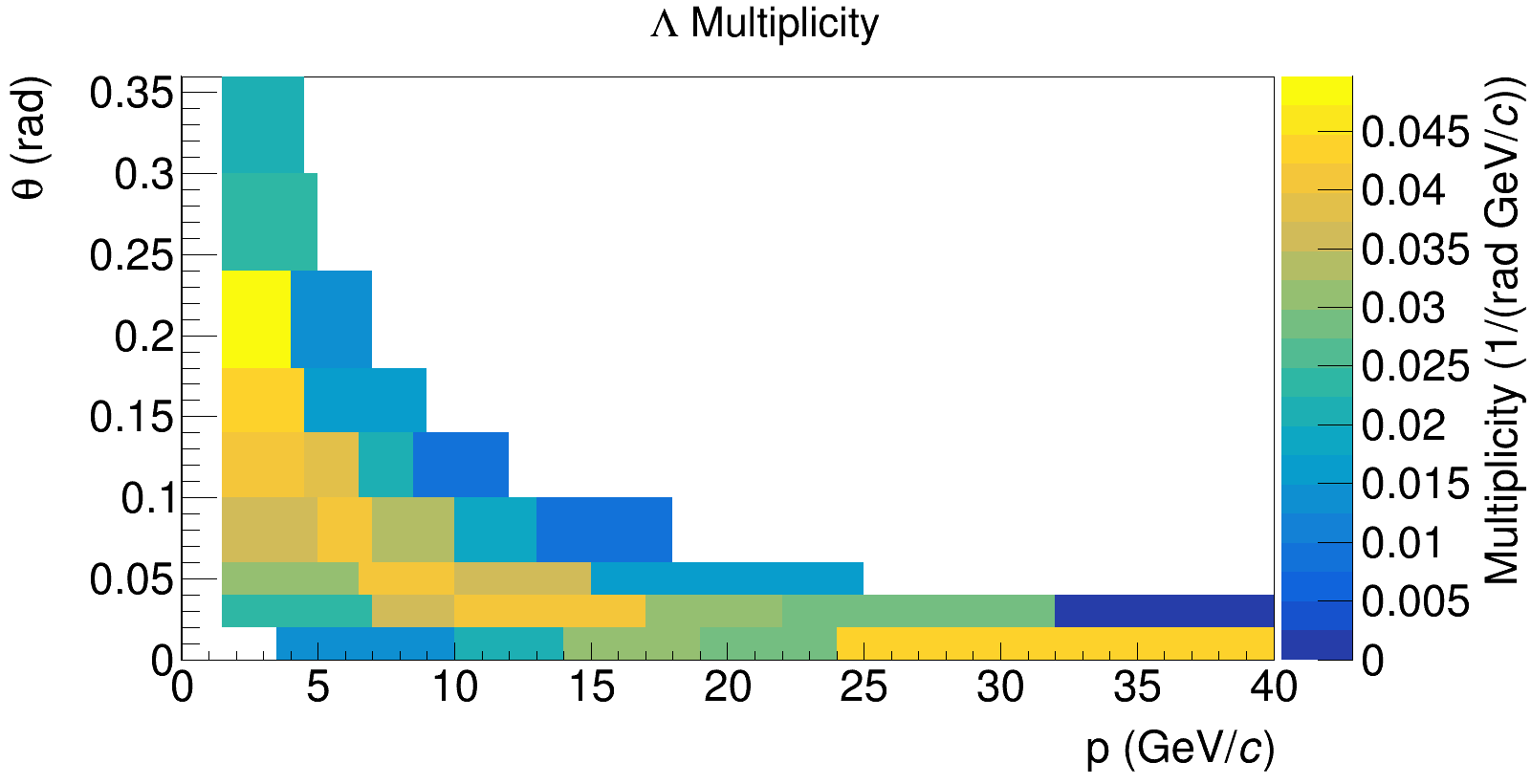}
\caption{Multiplicity measurements for the $\Lambda$ analysis. Numerical results can be found in \cite{pC90EDMS}.}
\label{fig:2DMultiplicitiesLam}
\end{figure*}

\begin{figure*}[t]
  \centering
    \includegraphics[width=0.50\textwidth]{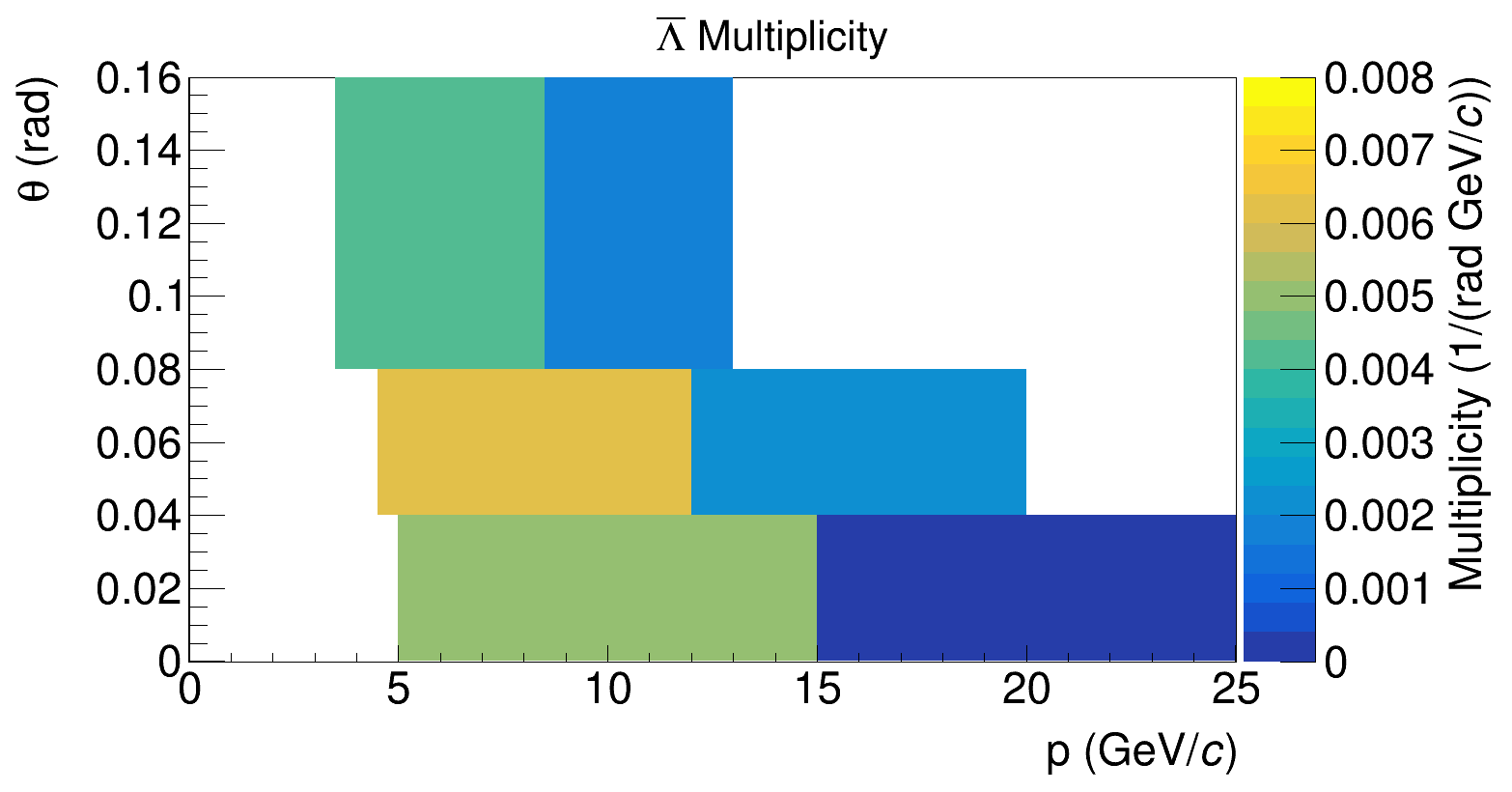}
\caption{Multiplicity measurements for the $\bar{\Lambda}$ analysis. Numerical results can be found in \cite{pC90EDMS}.}
\label{fig:2DMultiplicitiesALam}
\end{figure*}

\begin{figure*}[t]
  \centering
    \includegraphics[width=0.47\textwidth]{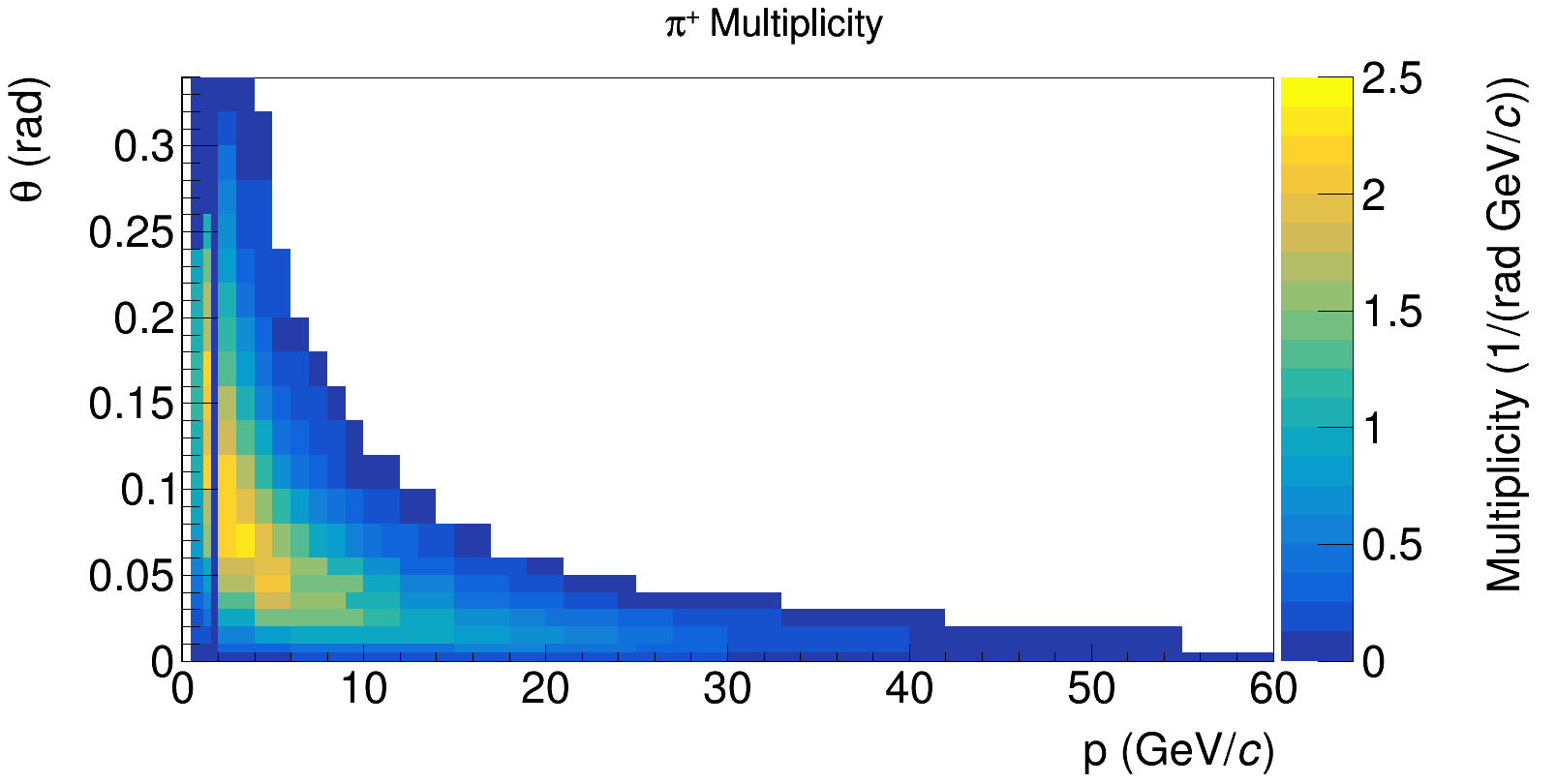}
    \hspace{2em}
    \includegraphics[width=0.47\textwidth]{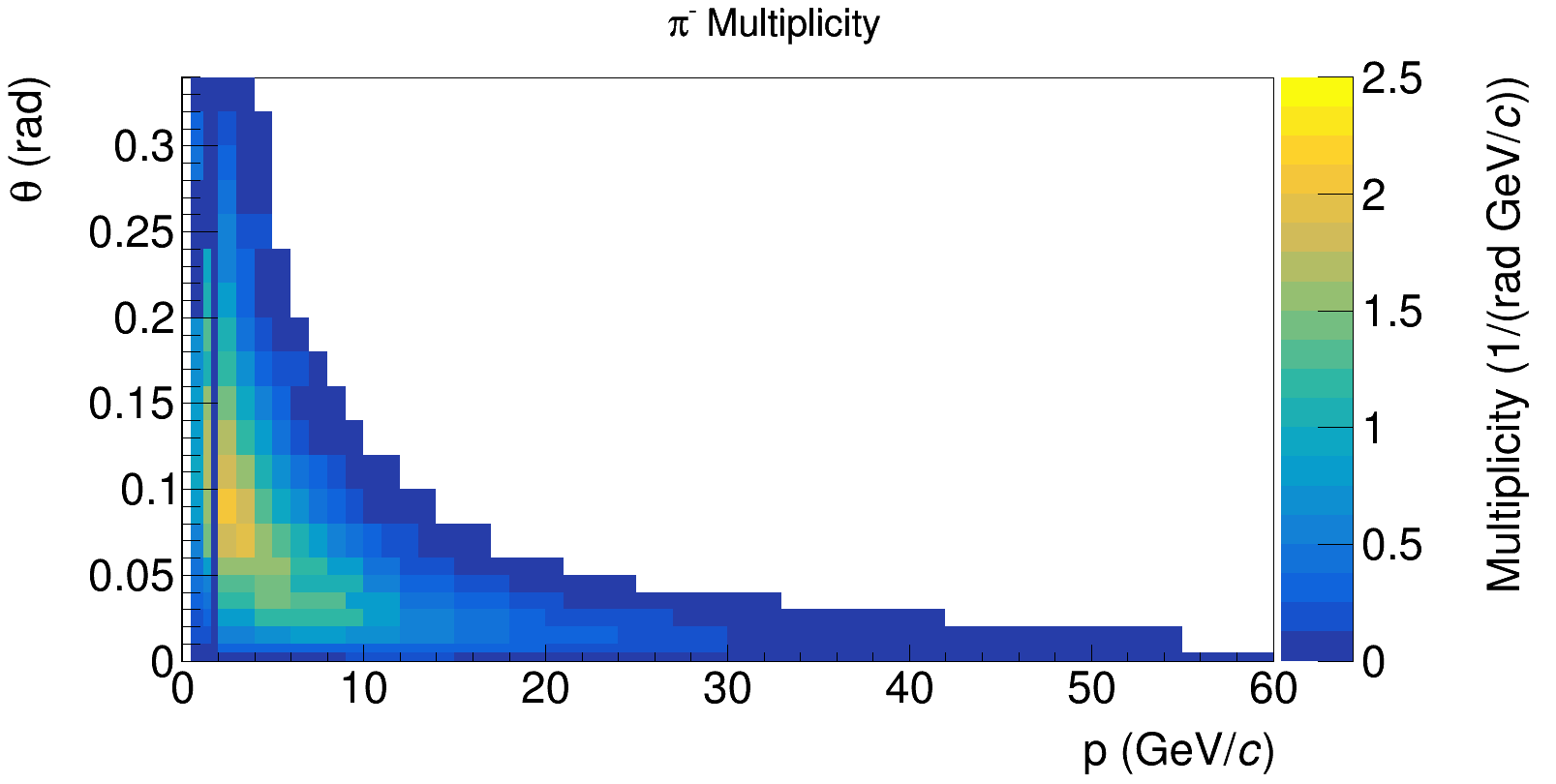}
\caption{Multiplicity measurements for the $\pi^{\pm}$ analyses. Numerical results can be found in \cite{pC90EDMS}.}
\label{fig:2DMultiplicitiesPi}
\end{figure*}

\begin{figure*}[t]
  \centering
    \includegraphics[width=0.47\textwidth]{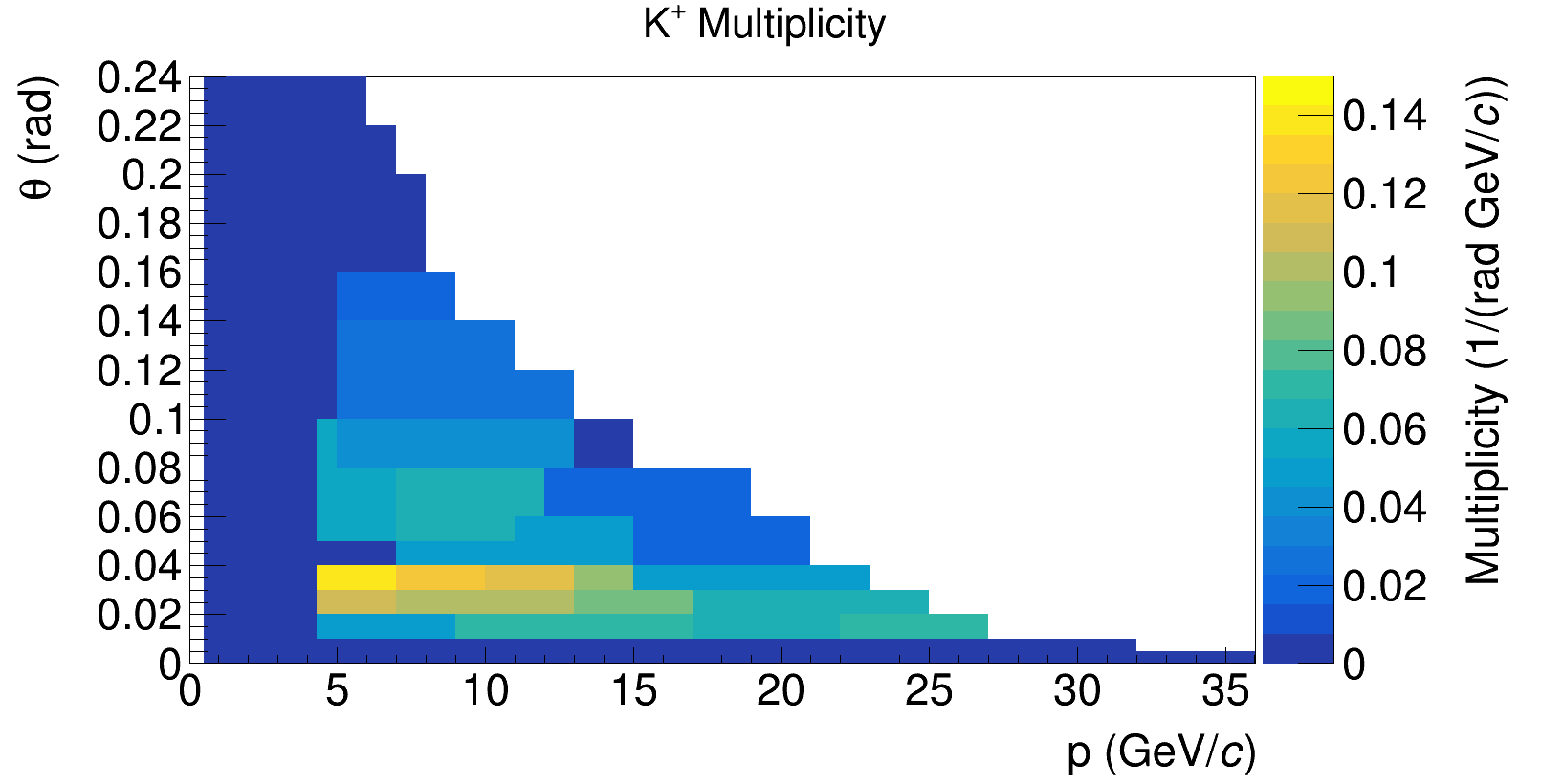}
    \hspace{2em}
    \includegraphics[width=0.47\textwidth]{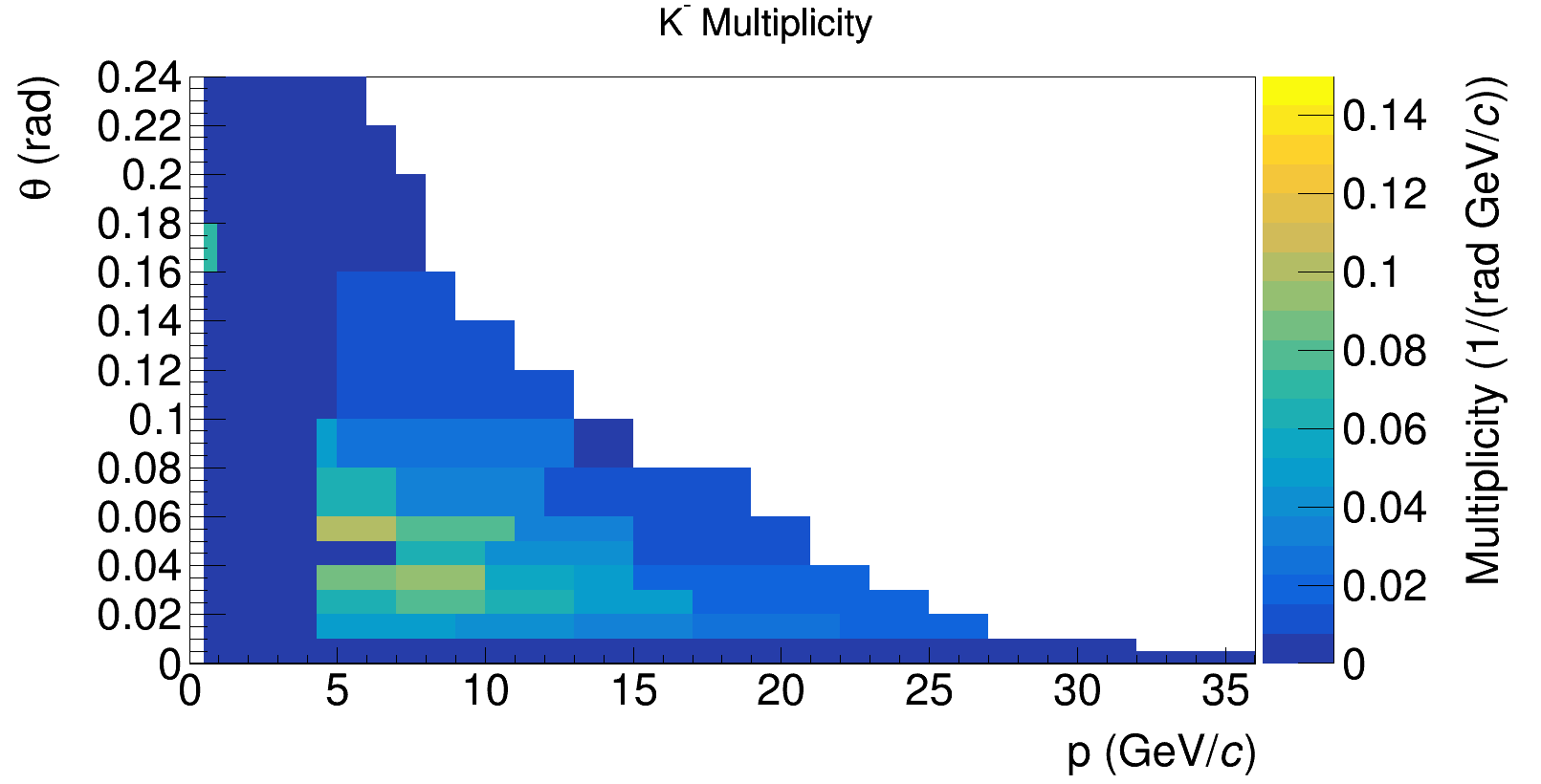}
\caption{Multiplicity measurements for the $K^{\pm}$ analyses. Numerical results can be found in \cite{pC90EDMS}.}
\label{fig:2DMultiplicitiesKaon}
\end{figure*}

\begin{figure*}[t]
  \centering
    \includegraphics[width=0.47\textwidth]{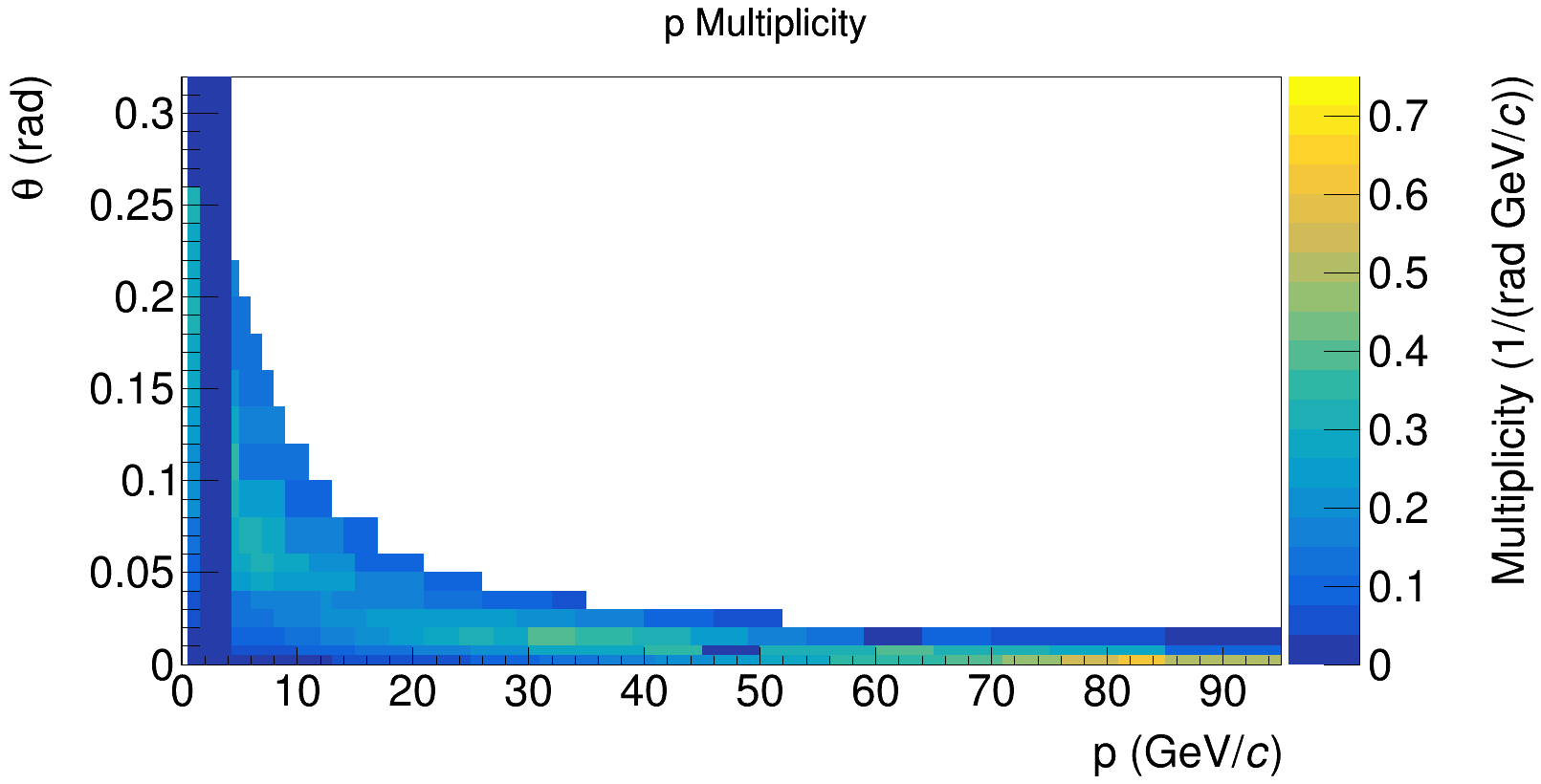}
    \hspace{2em}
    \includegraphics[width=0.47\textwidth]{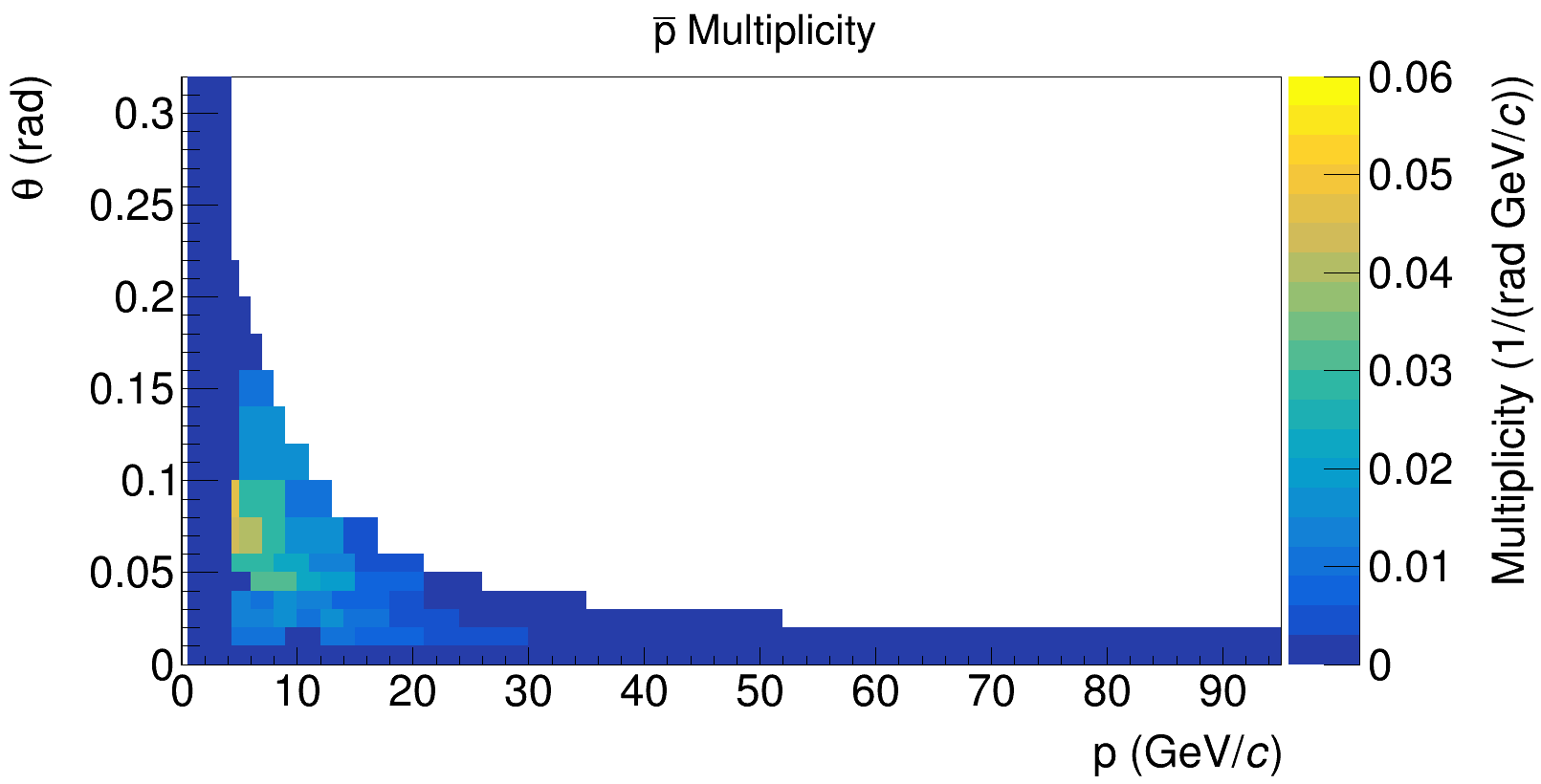}
\caption{Multiplicity measurements for the $p / \bar{p}$  analyses. Numerical results can be found in \cite{pC90EDMS}.}
\label{fig:2DMultiplicitiesProton}
\end{figure*}

Neutral- and charged-hadron production measurements for 90 GeV$/c$ proton-carbon interactions were presented, and example results were shown in Figures \ref{fig:sampleK0SMultiplicities}-\ref{fig:sampleALamMultiplicities} and Figures \ref{fig:samplePionMultiplicities}-\ref{fig:sampleProtonMultiplicities}. A two-dimensional overview for each particle species in all measured bins is shown in Figures \ref{fig:2DMultiplicitiesK0S}-\ref{fig:2DMultiplicitiesProton}. Bins with a total uncertainty exceeding 50\% or lying in the Bethe--Bloch overlap regions are excluded from the one-dimensional plots; in the two-dimensional plots these bins are shown with a multiplicity of 0. From Equation \ref{eq:neutralDifferentialMultiplicity}, the multiplicity measurements are normalized to the bin size.

\begin{figure*}[t]
  \centering
   \includegraphics[width=0.49\textwidth]{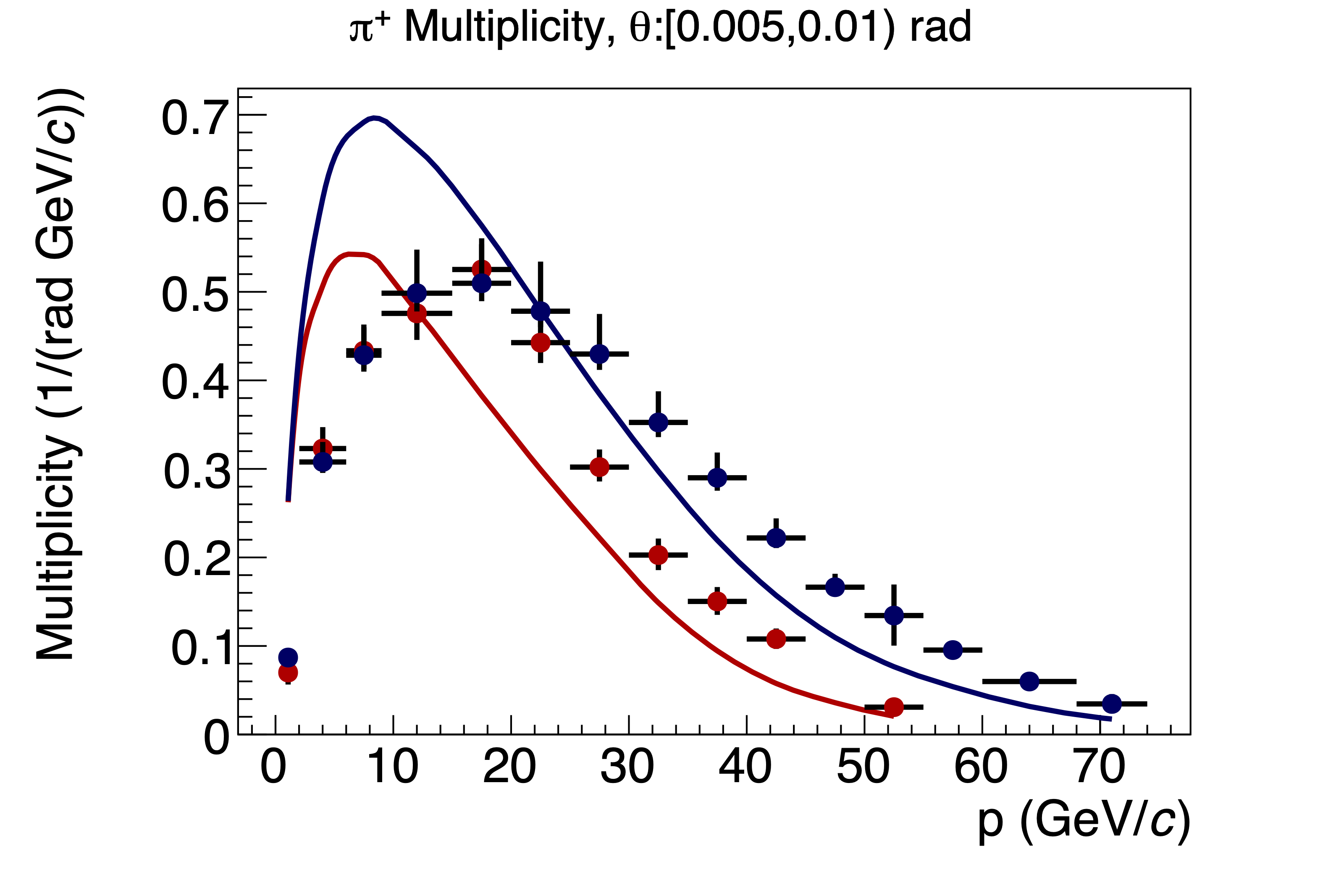}
    \includegraphics[width=0.49\textwidth]{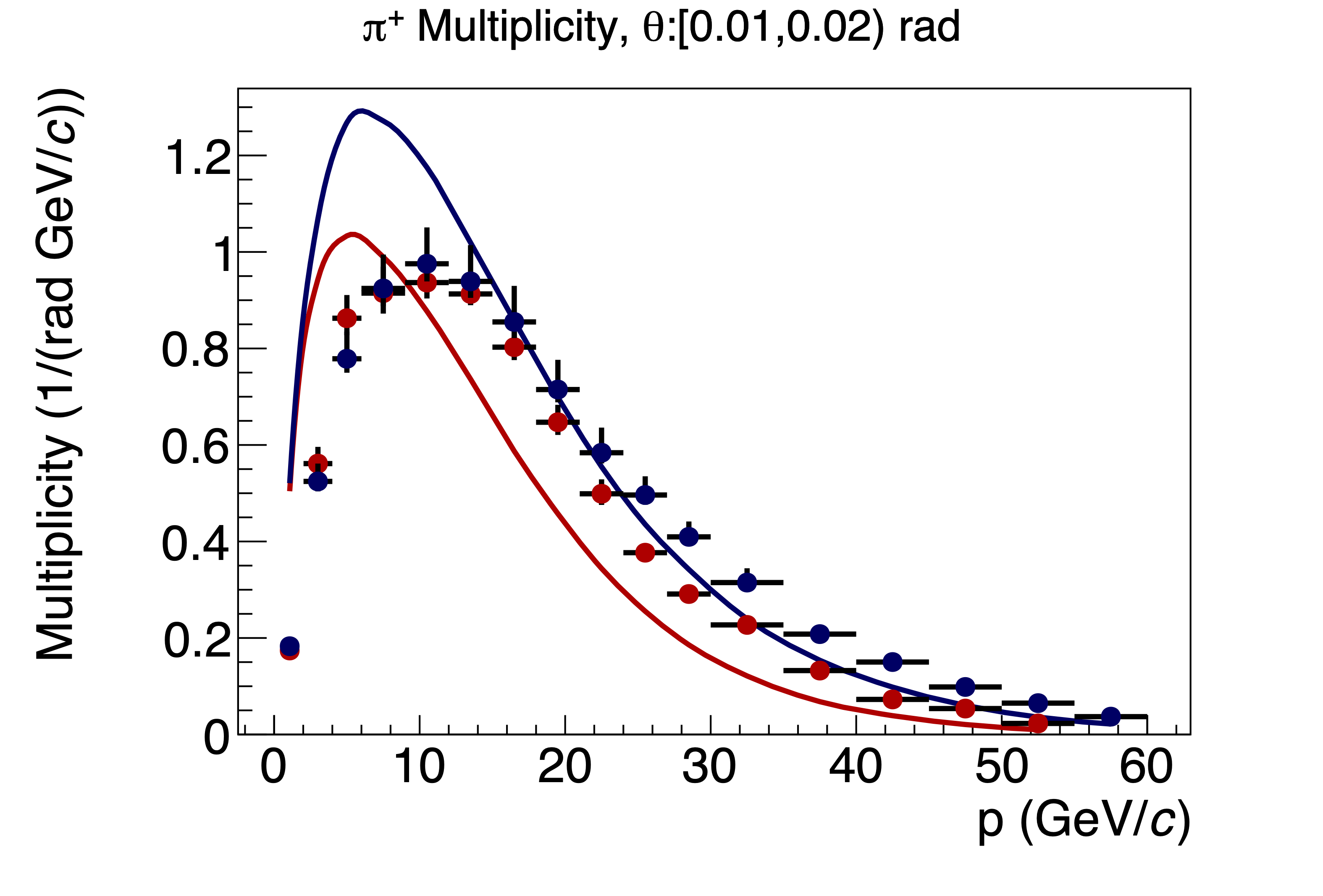} \\
    \includegraphics[width=0.49\textwidth]{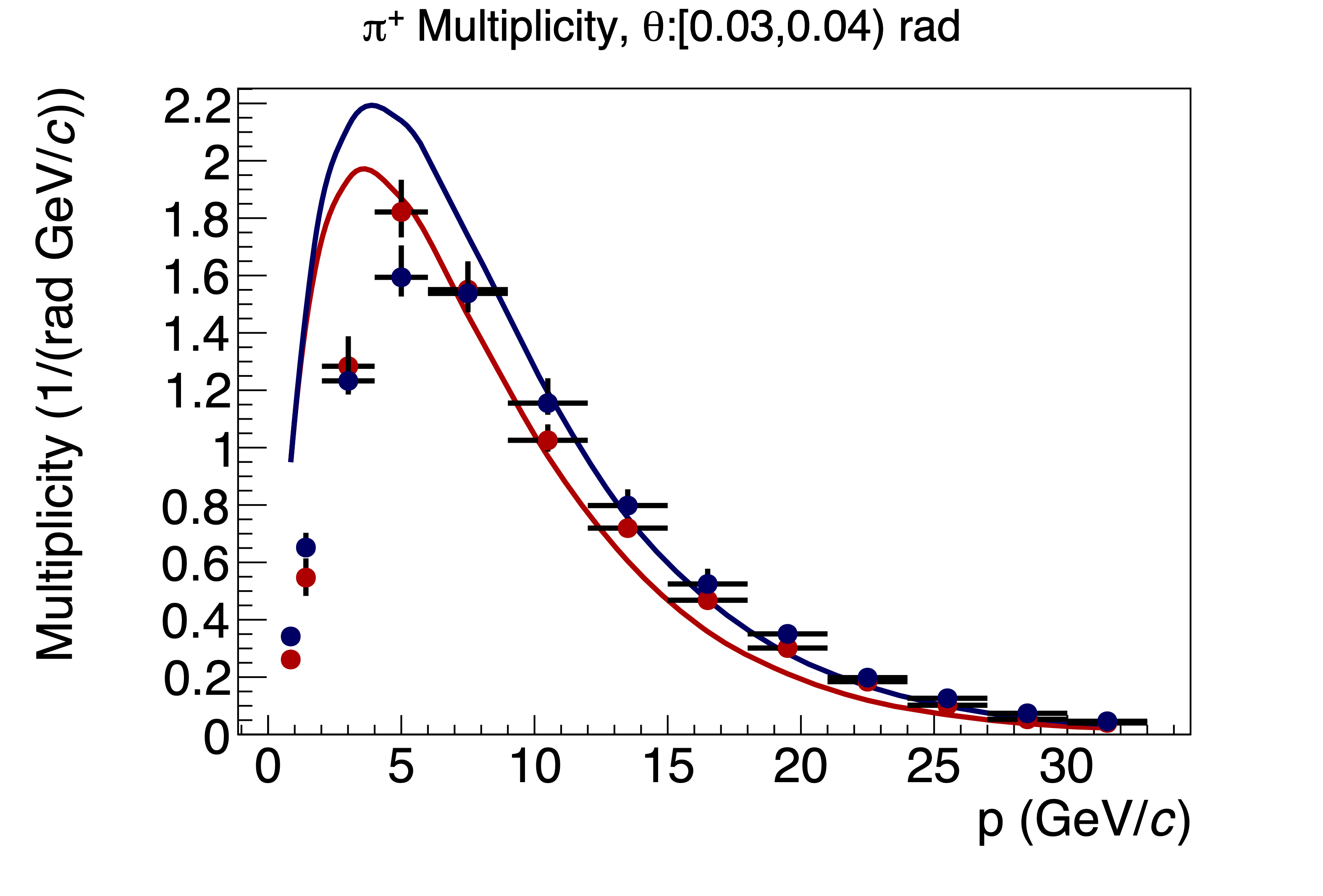}
    \includegraphics[width=0.49\textwidth]{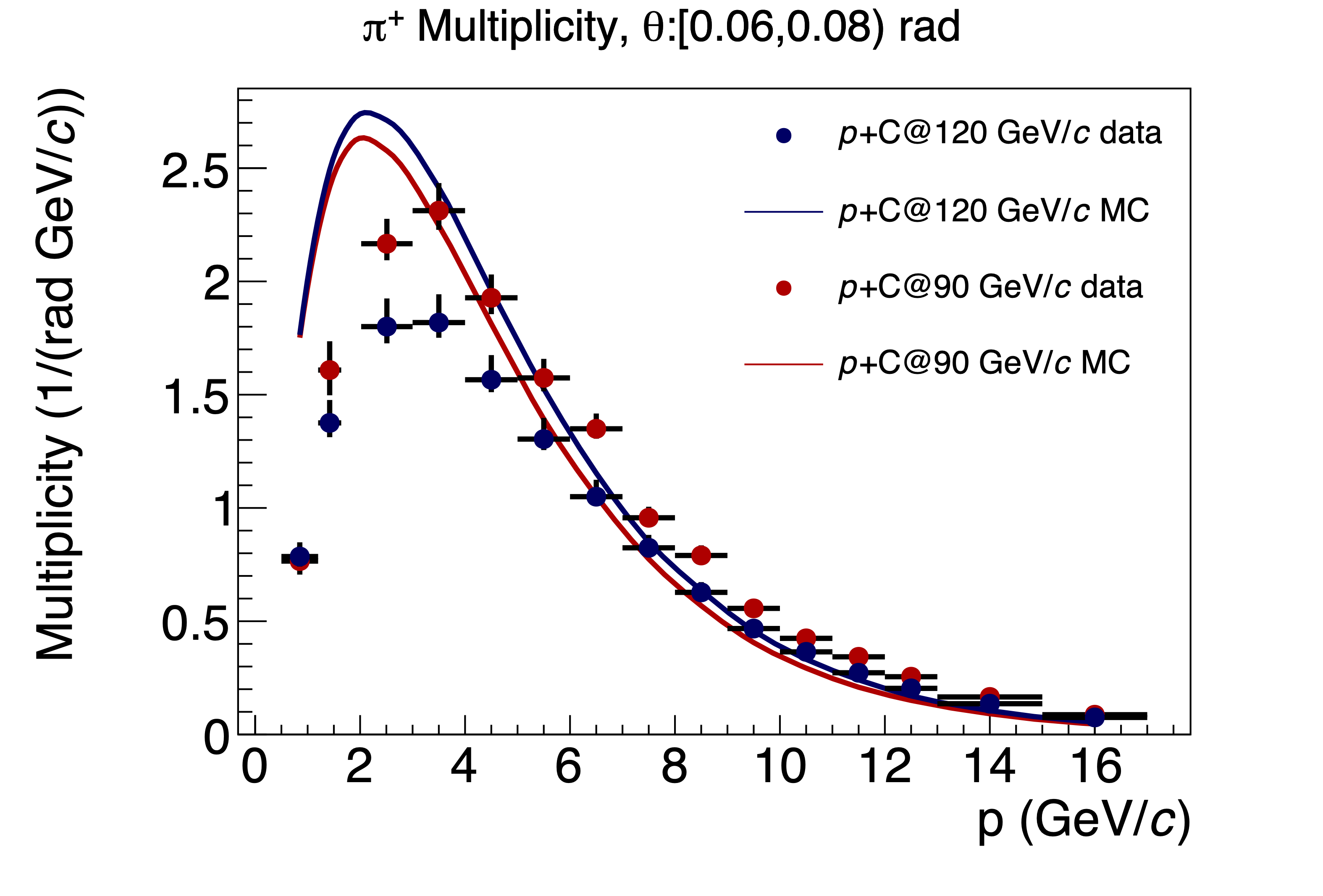}
\caption{Example comparison in the measured differential multiplicity between the 2017 90 GeV$/c$ proton-carbon data analyzed in this manuscript and the 2016 and 2017 120 GeV$/c$ proton-carbon data~\cite{adhikary2023measurementscharged}.  The Monte Carlo lines are \GeantFour version 10.7.0, using the FTFP\_BERT physics list.}
\label{fig:pionMultiplicityComparison}
\end{figure*}

Figure \ref{fig:pionMultiplicityComparison} shows an example comparison in the measured differential multiplicities for $\pi^{+}$ between the 2017 90 GeV$/c$ proton-carbon data analyzed in this manuscript and the 2016 and 2017 120 GeV$/c$ proton-carbon data \cite{adhikary2023measurementscharged}.
The Monte Carlo predicts that the differences between 90 GeV/$c$ and 120 GeV/$c$ are more significant at smaller angles.  In the comparison plots, the 120 GeV/$c$ data is generally above the 90 GeV/$c$ data in the forward direction. At higher angles the 120 GeV/$c$ data set is sometimes lower than the 90 GeV data set, which is not expected. NA61/SHINE collected a much higher statistics 120 GeV/$c$ dataset in 2023 with several different magnetic field configurations and plans to refine the 120 GeV/$c$ measurements in the future.

Numerical results of the multiplicity measurements, including statistical, systematic, and total uncertainties for each kinematic bin are summarized in CERN EDMS \cite{pC90EDMS}. Covariance matrices for each analysis are included as well.

The results presented can be used to improve the accuracy of neutrino beam flux simulations for current and planned long-baseline neutrino oscillation experiments, and they are particularly important for constraining secondary and tertiary interactions occurring in the creation of neutrino beams for experiments based in Fermilab. In addition, the results can be used to improve the modeling of proton-nucleus interactions.

There was no single uncertainty in the neutral- or charged-hadron analyses that was significantly larger than the other uncertainties in all kinematic bins. In most kinematic bins, the reconstruction, fit, and selection uncertainties
were the largest systematic uncertainties. For the charged-hadron analysis, the production cross-section uncertainty was also often one of the largest systematics; a precise measurement of the quasi-elastic component of the interaction cross section would significantly constrain this uncertainty.


\section*{Acknowledgments}
We would like to thank the CERN EP, BE, HSE and EN Departments for the
strong support of NA61/SHINE.

This work was supported by
the Hungarian Scientific Research Fund (grant NKFIH 138136\slash137812\slash138152 and TKP2021-NKTA-64),
the Polish Ministry of Science and Higher Education
(DIR\slash WK\slash\-2016\slash 2017\slash\-10-1, WUT ID-UB), the National Science Centre Poland (grants
2014\slash 14\slash E\slash ST2\slash 00018, 
2016\slash 21\slash D\slash ST2\slash 01983, 
2017\slash 25\slash N\slash ST2\slash 02575, 
2018\slash 29\slash N\slash ST2\slash 02595, 
2018\slash 30\slash A\slash ST2\slash 00226, 
2018\slash 31\slash G\slash ST2\slash 03910, 
2020\slash 39\slash O\slash ST2\slash 00277), 
the Norwegian Financial Mechanism 2014--2021 (grant 2019\slash 34\slash H\slash ST2\slash 00585),
the Polish Minister of Education and Science (contract No. 2021\slash WK\slash 10),
the European Union's Horizon 2020 research and innovation programme under grant agreement No. 871072,
the Ministry of Education, Culture, Sports,
Science and Tech\-no\-lo\-gy, Japan, Grant-in-Aid for Sci\-en\-ti\-fic
Research (grants 18071005, 19034011, 19740162, 20740160 and 20039012,22H04943),
the German Research Foundation DFG (grants GA\,1480\slash8-1 and project 426579465),
the Bulgarian Ministry of Education and Science within the National
Roadmap for Research Infrastructures 2020--2027, contract No. D01-374/18.12.2020,
Serbian Ministry of Science, Technological Development and Innovation (grant
OI171002), Swiss Nationalfonds Foundation (grant 200020\-117913/1),
ETH Research Grant TH-01\,07-3, National Science Foundation grant
PHY-2013228 and the Fermi National Accelerator Laboratory (Fermilab),
a U.S. Department of Energy, Office of Science, HEP User Facility
managed by Fermi Research Alliance, LLC (FRA), acting under Contract
No. DE-AC02-07CH11359 and the IN2P3-CNRS (France).

The data used in this paper were collected before February 2022.\\

\section*{Data Availability}
The data that support the findings of this article are openly available in~\cite{pC90EDMS}.

\bibliography{main}

\end{document}